\documentclass[twocolumn,useAMS,usenatbib]{mn2e}
\usepackage{graphicx,dcolumn,bm}
\usepackage{natbib}
\usepackage{multirow}
\usepackage{latexsym,amssymb}
\usepackage{amsmath}
\newcommand{\beq}{\begin{equation}}
\newcommand{\eeq}{\end{equation}}
\newcommand{\beqa}{\begin{eqnarray}}
\newcommand{\eeqa}{\end{eqnarray}}

\newcommand{\rmd}{\mathrm{d}}

\title[Lensing photoz calibration]{Precision photometric redshift
  calibration for galaxy-galaxy weak lensing\thanks{Based in part on
  observations undertaken at the European Southern Observatory (ESO)
  Very Large Telescope (VLT) under Large Program 175.A-0839.}}

\author[Mandelbaum et al.]
{R. Mandelbaum$^1$\thanks{{\tt rmandelb@ias.edu}, Hubble Fellow}, 
U. Seljak$^{2, 3}$\thanks{\tt seljak@itp.uzh.ch},
C.~M. Hirata$^4$, 
S. Bardelli$^5$, 
M. Bolzonella$^5$, 
\newauthor
A. Bongiorno$^{5,6}$, 
M. Carollo$^{7}$, 
T. Contini$^{8}$, 
C.~E. Cunha$^{9,10}$, 
B. Garilli$^{11}$, 
A. Iovino$^{12}$, 
\newauthor
P. Kampczyk$^{7}$, 
J.-P. Kneib$^{13}$, 
C. Knobel$^{7}$, 
D.~C. Koo$^{14}$, 
F. Lamareille$^{8}$, 
O. Le F\`{e}vre$^{13}$,
\newauthor
J.-F. Leborgne$^{8}$,  
S.~J. Lilly$^{7}$, 
C. Maier$^{7}$, 
V. Mainieri$^{15}$, 
M. Mignoli$^6$, 
J.~A. Newman$^{16}$, 
\newauthor
P.~A. Oesch$^{7}$,  
E. Perez-Montero$^{8}$, 
E. Ricciardelli$^{17}$, 
M. Scodeggio$^{9}$, 
J. Silverman$^{18}$, 
L. Tasca$^{13}$
\\$^1$Institute for Advanced Study, Einstein Drive, Princeton NJ
  08540, USA 
\\$^2$Institute for Theoretical Physics, University of Zurich, Zurich Switzerland
\\$^3$Department of Physics, University of California, Berkeley, CA 94720, USA
\\$^4$Mail Code 130-33, Caltech, Pasadena, CA 91125, USA; $^5$INAF Osservatorio Astronomico di Bologna, Bologna, Italy 
\\$^6$Dipartimento di Astronomia, Universit\`{a} degli Studi di
Bologna, Bologna, Italy 
\\$^{7}$Institute of Astronomy, Department of Physics, ETH Zurich,
CH-8093, Switzerland
\\$^{8}$Laboratoire d'Astrophysique de l'Observatoire
Midi-Pyr\'{e}n\'{e}es, Toulouse, France
\\$^{9}$Department of Astronomy and Astrophysics, University of
Chicago, Chicago, IL 60637, USA
\\$^{10}$Kavli Institute for Cosmological Physics, University of
Chicago, Chicago, IL 60637, USA 
\\$^{11}$INAF-IASF Milano, Milan, Italy; $^{12}$INAF Osservatorio
Astronomico di Brera, Milan, Italy; $^{13}$Laboratoire d'Astrophysique de Marseille, France 
\\$^{14}$UCO/Lick Observatory and Department of Astronomy and Astrophysics,
      University of California, Santa Cruz, CA 95064 USA
\\$^{15}$European Southern Observatory, Garching, Germany; $^{16}$Physics and Astronomy Dept., University of Pittsburgh,
Pittsburgh, PA, 15260
\\$^{17}$Dipartimento di Astronomia, Universita di Padova, Padova, Italy 
\\$^{18}$Max Planck Institut f\"{u}r Extraterrestrische Physik,
Garching, Germany  
}

\date{\today}

\begin{document}
\maketitle

\begin{abstract}
Accurate photometric redshifts are among the key
requirements  for precision weak lensing measurements.  Both the large size
of the Sloan Digital Sky Survey (SDSS) and the existence of 
large spectroscopic redshift samples that are flux-limited beyond its
depth 
have made it the optimal data source for developing methods to
properly calibrate photometric redshifts for
lensing.  Here, we focus on galaxy-galaxy
lensing  in a survey with spectroscopic lens redshifts, as in 
the SDSS.  We develop statistics that quantify the effect of source
redshift errors on the lensing calibration and on the weighting
scheme, and show how they can be used in the presence of redshift
failure and sampling  
variance.   We then demonstrate their use with  $ 2838$ 
source galaxies with spectroscopy from DEEP2 and zCOSMOS, evaluating 
several public photometric redshift
algorithms, in two cases including a full $p(z)$ for each object, and find
lensing calibration biases as low as $<1$\% (due to fortuitous
cancellation of two types of bias) or as high as 20\% for methods in
active use (despite the small 
mean photoz bias of these algorithms).  Our work
demonstrates that lensing-specific statistics must be used to
reliably calibrate the lensing signal, 
 due to asymmetric effects of (frequently non-Gaussian) photoz errors.
 We also demonstrate that large-scale structure (LSS) can 
strongly impact the photoz calibration and its error estimation, 
due to a correlation between the LSS and the photoz errors, 
and argue that at least two independent degree-scale spectroscopic 
samples are needed to suppress its effects.  Given the size of our
spectroscopic sample,   
we can  reduce the galaxy-galaxy lensing calibration error well below
current SDSS statistical errors. 
\end{abstract}

%
\vspace{-0.3in}

\section{Introduction}
\label{S:intro}

Galaxy-galaxy lensing is the deflection of light from distant source
galaxies due to the matter in more nearby lens galaxies.    
In the weak regime, gravitational lensing induces 0.1--10\% level 
tangential shear distortions of the shapes of background galaxies 
around foreground galaxies, allowing
direct measurement of the galaxy-matter correlation function around 
galaxies.  Due to the very small signal, typical
measurements involve stacking thousands of lens galaxies to get an
averaged lensing signal. 

Since the initial detections of galaxy-galaxy (g-g) lensing \citep{1984ApJ...281L..59T,1996ApJ...466..623B,1998ApJ...503..531H,2000AJ....120.1198F,2001ApJ...551..643S,2001astro.ph..8013M}, it has been used to address a wide variety of
  astrophysical questions using data from numerous sources. 
  These applications include (but are not limited to) determining the
  relation between stellar mass, luminosity, 
  and halo mass to constrain models of galaxy formation \citep{2005ApJ...635...73H,2006MNRAS.371L..60H,2006MNRAS.368..715M}; understanding
  the relation between halo mass from lensing and bias from galaxy
  clustering to constrain cosmological parameters
  \citep{2004AJ....127.2544S,2005PhRvD..71d3511S}; 
  measuring galaxy density profiles
  \citep{2004ApJ...606...67H,2006MNRAS.372..758M}; and understanding
  the extent of tidal stripping of the matter profiles of cluster
  satellite galaxies
  \citep{2002ApJ...580L..11N,2007A&A...461..881L}. In the future, 
galaxy-galaxy lensing will be used for geometrical tests that
  constrain the scale factor $a(t)$ and curvature $\Omega_K$ of the
  Universe \citep{2003PhRvL..91n1302J, 2004ApJ...600...17B,
  2006ApJ...637..598B}.  As data continue to pour in, and 
  future surveys are planned with even greater statistical 
  power, the time has come to place galaxy-galaxy lensing on a firmer
  foundation by addressing systematics to greater precision.

The g-g lensing signal calibration depends on several systematics,
including the calibration of the shear 
\citep{2006MNRAS.368.1323H,2007MNRAS.376...13M} and theoretical
uncertainties such as galaxy intrinsic alignments
\citep{2006ApJ...644L..25A,2006MNRAS.370.1422A,2006MNRAS.371..750H,2006MNRAS.372..758M,2007ApJ...662L..71F}, 
both areas in which there is significant ongoing work.  Here, 
we focus on the proper calibration of the source redshift distribution
for galaxy-galaxy lensing 
in the case where all lens redshifts are known.  
The SDSS has the rather unique capability of offering spectroscopic
redshifts for all lenses, which both removes any calibration bias due
to error in lens redshift estimation, and also allows us to compute
the signal as a function of physical transverse (instead of angular)
separation from the lenses, simplifying theoretical interpretation.
While several theoretical studies have estimated the effects of photoz
errors for shear-shear autocorrelations
\citep{2006MNRAS.366..101H,2006ApJ...636...21M, 2007arXiv0705.1437A, 2007arXiv0712.1562B}, we present the first 
such analysis for galaxy-galaxy lensing, in which we not only offer  
statistics to use to evaluate the calibration bias, but also carry 
out an analysis with attention to practical issues such as sampling
variance in the calibration sample.  This work will therefore enable
future g-g lensing analyses 
with other datasets to address other scientific questions, and reveal
potential issues with spectroscopic calibration of photoz's that are
more general than just g-g lensing.   
We also address the extension of these techniques to galaxy-galaxy lensing
without lens redshifts, and to cosmic shear, in
Appendix~\ref{S:extension}. 

Currently, there are two methods used for source redshift
determination in g-g lensing.  The first is the use of an average
redshift distribution for 
the sources.  The primary  difficulty with this method is
finding a sample of galaxies with
spectroscopy that has the same selection criteria as the source
galaxies.  Weak lensing requires well-determined shapes for each
source, so a lensing source catalog is not purely
flux-limited, and literature estimates of $\rmd N/\rmd z$ for
flux-limited samples may not be appropriate (we show in this paper
that for SDSS, the lensing-selected sample is at a higher mean
redshift  
than the corresponding flux-limited sample at fixed magnitude).  
The solution is to find a spectroscopic sample that
overlaps the source sample and is at least as deep, using it to 
determine the redshift 
distribution using only lensing-selected galaxies in the 
spectroscopic sample.  For deeper lensing surveys, no such
spectroscopic sample exists.  In other cases, it exists but may be
quite small, with large uncertainty in $\rmd N/\rmd z$ due to Poisson
error and, more 
significantly, large-scale structure.  The second difficulty is that
without individual redshift estimates for each source, 
there is no way 
to remove sources that are physically-associated with lenses from the
source sample, 
which can lead to dilution of the lensing signal by non-lensed
galaxies (a systematic 
that is easily controlled) and, more significantly, signal suppression
due to 
intrinsic alignments [which cannot yet be easily
controlled \citep{2006ApJ...644L..25A,2006MNRAS.372..758M}, and which
can cause contamination larger than the size of the statistical errors
for small 
transverse separations].

The second method is to use broad-band photometry to measure
photometric redshifts (photoz's) for each source galaxy.  Photoz
estimation exploits the fact that even with broad passbands, we
can still learn enough about the spectral energy distribution 
to estimate the redshift.  While photoz estimation that yields
accurate values over a wide range of redshifts for all galaxy types is
difficult, there have been several recent successes in this field
\citep{2006MNRAS.372..565F,2006A&A...457..841I}.  To fully constrain
the calibration of the g-g lensing 
signal, we must understand the full photoz error distribution as a function of
many parameters, particularly those relevant to galaxy-galaxy
lensing, such as brightness, colour, environment, and of course 
redshift.  Since the photoz error distributions will depend on a
complex interplay between the widths and shapes of the filter
functions, the set of filters used in the photoz estimates, the
photometry error distributions, and the spectral energy distributions of the
galaxies themselves, the photoz error distributions will not be
symmetric or Gaussian in general, even if the photometric errors
in flux are Gaussian (the magnitude errors are not in any case, and
some photoz methods use magnitudes instead of fluxes).  To be
accurate, this photoz error distribution must be determined with a
sample of galaxies with the same selection criteria (depth, colour,
etc.) as the source
sample.  This
is quite important because, as the photometry gets noisier, the
photoz error distribution can not just broaden, but can also develop
asymmetry, tails, and other non-Gaussian properties.  

So, as for
methods that use a statistical source
redshift distribution, we once again must find a
large  spectroscopic sample with the same selection criteria
as our source catalog.  (Some photoz methods also require a training
sample with the same selection criteria as the source sample.)  The
completeness and rate of spectroscopic redshift failure are both
potentially important, particularly if the spectroscopic redshift
failures all lie in a specific region of redshift or colour space.   If a
photoz method has a significant failure 
fraction, then we may be forced to eliminate
a large fraction of the source sample, thus increasing statistical
error significantly.  Three major advantages of photoz's for lensing are
that 
they (1) allow us to eliminate some fraction of the physically-associated
lens-source pairs, thus reducing the effects of intrinsic alignments, 
(2) allow us to optimally weight each galaxy by the expected signal,
and (3) allow us to reduce, if not eliminate, ``sources'' that are in
the foreground from 
the sample entirely (a special case of optimal weighting).

We present a method to obtain robust, 
percent-level calibration of the g-g lensing signal using a
sample of several thousand spectroscopic redshifts selected from the
source sample (i.e., with the same selection criteria). The sources of
spectroscopy we use to demonstrate this method are the DEEP2 and
zCOSMOS surveys (described in 
section~\ref{S:data}).  The use of two surveys in two areas of the sky
 carried out with  two different telescopes is important, because (a)
 they do 
 not have the same patterns of redshift failure, and (b) the 
large-scale structure in the two surveys is not correlated with each
other, so effects of sampling (cosmic) variance are reduced for the
combined sample.  In addition, 
we use space-based data for the  full COSMOS sample 
to quantify the efficacy of our star/galaxy separation scheme.

We then use this method to analyze the redshift-related
calibration bias of the lensing signal in previous g-g lensing
analyses that used our 
SDSS source catalog
\citep{2004MNRAS.353..529H,2005MNRAS.361.1287M,2005PhRvD..71d3511S,2006MNRAS.368..715M,2006MNRAS.372..758M,2006MNRAS.370.1008M,2007JCAP...06...24M}. 
Our calibration bias analysis is quite 
important, as our statistical error for some applications has dropped
below 5\%, making our systematics requirements more stringent.  

More importantly, we take a broad
view, testing not just the redshift determination methods that we have
used in the past, but also several new ones that have been developed
in the past few years, in order to determine which ones are most
useful for lensing.  In the process, we determine which common photoz
failure modes and error distributions are most problematic for g-g
lensing.  The results of our
analysis will be useful not only for SDSS g-g lensing,
and the method we present is generally useful for future
weak lensing analyses (and generalizable to scenarios without
spectroscopy for lenses and to shear-shear autocorrelations),
particularly as larger, deeper spectroscopic datasets are becoming
available. 

In section~\ref{S:data}, we describe the lensing source
catalog and the spectroscopic redshift samples.
Section~\ref{S:zdetermination} includes a description of the source
redshift determination algorithms that we will test in this work.  In
section~\ref{S:method}, we describe our method for determining the
source redshift-related calibration bias, including handling
complexities such as 
large-scale structure.  We present the results of our analysis in
section~\ref{S:results}, and discuss the implications of these results
in section~\ref{S:discussion}.

When computing angular diameter distances, we assume a flat cosmology
with $\Omega_m=0.27$ and $\Omega_{\Lambda}=0.73$. 

\section{Data}
\label{S:data}

%
\subsection{SDSS}

The data used for the lensing source catalog are obtained from the SDSS
\citep{2000AJ....120.1579Y}, an ongoing survey to image roughly
$\pi$ steradians of the sky, and follow up approximately one million of
the detected objects spectroscopically \citep{2001AJ....122.2267E,
2002AJ....123.2945R,2002AJ....124.1810S}. The imaging is carried out
by drift-scanning the sky 
in photometric conditions \citep{2001AJ....122.2129H,
2004AN....325..583I}, in five bands ($ugriz$) \citep{1996AJ....111.1748F,
2002AJ....123.2121S} using a specially-designed wide-field camera
\citep{1998AJ....116.3040G}. These imaging data are used to create the
source catalog that we use in this paper. In addition, objects are targeted for
spectroscopy using these data \citep{2003AJ....125.2276B} and are observed
with a 640-fiber spectrograph on the same telescope
\citep{2006AJ....131.2332G}. All of these data are 
processed by completely automated pipelines that detect and measure
photometric properties of objects, and astrometrically calibrate the data
\citep{2001ASPC..238..269L, 2003AJ....125.1559P,2006AN....327..821T}. The SDSS is well
underway, and has had seven major data releases \citep{2002AJ....123..485S,
2003AJ....126.2081A, 2004AJ....128..502A, 2005AJ....129.1755A,
2004AJ....128.2577F, 2006ApJS..162...38A, 2007ApJS..172..634A,2007arXiv0707.3413A}. 

The source sample we describe was originally presented in
\cite{2005MNRAS.361.1287M}, hereinafter M05.  It 
includes over 30 million galaxies from the SDSS imaging data with
$r$-band model magnitude brighter than 21.8. 
Shape measurements are obtained using the REGLENS pipeline, including PSF
correction done via re-Gaussianization \citep{2003MNRAS.343..459H} and
with selection criteria designed to avoid various shear calibration
biases.  A full 
description of this pipeline can be found in M05.

\subsection{DEEP2}

The DEEP2 Galaxy Redshift Survey  \citep{2003SPIE.4834..161D,2003ApJ...599..997M,2004ApJ...609..525C,2005ASPC..339..128D} consists of spectroscopic
observation of four fields using the DEep Imaging Multi-Object
Spectrograph (DEIMOS, \citealt{2003SPIE.4841.1657F}) on the Keck
Telescope.  This paper uses data 
from field $1$, the 
Extended Groth Strip (EGS), centered at RA $14^{\mathrm h}17^{\mathrm m}$, 
Dec. $+52^{\circ}\,30'$ (J2000) and with dimensions $120'\times 
15'$ \citep{2007ApJ...660L...1D}.   
Galaxies brighter than $R_{AB}=24.1$ were observed in all four DEEP2
fields, but in the other three fields besides EGS, two colour cuts were
made to exclude galaxies with redshifts below $z\sim 0.7$.  The DEEP2
EGS sample, in contrast, includes objects of all colours with $R_{AB} <  
24.1$, although colour-selected $z<0.75$ objects with $21.5 < 
R_{AB} < 24.1$ receive 
slightly lower selection weight.  This is the sample from which a
bright  subset, $r < 
21.8$, was extracted for this paper.  The selection probabilities for 
all objects are well-known, allowing us to account for this 
deweighting directly, though this has little impact for this study, since only a 
small fraction of galaxies with useful SDSS shape measurements are 
fainter than $R=21.5$, and they have little statistical weight  due to 
their larger shape measurement errors. Due to
saturation of the CFHT detectors used for 
target selection, no galaxies brighter than $R_{AB}\approx 17.6$ were
targeted; these galaxies  constitute a very small fraction of our source sample.

For this paper, we use all EGS data collected through the spring of
2005, a parent catalog of more than 13~000 spectra \citep{2007ApJ...660L...1D}.
 The $155$ DEEP2 EGS objects with $r < 21.8$ (the
        limit of our source catalog) that failed to yield redshifts in
        initial DEEP2 analyses were reexamined in detail; after this
        effort, the net redshift success rate (defined as DEEP2 quality 3
        or 4) was 96\%, significantly higher than for the full EGS
        sample. The positions of the DEEP2 EGS matches in our source
        catalog are shown in the right panel of
        Fig.~\ref{F:radec}. There are $\sim 1530$ 
        SDSS galaxies in this region with matches in DEEP2 at $r <
        21.8$. Roughly 65\% of those pass the lensing selection, leaving us
        with a sample of $1013$. 

\subsection{zCOSMOS}

The other redshift survey used for this work is zCOSMOS
\citep{2007ApJS..172...70L}, which uses
the Visible Multi-Object Spectrograph (VIMOS,
\citealt{2003SPIE.4841.1670L}) on the 8-m European Southern Observatory's
Very Large Telescope (ESO VLT) to obtain spectra for galaxies
in the COSMOS field, which is 1.7 deg$^2$ centered at RA $10^{\mathrm
  h}$, Dec. $+2^{\circ}\,12'\,21''$.  We
use data from the zCOSMOS-bright survey, which is purely flux-limited
to $I_{AB}=22.5$, well beyond the flux-limit of our source catalog,
and currently contains $\sim 10^4$ galaxies (Lilly et~al., in prep.).
Observations began in 2005 and will take at least three years to
complete.  

One important benefit of the zCOSMOS data is that due to its
location in the Cosmological Evolution Survey (COSMOS) field
\citep{2007ApJS..172...99C,2007ApJS..172....1S,2007ApJS..172...38S,2007ApJS..172....9T},
there is very deep broadband 
observing data from a variety of telescopes in addition to a single
passband observation from the Advanced Camera for Surveys (ACS) on the Hubble
Space Telescope (HST).  This photometry has been
used to generate extremely high-quality photometric redshifts using
the Zurich Extragalactic Bayesian Redshift Analyzer (ZEBRA,
\citealt{2006MNRAS.372..565F}), which will be described further in
Section~\ref{S:zdetermination}, and several other photoz codes
\citep{2007ApJS..172..117M}.  Using data with $u*$, $B$, $V$, $g'$, 
$r'$, $i'$, $z'$, and $K_s$ photometry,
the photometric redshift 
accuracy for the bright, $I$-selected sample is remarkable,
$\sigma_{\Delta z/(1+z)} < 
0.03$. This accuracy is achieved using $10$\% of the zCOSMOS sample as
a training set.  In cases of spectroscopic redshift failure, these
nearly noiseless photoz's can be 
used instead. We will demonstrate explicitly that the effect on the
estimated lensing redshift
calibration bias of using their photoz's for redshift failures is
within the statistical error. 
Consequently, the nominal 8\% spectroscopic redshift failure rate for zCOSMOS
galaxies in our source catalog is effectively zero for
our purposes.

The HST imaging in the full COSMOS field was also used for another
test because it enables 
star/galaxy separation to be performed more accurately than in SDSS.
Consequently, we use the full COSMOS galaxy sample to match
against our source catalog and identify the stellar contamination fraction to
high accuracy. 

\begin{figure}
\begin{center}
\includegraphics[width=3.0in,angle=0]{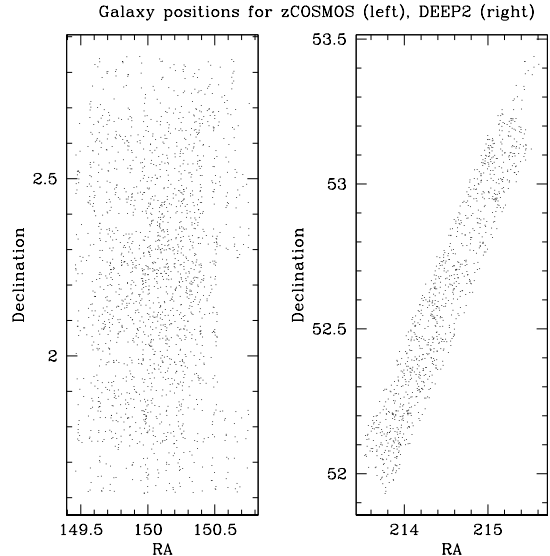} 
\caption{\label{F:radec}Positions of the zCOSMOS (left)
  and DEEP2 (right) spectroscopic galaxies used in this work.}
\end{center}
\end{figure}

The positions of the zCOSMOS 
matches in our source catalog is shown in the left panel of
Fig.~\ref{F:radec}.  We  have spectra in an area covering 
$\sim 1.5$ square degrees, 88\% of the eventual area of the zCOSMOS
survey.  The sampling is denser in some regions than in others
(and will eventually be filled out evenly in the full area). In this
region, there are $\sim 3000$ SDSS galaxies with $r<21.8$; roughly 65\%
pass our lensing selection cuts, leaving us with $1825$ matches in the
source catalog.

\section{Redshift determination algorithms}\label{S:zdetermination}

Here we describe the source redshift determination algorithms in more
detail.  We begin with those used in our current lensing source
catalog, for which we want to assess calibration biases in past works,
then describe methods that have more recently become available. 

\subsection{Previous methods}

In our catalog, which was created in 2004, we used three
approaches to source redshift determination, all described in detail
in M05. 
For the $r<21$ sources, we used photometric redshifts from 
  kphotoz v3\_2 \citep{2003AJ....125.2348B} and
their error distributions determined using a sample of $162$ 
galaxies in the DEEP2 EGS.  We also required 
$z_p>z_l+0.1$ to avoid contamination from physically-associated
lens-source pairs.  For the $r>21$ sources, we used a source redshift
distribution from DEEP2 EGS (from fitting to $116$ redshifts), which
means that we lack individual redshift estimates for each
source.  The sample of redshifts used for this early work
with the EGS was a factor of $3.5$ smaller than the EGS
sample used for this work, or a factor of $10$ smaller than the combined
EGS $+$ zCOSMOS sample used here.  For the high-redshift LRG source
sample (see selection criteria in M05), we used well-calibrated 
photometric redshifts and their error distributions determined using
data from the 2dF-SDSS LRG and Quasar Survey (2SLAQ), as presented in
\cite{2005MNRAS.359..237P}.  

\subsection{New options}

There are several relatively new photoz options for SDSS data, all of
which have relatively low failure rates of $\sim5$\%.  The
first is available in the SDSS DR5 (data release 5) skyserver ``Photoz''
table  
\citep{2000AJ....120.1588B,2003AJ....125..580C}. 
The photoz's for this template method are determined by fitting observed galaxy colours to
empirical templates 
from \cite{1980ApJS...43..393C} extended using spectral synthesis
models. There is an additional step (not used for all template
methods) in which the templates are iteratively adjusted using a
training sample.  We have
performed our tests on both the DR5 and DR6 template photoz's, and
found no significant differences in performance between the two.

The second new option is available in the SDSS DR6 skyserver in the
``Photoz2'' table.  These photoz's were computed using a neural net
(NN) 
algorithm similar to that of \cite{2004PASP..116..345C} trained using
a training set from many data sources 
combined: SDSS spectroscopic samples, 2SLAQ, CFRS, CNOC2, DEEP,
DEEP2, and GOODS-N.  A more complete description of both NN photoz's
in the DR6 database can be found in
\cite{2007arXiv0708.0030O}: the ``CC2'' photoz's use colours and
concentrations, while the ``D1'' photoz's use magnitudes and
concentrations.  In the text, we will describe any difference between
the DR5 and DR6 results; \cite{2007arXiv0708.0030O} recommends against
using the DR5 photoz's for science applications now that the improved
DR6 versions exist.  

The third new option we test is the ZEBRA 
\citep{2006MNRAS.372..565F} algorithm, which has already been
successfully used with much deeper imaging data in the
COSMOS field.   This method involves template-fitting, but also takes
a flux-limited sample of galaxies (without spectroscopic redshifts)
from the data source for which we want photoz's.  These data are used
to create a Bayesian modification of the likelihoods based on the
$N(z)$ for the full sample \citep{2006ApJS..162...20B} and on its
template distribution.  In practice,
this prior helps avoid scatter to low 
redshifts.    A key question we will address is how this algorithm 
behaves with the significantly noisier SDSS photometry.  To avoid
confusion, we will refer to the high-quality ZEBRA photoz's derived
using the deep photometry in the COSMOS field as ``ZEBRA'' photoz's,
and the ZEBRA photoz's using the much shallower SDSS photometry as
``ZEBRA/SDSS'' photoz's.

To be specific about the training method, to get the ZEBRA/SDSS
photoz's, half of a flux-limited 
sample of SDSS galaxies with zCOSMOS redshifts are used for template
optimization.  This part of the analysis includes fixing the redshifts
of those galaxies to the spectroscopic redshift, finding the
best-fitting template, and optimizing it as described in
\cite{2006MNRAS.372..565F}.  Then, a sample of $10^5$ SDSS galaxies
(flux-limited to $r=22$) without spectra were used to iteratively
compute the template-redshift prior. 

\subsection{Effects of photoz error for lensing}\label{SS:biasscatter}

Finally, we clarify the effects of photoz error on the lensing
calibration:
\begin{itemize}
\item A positive photoz bias, defined as a nonzero $\langle
  z_p-z\rangle$, will lower the signal (because the critical surface
  density, defined below in Eq.~\ref{E:Sigmac}, 
  will be underestimated).
\item A negative photoz bias will raise the signal.
\item Photoz scatter will usually lower the signal due to the shape of
  the critical surface density near $z_l$.  This effect can be very
  significant for sources at redshifts below $\sim z_l+1.5\sigma$, 
  where $\sigma$ is the size of the scatter.
\end{itemize}

The last point is very important for a shallow survey like SDSS when
the lens redshift is above $z_l\sim 0.1$, because of the large number of
sources within a few $\sigma$ of the lens redshift.  For a deeper
survey such as the Canada-France-Hawaii Telescope Legacy Survey (CFHTLS),
with lenses and sources separated by $\Delta z \sim 0.5$ on average
this effect may in fact be negligible.  The effects of
photoz bias are important not just in the mean, but as a function of
redshift.  If low redshift sources have nonzero photoz bias, and
high redshift sources have nonzero photoz bias in the opposite
direction,  so that the mean photoz bias for the full sample is zero,
the effect of the opposing photoz biases on lensing calibration will
not, in general, cancel out since the effect on lensing calibration 
tends to be more significant for the sources that are closer to the lenses.

Catastrophic photoz errors are those that are well beyond the typical
scatter, typically occurring due to some systematic error, colour-redshift
degeneracy, or other problem (and by definition, these photoz's are
not flagged as problematic by the algorithm, so they can only be
identified using a spectroscopic sample with similar selection to the
target sample).  The catastrophic error rate may be
important, depending on the type of 
catastrophic error.  For example, sending a few percent of the sources to $z_p=0$
will not lead to calibration bias, it will simply lead to that
fraction of the sources not being included because they have
$z_p<z_l$, causing a percent-level increase in the final error.  In
short, it is clear that the three metrics often used to quantify
the accuracy of photoz methods -- the mean bias, scatter, and
catastrophic failure rate -- are not sufficient to quantify the
efficacy of a photoz method for lensing. In this paper, we will
introduce a metric that is optimized towards understanding the effects
of photoz's on galaxy-galaxy lensing calibration, and present results
for the photoz mean bias, scatter, and catastrophic failure rate only
as a means of understanding the results for our lensing-optimized
metric. For other science applications, the optimal metric may be
quite different from what we present here.

\section{Methodology}
\label{S:method}

\subsection{Theory}\label{SS:theory}

Galaxy-galaxy lensing measures the tangential shear 
distortions in the shapes of background galaxies 
induced by the mass distribution around foreground galaxies (for a
review, see \citealt{2001PhR...340..291B}). 
The result is a measurement of the shear-galaxy cross-correlation as a 
function of relative foreground-background separation on the sky. 
We will assume that the redshift of the foreground galaxy 
is known, so we express the relative 
separation in terms of transverse comoving scale $R$. 
One can relate the shear distortion $\gamma_t$ to
$\Delta\Sigma(R)=\bar{\Sigma}(<R)-\Sigma(R)$, 
where $\Sigma(R)$ is the 
surface mass density at the transverse separation $R$ and $\bar{\Sigma}(<R)$ 
its mean within $R$, via
\begin{equation}\label{E:defgamma}
\gamma_t=\frac{\Delta\Sigma(R)}{\Sigma_{c}}.
\end{equation}
Here we use the critical mass surface density, 
\begin{equation}
\Sigma_{c}=\frac{c^2}{4\pi G} \frac{D_S}{(1+z_L)^2 D_L D_{LS}},
\label{E:Sigmac}
\end{equation}
where $D_L$ and $D_S$ are angular diameter distances to the lens and
source, $D_{LS}$ is the angular diameter distance between the lens
and source, and the factor of $(1+z_L)^{-2}$ arises due to our use of
comoving coordinates.  For a given lens redshift,
$\Sigma_c^{-1}$ rises from zero at $z_s = z_L$ to an asymptotic value
at $z_s \gg z_L$; that asymptotic value is an increasing function of
lens redshift.  

In this work, we focus on calibration bias in $\Delta\Sigma$ due
to bias in $\Sigma_c$ arising from source redshift uncertainty.

\subsection{Redshift calibration bias determination}

Here, we present a method for testing the accuracy of source redshift
determination that is optimized towards g-g lensing.   
Formally, we wish to calculate the differential surface density
$\Delta\Sigma$ using our estimator $\widetilde{\Delta\Sigma}$, which
is  defined as a weighted sum over lens-source pairs $j$,
\begin{equation}\label{E:defhatds}
\widetilde{\Delta\Sigma} = \frac{\sum_j \tilde{w}_j \tilde{\gamma}_t^{(j)} \tilde{\Sigma}_{c,j}}{\sum_j \tilde{w}_j}.
\end{equation}
To isolate the
dependence of calibration on redshift-related quantities, we will
assume that the estimated 
tangential shear, $\tilde{\gamma}_t$, is 
unbiased.  $\tilde{\Sigma}_{c,j}$ (derived from
our source redshift estimator) is the critical surface density
estimated for a given
lens-source pair $j$.  The weights for each
lens-source pair are determined using redshift information as well: 
\begin{equation}\label{E:wj}
\tilde{w}_j = \frac{1}{\tilde{\Sigma}_{c,j}^{2}(e_{\rm rms}^2 + \sigma_e^2)}
\end{equation}
where $e_{\rm rms}$ is the rms ellipticity per component for the
source sample (shape noise), and $\sigma_e$ is the ellipticity
measurement error per component. 

We want to relate our estimated $\widetilde{\Delta\Sigma}$ 
to the true $\Delta\Sigma$.  To do so, we use the relation between the
measured shear and $\Delta\Sigma$, Eq.~(\ref{E:defgamma}).  Putting
equation~\ref{E:defgamma} into 
equation~\ref{E:defhatds} (assuming $\langle \tilde{\gamma}\rangle =
\gamma$), we define the redshift calibration bias $b_z$ via  
\begin{equation}\label{E:defbz}
b_z + 1 = \frac{\widetilde{\Delta\Sigma}}{\Delta\Sigma} = \frac{\sum_j \tilde{w}_j
  \left(\Sigma_{c,j}^{-1}\tilde{\Sigma}_{c,j}\right)}{\sum_j \tilde{w}_j}, 
\end{equation}
a weighted sum of the ratio of the estimated to the true critical
surface density.

This expression must be computed as a function of lens redshift.  In
the limit that the sources are at much higher redshift than the
lenses, $\Sigma_c$ does not depend as strongly on the source redshift,
so (for a given photometric redshift bias) $|b_z|$ will be smaller
than if the lens redshift is just below 
the source redshift.  For a
lens sample with redshift distribution $p(z_l)$, the average
calibration bias $\langle b_z\rangle$ can be computed as a
weighted average over the redshift distribution,
\begin{equation}\label{E:avgbz}
\langle b_z\rangle = \frac{\int \rmd z_l \, p(z_l)\,\tilde{w}_l(z_l)\, b_z(z_l)}{\int
  \rmd z_l \, p(z_l)\, \tilde{w}_l(z_l)} 
\end{equation}
where the redshift-dependent lens weight $\tilde{w}_l(z_l)$ is defined as the
total weight derived from all sources that contribute to the lensing
signal for a given lens redshift, $\sum_j \tilde{w}_j$.

In the ideal case, we would do this calculation with a large, complete
spectroscopic sample drawn at random from our source sample, sparsely
sampled on the sky and therefore lacking features in the redshift
distribution due to large-scale structure. 
We can then find $b_z(z_l)$ on a grid of lens redshifts by forming
the sums in equation~\ref{E:defbz} using all sources with spectra.
Finally, we can use the total weight as a function of lens redshift
and the lens redshift distribution to estimate the
average redshift bias of the lensing signal.

To get the errors on the bias in this simple scenario, we can simply
bootstrap resample our sample of source galaxies with spectroscopy.  
For a sample of $N_{\rm gal}$ galaxies,  bootstrap
resampling requires us to make many ``new'' galaxy samples consisting 
of 
$N_{\rm gal}$ galaxies drawn from the original sample {\em with
  replacement}.  Assuming that the observed galaxy redshifts accurately
reflect the underlying redshift distribution, and the redshifts are
uncorrelated,
the mean best-fit redshift 
distribution will reflect the true one, and the errors in the
redshift calibration bias can be 
determined from the variance of the calibration biases for each
bootstrap resampled dataset.  Since the bootstrap 
depends on the assumption that the objects we are bootstrapping are
independent,  this method only gives proper errors in the
case where LSS is unimportant. 

In general, there are several problems that mean we are no longer
dealing with the ideal case.  The first problem is sampling variance, since most
redshift surveys are completed in a well-defined, small region of the
sky.  The second is the fact that most redshift surveys suffer from
some incompleteness, and that incompleteness may be a function of
apparent magnitude or colour, which means
that the loss of those redshifts can make the spectroscopic sample
no longer comparable to the full source sample.  We
attempt to ameliorate these problems  by using two sources
of spectroscopy on different areas of the sky and with different
spectrographs and analysis pipelines, so that the LSS and
incompleteness tendencies in each sample are different.  Below, we
address these deviations from the ideal case in more detail.  

\subsection{Effects of sampling variance}\label{SS:lss}

Large-scale structure can be problematic when using surveys on small
regions of the sky to determine bias in the lensing signal
due to photometric redshift error.  The
LSS may emphasize particular regions of the source
redshift distribution that have unusual features in the photometric
redshift errors.  To avoid
this problem, we would like to fit for a redshift distribution
in a way that accounts properly for uncertainties due to sampling
variance.  There are many approaches to this problem in the
literature, such as  that demonstrated in \cite{2006ApJS..162...20B}.

The simplest way around our aforementioned problem, that LSS causes
the redshifts to be correlated so that the assumption behind the
bootstrap is violated, is to bootstrap the bins in the redshift
histogram instead.  In the limit that the bins are significantly wider than
the typical sample correlation length, the
correlations within the bins will be far more important than the
correlations between adjacent bins.  Thus, the requirement that the
bootstrapped data points be independent is much closer to being
fulfilled.  Here, we will use redshift bins with size $\Delta
z=0.05$, where each bin is considered as a pair of points $(z_i,
N(z_i))$.    In a given bootstrapped histogram, some redshift bins
$(z_i, N(z_i))$ will
be included multiple times, others not at all, but each time a given
bin is used, it has the same number of galaxies as in the real data.
While this method is simplistic, it has the advantage of not requiring
us to understand the details of the sample selection, since the
lensing selection is a very non-trivial cut to understand and
simulate.  The resulting errors on the best-fit $N(z)$ from this
bootstrap will include the effects of both Poisson error (which is
non-negligible given the size of the samples used) and large-scale
structure.  The errors are valid assuming that there are no
correlations between the $150h^{-1}$Mpc-wide bins.  We discuss this
assumption, which depends not just on straightforward integration of
the matter power spectrum but also redshift-space distortions, galaxy bias, and
magnification bias, further in section~\ref{SS:errorsize}.   

%
%
For each bootstrapped histogram with bins centered at $z_i$ containing
$N_i$ galaxies each, we minimize the function
\begin{equation}\label{E:defdelta}
\Delta^2 = \sum_i w_i^{(\Delta)} [N_i - N_i^{\rm (model)}]^2.
\end{equation}
via summation over redshift bins $i$. 
$N_i^{\rm (model)}$ is the number of galaxies predicted to lie in
bin $i$ given the model for $\rmd N/\rmd z$, i.e.
\begin{equation}
N_i^{\rm (model)} = \int_{z_i-\Delta z/2}^{z_i+\Delta z/2} \frac{\rmd
  N}{\rmd z} dz.
\end{equation}
For each bootstrapped histogram, we also imposed a normalization
condition on the fit that $\int_{0}^{\infty} \rmd z (\rmd N^{\rm (model)}/\rmd
z)=N_{\rm gal}$ (the total number of galaxies in the spectroscopic
sample).  In the case of Poisson error, the natural choice for $w_i^{(\Delta)}$
is $1/N_i^{\rm (model)}$.  However, in the presence of LSS, which
contributes significantly to the variance in each bin, the
distribution of values in each bin is, in fact, unknown, so the optimal
weighting scheme is unclear.   Consequently, we
use the simplest possible weighting scheme, $w_i^{(\Delta)}=1$ for all $i$. We
have, however, confirmed that if we do use
$w_i^{(\Delta)}=1/N_i^{\rm (model)}$, then the changes in the best-fit
redshift distribution parameters, and the implied changes in redshift
calibration bias, are well below the $1\sigma$ level.

Our 2-parameter model for the redshift distribution is
\begin{equation}\label{E:zdist}
\frac{\rmd N}{\rmd z} \propto \left(\frac{z}{z_*}\right)^{\alpha-1}
  \exp{\left[-0.5(z/z_*)^2\right]}
\end{equation}
which has mean redshift
\begin{equation}\label{E:meanz}
\langle z \rangle = \frac{\sqrt{2}\, z_*\Gamma\left[(\alpha+1)/2\right]}{\Gamma\left(\alpha/2\right)}.
\end{equation}
This choice is based purely on the empirical observation that it
 describes the shape of the redshift distribution better than
the many other functional forms that we tried, and addition of extra
 parameters did not significantly improve the best-fit $\Delta^2$.  In
 particular, allowing the power-law inside the exponent to vary from
 $2$ (a common choice) did not lead to any significant change to the
 best-fit redshift distribution below
 $z=0.8$, where the vast majority of the galaxies are located.  The
 changes above that redshift are marginally 
 statistically significant, but there are so few sources above that
 redshift that our final results for the redshift bias that we eventually
 want to calculate do not change within the statistical error.

We will present best-fit redshift distributions for zCOSMOS and DEEP2
EGS separately to demonstrate that the results are
consistent within the errors.  We then use both samples combined to
create an overall redshift distribution.  

This distribution is crucial to our scheme to avoid sampling variance
effects in the determination of the redshift calibration bias.  To
counterbalance regions of source redshift space that are over- or 
under-represented in our spectroscopic sample 
due to LSS fluctuations, we incorporate an additional
weight into the calculation of the redshift bias in
Eq.~(\ref{E:defbz}).  For a galaxy in redshift bin $i$ in our histogram,
the LSS weight ($w_{\rm LSS}$) is the ratio of the number of galaxies
predicted to lie in bin 
$i$ from our best-fit redshift distribution,  to
the number actually found in that bin ($N_{i}^{\rm (model)}/N_i$).  Thus,
those regions in 
redshift space with too many/few galaxies due to LSS or Poisson
fluctuations will be
down/up-weighted appropriately.  We can then get errors on
the average redshift bias $\langle b_z\rangle$ using the best-fit
redshift histograms for each bootstrap resampled histogram to derive
the LSS weights.  This procedure incorporates uncertainty in the
source redshift distribution appropriately, since we never need to
bootstrap the galaxies themselves.  

In an analysis containing many patches of sky, the size of the errors
can be verified by comparing the redshift bias computed in each patch
of sky.  Unfortunately, with only two patches of sky, this method is
not an option for this work.

\subsection{Redshift incompleteness and failures}\label{SS:failures}

For precision results, we require a high redshift completeness and
quality.  
There are several tests that we can carry out to ensure that the sample is
of high quality.  We consider the redshift failures separately for the
DEEP2 and zCOSMOS samples.  In both cases, we will determine the
magnitude and colour distribution of the failures relative to the full
sample, to see if a particular region of redshift space is causing the
problems.  

For zCOSMOS, there are high-quality photoz's derived from very deep
photometry which we can use in the case of spectroscopic redshift
failure.  To control for any effect on the computed redshift
calibration bias, we also check the results using the zCOSMOS photoz's
for a larger portion of the full sample, to ensure that 
noise in these photoz's has a negligible effect on the results.

For DEEP2 EGS, we lack redshift estimates for the failures.  To place
a very conservative bound on the effect of 
failures on the estimated calibration bias, we estimate the redshift
bias with all the 
failures forced to $z=0$, and then to $z=1.5$.  
For both surveys, we will compare the ranges of colours and redshifts
spanned by the successes and failures, to ensure that our
procedures for handling redshift failure are justified.

The next issue is the quality of the non-failed redshifts, which in DEEP2 are
assessed by visual inspection and repeat observations, and in
zCOSMOS using the photoz's as 
well.  For DEEP2, we
have used only $Q=3$ and $Q=4$ redshifts, which are 96\% of our
sample, and are estimated to be $95$\% and $99.5$\% reliable.  For
zCOSMOS, the reliabilities for $Q=3$ and $4$ objects (92\% of our
sample) are $>99$\%.  For
this survey we also use $Q=2.5$, those with slightly lower  
quality in principle but with extremely good matches between the
spectroscopic and photometric redshift, and $Q=9.5$
(single-line redshifts with good matches between the spectroscopic and
photometric redshifts, which in this apparent magnitude and redshift
range are usually from H$\alpha$), both of which also are $>99$\%
reliable as determined from repeat observations. 

%
In the DEEP2 EGS, there are also minor selection effects to control for.
The first effect is the fact that no galaxies brighter than $r\sim 18.5$
were targeted.  Galaxies brighter than that limit constitute only 4\%
of the source sample, but we nonetheless include tests of the effect
this has on the result.

The other selection effect in DEEP2 EGS occurs at magnitudes fainter
than $R=21.5$, where $z<0.75$ objects are given slightly
       lower selection weights than higher-z galaxies.  While the 
fraction of source galaxies fainter than this magnitude is only $\sim 12$\%, we
use their selection probabilities $p_{\rm sel}$ to properly
compensate for this effect.  To be explicit, the total weight for each
source is thus a product of lensing weight $\tilde{w}_j$, the LSS weight
$w_{\rm LSS}$, and $\textrm{max}(p_{\rm sel})/p_{\textrm{sel},j}$ (or 1
for the zCOSMOS galaxies).

Finally, we clarify our statement that our method requires the
spectroscopic sample used to evaluate photoz's to be comparable to the
source sample.  As demonstrated above, it is possible to use weights to
account for well-defined targeting priorities that might make the
spectroscopic sample slightly non-representative of the source
catalog.  Thus, our statement that we 
require the spectroscopic sample to be comparable to the source sample
is really a statement that it must contain all galaxy types (spectral
types, magnitudes, etc.) in the source sample with representation
levels that are sufficient to overcome the noise.  If some reweighting 
is necessary to account for under- or over-representation of a given
population, then for our purposes, 
this is sufficient to fulfill our requirements.  Thus, one could {\it not}
use a spectroscopic sample with a strict cutoff two magnitudes
brighter than the flux limit of the source catalog.  One could use
a spectroscopic sample that has a lower redshift success rate for fainter
galaxies, as long as that lower success rate is due to statistical
error, so that the failures have the same redshift distribution as the
successes, rather than some systematic error (e.g. inability to determine
redshifts for any object of a particular spectral type above some
cutoff redshift).  Reweighting 
schemes to account for different fractions of various galaxy
populations in the training and photometric samples are being
successfull used by
the SDSS neural net photoz group to predict redshift distributions and
photoz error distributions in the photometric samples.\footnote{Lima, Cunha, Oyaizu, Lin,  Frieman, 2007, in prep.}

\subsection{Direct use of photoz's}

Here, we explain our use of photoz's directly for $\Sigma_c$ estimation.  One
might argue that since we have a spectroscopic sample, we should
estimate $\Sigma_c$ using a deconvolved photoz error
distribution.  However, in this paper we test the use of photoz's
directly, for several reasons.

First, as we have argued previously, a key advantage of using photoz's
is that we can  eliminate intrinsically-aligned sources.
Once we start eliminating sources from the sample on the basis of
detailed cuts on photoz, colour, or apparent magnitude, we would have
to re-estimate the photoz error distribution for the sample that
passes these cuts and redo the deconvolution procedure.  This is 
computationally expensive and potentially difficult to do robustly, if
the cuts result in our photoz error distribution being
poorly-determined due to insufficient spectroscopic galaxies that pass
the cuts to properly sample the distribution.  We would therefore like
to find a photoz method that can lead to accurate lensing calibration
on its own.

There is, in principle, one simple option that might improve the lensing
calibration and that can be done without full deconvolution: we can
correct each photoz for the mean photoz bias.  To be accurate, this
should be done as a function of galaxy colour and magnitude.  We will test
the results of doing so for one of the photoz methods when we present
the results of our analysis.

The final reason to use photoz's directly is because that is the
approach taken in many lensing papers to date, and we would like to
test the accuracy of what is currently done in the field to see what
improvements need to be made.  In section~\ref{SS:posterior}, we will
consider using a full  $p(z)$  as a
new alternative approach to using the photoz alone.

\section{Results: application to SDSS lensing}\label{S:results}

%
\subsection{Matching results}\label{SS:matching}

There are
1013 and 1825 galaxies in our source catalog with spectra from DEEP2
EGS and zCOSMOS,  
respectively (including redshift failures).  We  now characterize
these matches relative to  the entire source catalog and compared to
each other.

\begin{figure}
\begin{center}
\includegraphics[width=3.0in,angle=0]{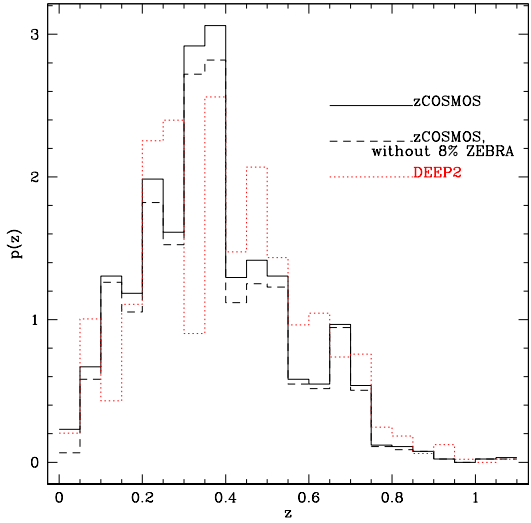} 
\caption{\label{F:bothstatsz}Redshift histogram for the matches
  between the source catalog and 
  the spectroscopic samples.}
\end{center}
\end{figure}

Figure~\ref{F:bothstatsz} shows the redshift histograms for matches
between the source catalog and the zCOSMOS and DEEP2 samples.  The
zCOSMOS histogram is shown both with and without precision photometric
redshifts for the  redshift failures, whereas for DEEP2, the failures
(4\%) were excluded entirely.  As shown, there is 
significant large-scale structure in the redshift histograms, but not
correlated between the two samples.  Visually, the redshift histogram for
DEEP2 appears to be at slightly higher redshift on average.  We 
assess the statistical significance of any differences below.

\begin{figure}
\begin{center}
\includegraphics[width=3.0in,angle=0]{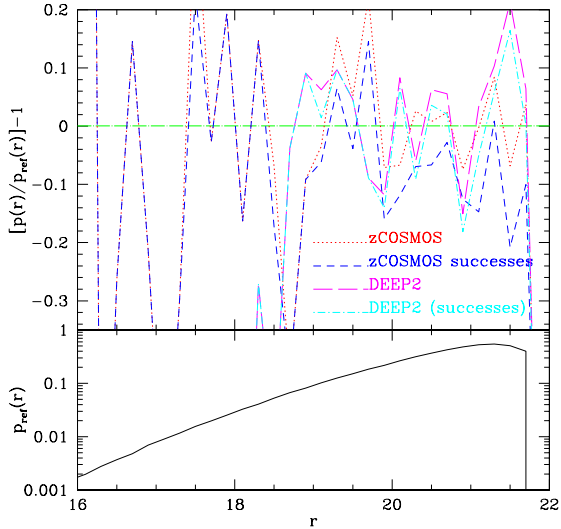} 
\caption{\label{F:bothstatsmr}Bottom: $r$-band apparent magnitude
  histogram for the full source catalog.  Top:
  Difference between the apparent magnitude histogram for the
  zCOSMOS and DEEP2 samples relative to that for the full source
  catalog. }
\end{center}
\end{figure}

Figure~\ref{F:bothstatsmr} shows the distribution of apparent $r$-band
magnitude $p(r)$ for the zCOSMOS and DEEP2 matches relative to that of the
entire source catalog, $p_{\rm ref}(r)$.  The apparent magnitude histogram for zCOSMOS is 
quite similar to that for the full source catalog (within the noise),
and the failures 
are predominantly at the faint end.  The apparent magnitude histogram
for DEEP2 shows the deficit at $r< 18.5$ (4\% of the sample) due to
targeting constraints.   
\begin{figure}
\begin{center}
\includegraphics[width=3.0in,angle=0]{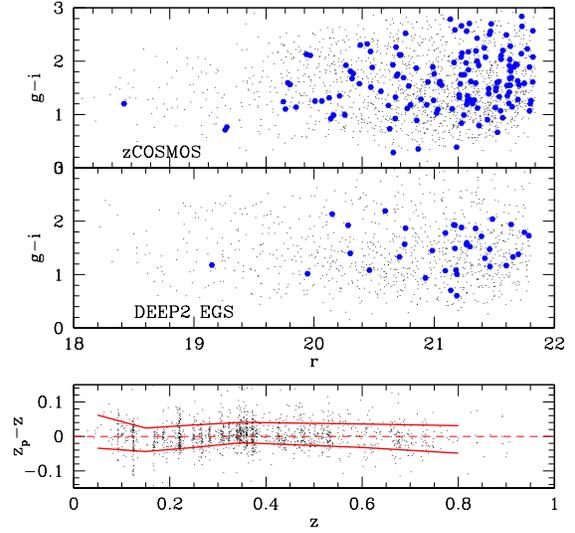} 
\caption{\label{F:fail}Colour-magnitude scatter plots  for
  redshift successes and failures in zCOSMOS (top) and DEEP2
  (middle).  Successes are shown as black points and failures as blue
  hexagons.  The bottom panel shows the zCOSMOS photoz error
  as a function of redshift for the redshift successes, including the
  68\% CL 
  errors as a function of redshift (red lines).}
\end{center}
\end{figure}

Of the matches, $151$ of those in zCOSMOS (8\%) and $38$ of those in
  DEEP2 (4\%) are redshift failures (where failures are defined as
  having redshift success rates below 99\%).  In 
  Fig.~\ref{F:fail}, we show the distributions of various
  quantities for the zCOSMOS and DEEP2 failures as compared with the
  full sample.  Fig.~\ref{F:bothstatsmr} shows the relation of the
  failures to the general sample as a function of apparent magnitude;
  the top part of Fig.~\ref{F:fail} shows that the colour distribution
  for the failures is similar to the colour distribution for the
  successes.  We thus have no reason to believe the failures lie in a
  particular region of redshift space. The DEEP2 failures lie in the
  $0<z<0.75$ colour locus, just like the majority 
  of the successes in this bright subsample of the EGS data.  (This is
  not true for deeper 
  redshift samples, such as the other DEEP2 fields, where failures
  typically occur for blue, 
  $z>1.5$ galaxies.  The flux and apparent size cuts imposed on our
  sample essentially remove any such galaxies.)   Inspection of the
  $38$ DEEP2 spectra suggests that the 
  redshift distribution is similar to that for the successes, with
  failures due to bad astrometry, a bad
        column running through the spectrum, or similar failures that
  do not correlate with redshift.  We also show the zCOSMOS
  photoz error distribution as a function of redshift in the bottom of
  Fig.~\ref{F:fail} for spectroscopic redshift successes.  The photoz errors
  for this sample are indeed as small as, or even smaller than, 
  those presented elsewhere for these photoz's
  \citep{2006MNRAS.372..565F}.  We may view this error as a
  ``systematic floor'' to the error, with the increase in error for
  the ZEBRA/SDSS photoz's being ascribed to the much noisier
  photometry.  We will see that this statistical error 
  dominates the error budget.

Next, we present redshift
distributions for each survey separately, with two purposes: (1) to
demonstrate that they are 
consistent with being drawn from the same underlying redshift
distribution, and (2) to determine the weights to compensate for
sampling variance 
as described in section~\ref{SS:lss}.  

\begin{figure*}
\begin{center}
\includegraphics[width=5.8in,angle=0]{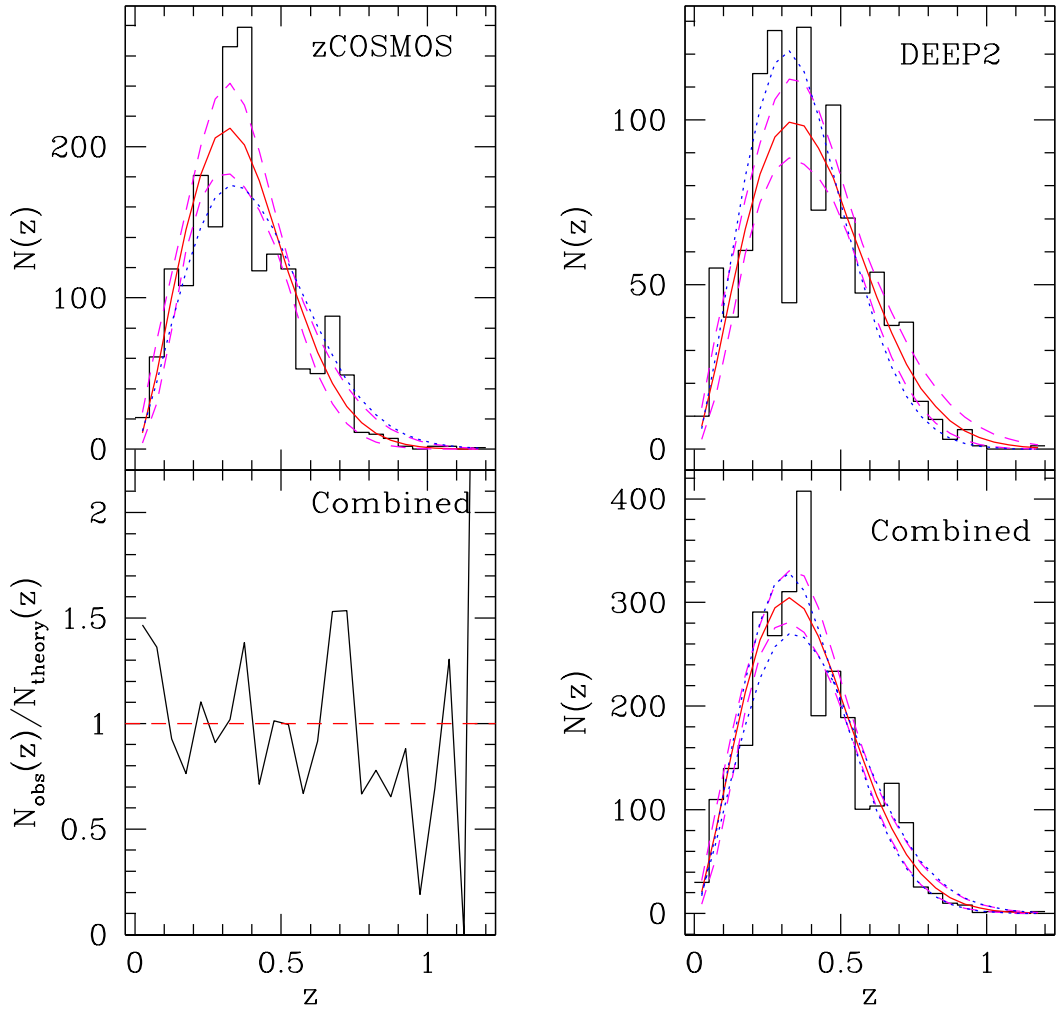} 
\caption{\label{F:showfit}Top: Rescaled redshift 
  histograms for the matches between the source catalog and
  the zCOSMOS (left) and DEEP2 (right) sample with best-fit
  histograms.  The black 
  histogram is the observed data, the smooth red curve is the best-fit
  histogram, 
  the dashed magenta lines are the $\pm 1\sigma$ errors, and the dotted blue line
  is the best-fit redshift histogram for the other survey.  Bottom right: Same
  as above, for combined sample, with the dotted blue lines showing
  the results for each survey separately.  Bottom left: ratio of observed to
  best-fit $N(z)$ for the combined sample.}
\end{center}
\end{figure*}

Fig.~\ref{F:showfit} shows the observed and best-fit redshift histograms for zCOSMOS, 
DEEP2, and both surveys combined.  Table~\ref{T:zdistparam} shows the
corresponding best-fit parameters from Eq.~(\ref{E:zdist}).  The
weighting to account for the DEEP2 selection at $R>21.5$ causes a
negligible change in the results.  By bootstrapping the redshift histogram as
described in section~\ref{SS:lss}, we have determined the median
predicted number of galaxies in each bins, and the 68\% confidence
limits on that number, as shown on the plot. Because we have imposed a
normalization condition on the fit, the 
errorbars are correlated between various parts of the histogram.  We
can see from the plot and table~\ref{T:zdistparam} that while the
DEEP2 sample is at slightly 
higher redshift on average, the redshift distributions from zCOSMOS
and DEEP2 are consistent with each other within the (Poisson plus LSS)
errors.  While it is difficult to compare the curves for $z>0.7$,
where the number of galaxies has declined sharply, we can compare the
total fraction of the sample with $z>0.7$ to show that they are
consistent: for DEEP2 EGS, this fraction lies between $[0.05,0.12]$ at
the 68\% CL; for zCOSMOS, between $[0.02,0.08]$. 
These limits were determined using the fraction above $z>0.7$ for the 
best-fit $N(z)$ for 200 bootstrap-resampled redshift histograms, and
therefore include both Poisson error and sampling variance.  It is
clear that any discrepancy between the best-fit zCOSMOS and DEEP2
redshift histograms with respect to the fraction of the sample above
$z>0.7$ are not significant at the 68\% CL.

 As shown in the lower left
panel of Fig.~\ref{F:showfit}, there is no systematic tendency for the
observed and  
best-fit $N_i(z)$ for the full sample to deviate from each other, only
Poisson and LSS fluctuations, so the form we have chosen for $\rmd
N/\rmd z$ is acceptable.  (The 
fluctuations are quite large for $z>1$ because the best-fit
$N_i^{\rm (model)}$ drops below $1$, so discreteness will cause the ratio of
$N_i/N_i^{\rm (model)}$ 
to be either zero or some large number.)  
It is important to note that this plot is the
{\it unweighted} redshift distribution; inclusion of the lensing
weights in Eq.~(\ref{E:wj}) will change the effective source redshift
distribution.
\begin{table}
\begin{center}
\caption{Parameters of fits to redshift distribution from
  Eq.~\ref{E:zdist}.\label{T:zdistparam}} 
\begin{tabular}{cccc}
\hline\hline
Sample & $z_*$ & $\alpha$ & $\langle z \rangle$ \\
\hline
zCOSMOS & $0.259\pm0.040$ & $2.58\pm 0.58$ & $0.369\pm0.018$ \\
DEEP2 EGS & $0.300\pm0.041$ & $2.35\pm 0.41$ & $0.408\pm0.025$ \\
Both & $0.275\pm0.025$ & $2.42\pm 0.36$ & $0.382\pm0.012$ \\
\hline
\end{tabular}
\end{center}
\end{table}

\subsection{Photoz error distributions}

\begin{figure*}
\begin{center}
\includegraphics[width=6.0in,angle=0]{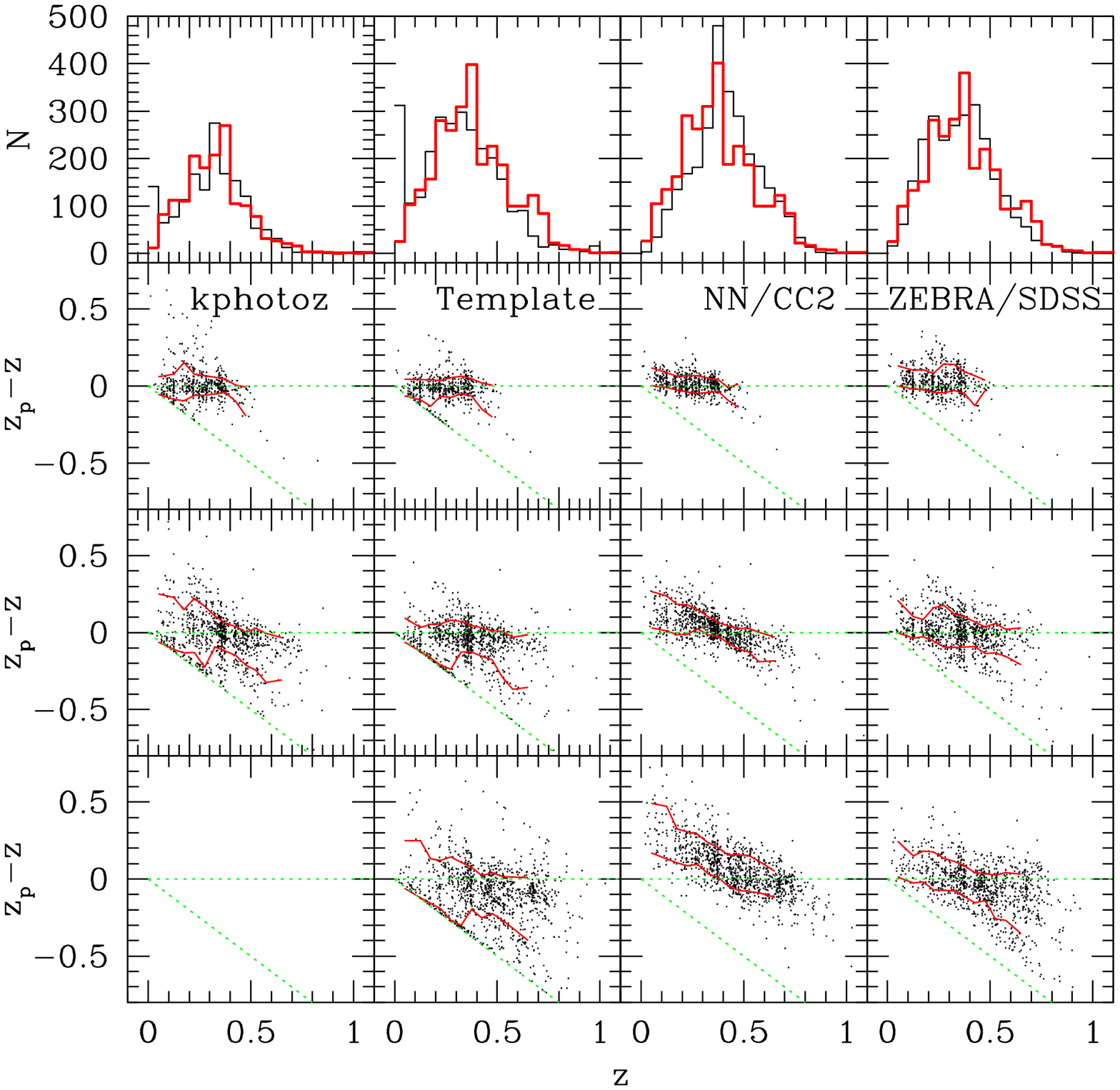}
\caption{\label{F:zperr}
For each photoz method described in the text (in columns labeled
according to the method), the top row shows the redshift histogram
determined using the photoz (thin black line) and using the
spectroscopic redshift (thick red line).  The spectroscopic redshift
histograms are not quite identical for all methods because we exclude photoz
failures for each method and because kphotoz was only used for those galaxies with
$r<21$.  The lower three panels show photometric redshift errors, for galaxies divided by apparent magnitude: $r<20$ in the
second row, $20\le r < 21$ in the third row, and $r\ge 21$ in the
fourth row.  The points correspond to individual galaxies in the
source catalog with spectra; the 68\% confidence limits on the photoz
error are shown as red solid lines.  There are also green dashed lines
indicating zero error and the lower limit on the error given that the
photoz must exceed zero.}
\end{center}
\end{figure*}
As a way of understanding the trends in our lensing-optimized photoz
error statistic $b_z$, we first examine the photoz error distribution
as a function of redshift. 
Figure~\ref{F:zperr} shows the photoz error  as a function of the
(true) redshift for the lensing-selected galaxies 
from zCOSMOS and DEEP2 for the photoz algorithms tested in this
work.  The galaxies are divided by apparent magnitude into three
samples with $r<20$, $20\le r<21$, and $r\ge 21$, and we show 
the 68\% CL errors determined in bins of size $\Delta 
z = 0.05$ for each apparent magnitude bin.   For all methods,  the error 
distributions tend to be highly non-Gaussian, often skewed and
with significant tails.  While the requirement that $z_p>0$ makes
skewness 
inevitable at low $z$ even for a well-behaved photoz estimator,
the effect persists to such high redshift for all methods that this
constraint is clearly not the cause.  Thus, the 68\% confidence limits
as a function of redshift are more useful than a calculation of
the average photoz bias and scatter.  Nonetheless, we do tabulate the
mean bias $\langle z_p-z\rangle$ and the overall scatter $\sigma(z_p)$ in
Table~\ref{T:photozerror} for each method, for the full sample and the
$r<21$ subset (to facilitate comparison between kphotoz, used only for
$r<21$, and the other methods).
\begin{table}
\begin{center}
\caption{Mean properties of the photoz algorithms, for the full
  sample and for $r<21$ only in parenthesis.\label{T:photozerror}} 
\begin{tabular}{crr}
\hline\hline
Method & Mean bias & Scatter \\
\hline
kphotoz & $\qquad$ ($-0.015$) & $\qquad$  ($0.14$) \\
Template & $-0.064$ ($-0.043$) & $0.16$ ($0.12$) \\
NN/CC2 & $0.034$ ($0.013$) & $0.14$ ($0.11$)\\
NN/D1 & $0.038$ ($0.020$) & $0.13$ ($0.10$) \\
ZEBRA/SDSS & $-0.014$ ($0.012$)& $0.15$ ($0.12$)\\
\hline
\end{tabular}
\end{center}
\end{table}

For the kphotoz method, there is a
clear tendency to fail towards very low redshift, as demonstrated by the
peak in $p(z_p)$ for $z_p<0.05$.  For lensing, such failures will be
flagged as being below the lens redshift for nearly all relevant
lens redshifts, thus excluding them from the source sample.
Consequently, the only effect of this failure mode is to reduce the
number of available sources, not to bias the weak lensing
results. However, it is apparent that this method is as noisy for
$r<21$ as the other photoz algorithms are for $r<21.8$, and that the
photoz error tends to be positive for $z\lesssim 0.4$ and negative
above that.

For the template-based database photoz's, there is an even 
stronger failure mode towards $z_p=0$ than for kphotoz (because the
template method goes fainter than the kphotoz sample).
This failure mode contributes to the significantly negative 68\% CL
limits on the photoz error, since the points suggest that ignoring
these failures leads to a more symmetric error distribution.   
We must quantify the effect this has in reducing the total weight;
even if the bias in the lensing signal due to the strong failure mode
is small, the increased statistical error due to loss of sources may be
problematic.  This failure mode is the cause of the large mean photoz bias in
Table~\ref{T:photozerror}.   

For the neural network algorithm, the plot shows the CC2 (colour- and
concentration-based) photoz's, but the trends are qualitatively
similar for the D1 (magnitude- and concentration-based) photoz's.
There are entries for both versions in Table~\ref{T:photozerror}.  
As shown, the method has a reasonably small overall scatter and no
major failure modes.  We caution the reader that the same is not true
for the NN photoz's in the DR5 database, for which 
there is a significant scatter to 
redshifts $0.75<z_p<1$ that more than doubles the number of sources
estimated to be in this redshift range.  The scatter is also larger
for the DR5 NN photoz's.  In both the DR5 and the DR6 versions, there
is a tendency towards positive photoz bias at low-intermediate
redshifts ($0<z<0.4$) that may bias the lensing signal low.  

Finally, the ZEBRA/SDSS method also lacks a major catastrophic
failure mode and has reasonably small overall photoz bias.  
The  redshift histograms derived from the
spectroscopic and photometric redshifts agree remarkably well.  As for
the NN/CC2 photoz's, there is a trend towards positive photoz error at
low redshift and negative error at high redshift.  Because of the
overall lower number of 
sources above $z\gtrsim 0.4$, and the decreased dependence of
$\Sigma_c$ on source redshift at higher redshift, we have no reason to
believe that the 
effects of the different direction of the calibration biases in the
lensing signal will cancel out.  
We can also conclude, in comparison with the ZEBRA photoz errors in
the lower panel of Fig.~\ref{F:fail}  
(using the far deeper COSMOS photometry) for
the same exact set of sources, that for the redshifts and magnitudes
dominated by 
this source sample, statistical error due to noisy SDSS photometry
dominates over systematic error in this photoz method. 


%

\subsection{Redshift bias}\label{SS:redshiftbias}

\begin{figure}
\begin{center}
\includegraphics[width=3.0in,angle=0]{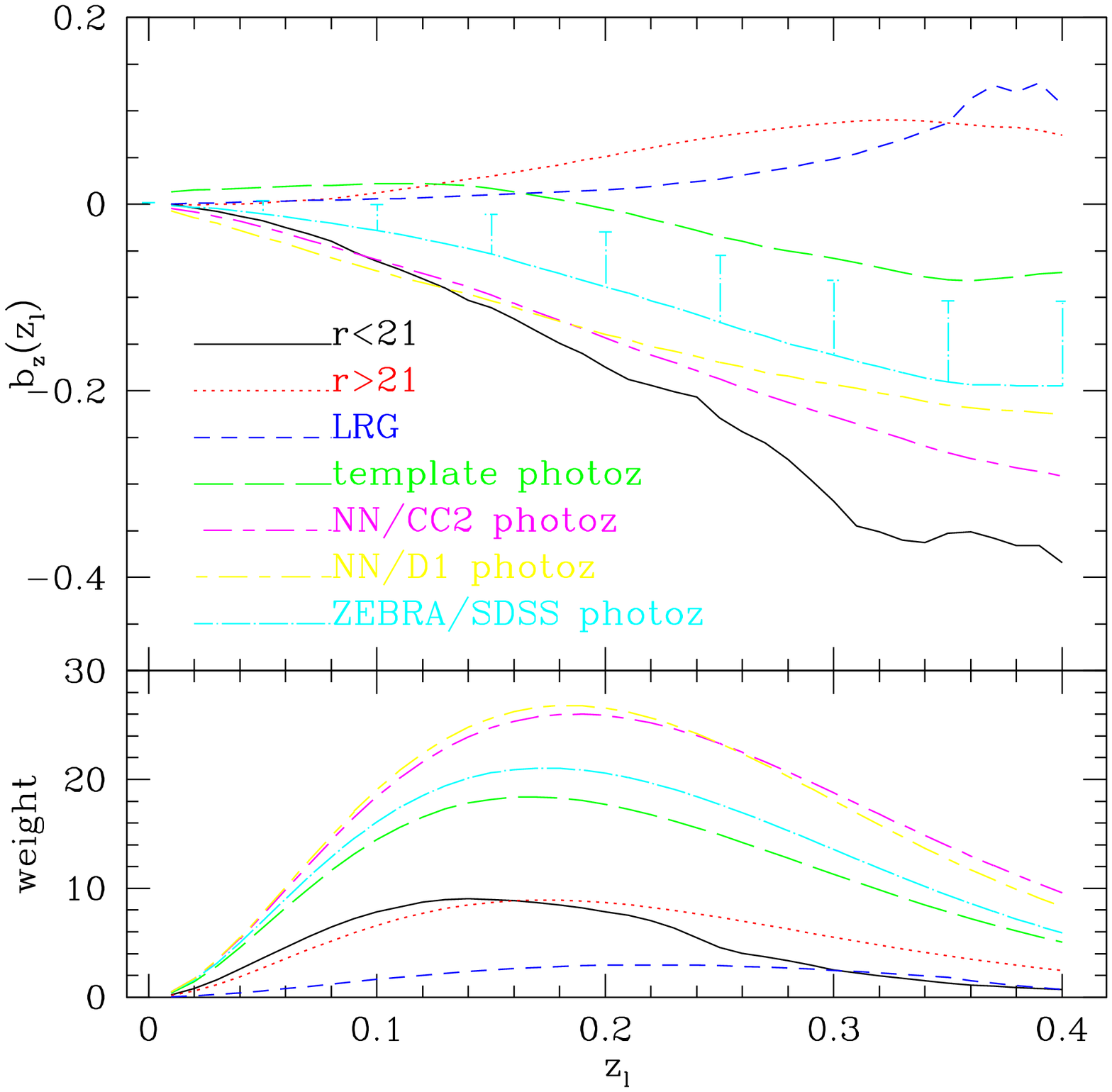}
\caption{\label{F:biaszl}Redshift bias $b_z(z_l)$ (top) and weight
  (bottom, arbitrary units) for many methods of
  source redshift determination as described in the text. To make the
  plot simpler to read, we have left off errorbars except for in one
  case, the ZEBRA/SDSS method, which is shown with an errorbar in one
  direction to indicate the typical size of the uncertainty in
  $b_z(z_l)$ for all the methods.}
\end{center}
\end{figure}

In Figure~\ref{F:biaszl}, we show the lensing calibration bias
$b_z(z_l)$ for different source redshift 
determination methods, using the full lensing-selected spectroscopic
redshift sample.  The bottom panel shows the total lensing weight
ascribed to the source sample for that lens redshift, determined via
summation over the lensing weights described in
Section~\ref{SS:failures}.    
Note that the $r<21$ and LRG samples use photoz's with the requirement that
$z_p>z_l+0.1$, to reduce contamination by physically-associated
sources (for consistency with our previous analyses).  However,
for the new photoz methods,  we have not imposed any such condition
(we will revisit this choice later).

As shown, the $r<21$ sample with photoz's from kphotoz has a
significant negative calibration bias that 
increases with lens redshift to $-35$\% at $z_l=0.35$. As for all
methods, the bias worsens with lens redshift because, for a given
source with some photoz error, a higher lens redshift leads to a
higher relative error in $\tilde{\Sigma}_c^{-1}$.  The $r>21$
sample (using $\rmd N/\rmd z$ from DEEP2 EGS) has a small positive
bias that increases to 10\% at 
$z_l=0.35$. We assess the significance of these biases for our
previous work in section~\ref{SS:prevwork}.  The results for the LRG
source sample confirm our assertion in previous works that for $z_l <
0.3$, this sample is essentially free of redshift bias.  

The lack of significant redshift calibration bias for the template
photoz code for $z_l<0.25$ can be explained by the trends in
Fig.~\ref{F:zperr}: the calibration bias due to the slight negative
photoz bias balances out 
the calibration bias due to photoz scatter.  Even at higher redshift,
the redshift calibation bias, while nonzero, is less 
significant than for the other photoz methods.  The neural net and
ZEBRA/SDSS photoz's, 
however, have significant negative bias ($-30$\% to $-20$\%,
at $z_l=0.4$), presumably because of the aforementioned
tendency to positive photoz bias for $z_s<0.4$.  This difference
between the three methods is also the reason why the latter two
methods have high total weight for the range of lens redshift
considered here, whereas the template photoz code has lower weight (a) 
 because of its scatter to low photoz (which eliminates possible
sources from the sample) and (b)  because it does not tend  to scatter
sources to higher photoz, which 
increases the weight artifically at the expense of biasing the
signal. We emphasize that this higher weight for the two photoz
methods does {\em not} mean that
the error on $\Delta\Sigma$ is lower with these methods, because it
may be due purely to the overestimate of $\tilde{\Sigma}_c^{-1}$.  In
section~\ref{SS:puritycomp}, we will address the effect of
using photoz's on the statistical error in  $\Delta\Sigma$.

Given that kphotoz has a similarly sized 
photoz error ($r<21$ only) as 
the other photoz methods for the full source sample (all magnitudes),
it is important to understand why the lensing calibration bias is so
much worse for this method.  The reason this occurs is that the $r<21$
sample is at lower mean redshift.  Since those sources are closer on
average to the lens redshift, the same size photoz error translates to
a larger error in $\Sigma_c$. 

\begin{figure}
\begin{center}
\includegraphics[width=3.0in,angle=0]{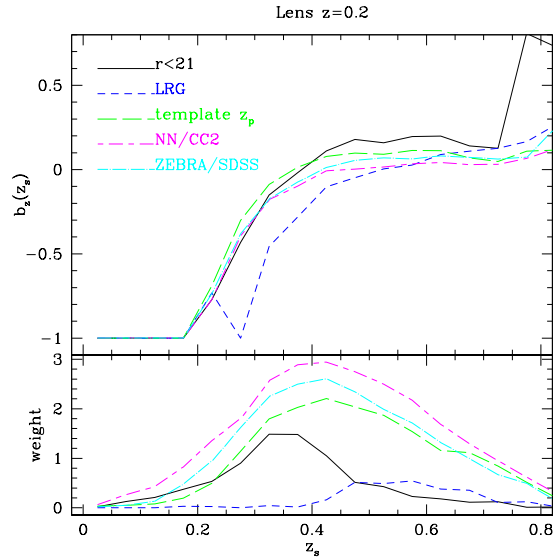}
\caption{\label{F:biaszs}Redshift bias $b_z(z_s)$ (top) and weight
  (bottom, arbitrary units) for fixed $z_l=0.2$ with many methods of
  source redshift determination as described in the text. Errorbars
  are not shown here to make the plot simpler to read.}
\end{center}
\end{figure}

To understand the results, we consider fixed lens redshift
of $z_l=0.2$, and show the redshift bias as a function of true source
redshift for each method in Fig.~\ref{F:biaszs} (again, with lensing
weight as a function of source redshift as in
Section~\ref{SS:failures}).  Clearly, all source redshift bins with 
$z_s<0.2$ must give $b_z=-1$, because the sources are not lensed.
Above $z_s=z_l=0.2$, the calibration bias is no longer identically
zero, but may be significantly negative due to scatter in the
estimates of 
source redshift (near $z_s=z_l$, the derivative $\rmd \Sigma_c/\rmd
z_s$ is large so photoz errors are very important).  As the source
redshift increases, the same photoz error becomes less important
because that derivative decreases, so the calibration bias approaches
zero.  The other important quantity to consider is the weight 
in each source redshift bin; if those source redshift bins with
significant bias are given little weight, then the bias does not
matter. If there is no weight for $z_s<0.2$ that means that 
none of the galaxies with true $z_s<0.2$ have had photoz misestimated 
to be above that.  
This plot makes it
clear that part of the reason for the significant bias for the NN,
kphotoz and ZEBRA/SDSS photoz's is that they give too much 
weight to $z_s\lesssim 0.3$.  This is less of
a problem for the template photoz's, so the
calibration bias for this method is much less. 

\begin{figure}
\begin{center}
\includegraphics[width=3.0in,angle=0]{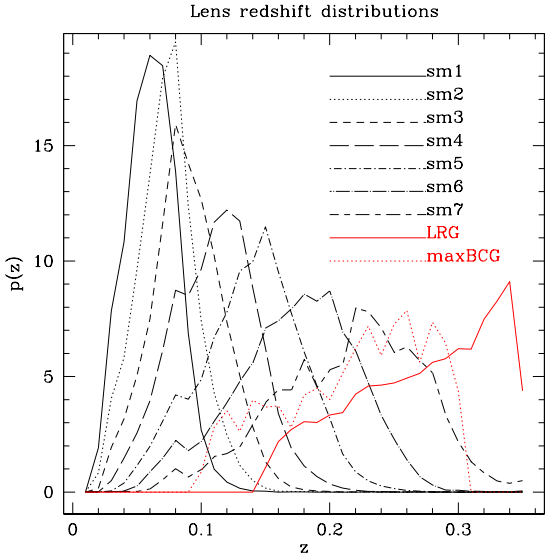}
\caption{\label{F:zdistlens}Lens redshift distributions for the lens
  samples described in the text.}
\end{center}
\end{figure}

Finally, we show the resulting mean calibration bias when these
results are averaged over a lens 
redshift distribution using Eq.~(\ref{E:avgbz}).  Errors are determined
using the prescription in section~\ref{SS:lss}.  The lens redshift
distributions that we 
consider are as follows: ``sm1''--''sm7'' are the redshift
distributions for the seven stellar mass bins from
\cite{2006MNRAS.368..715M}; ``LRG'' 
is the redshift distribution for the spectroscopic LRGs, a
volume-limited sample, used for lensing in
\cite{2006MNRAS.372..758M}; and ``maxBCG'' is the 
redshift distribution of the SDSS maxBCG clusters
\citep{2007ApJ...660..221K,2007ApJ...660..239K}.  
These nine lens redshift distributions 
are plotted in figure~\ref{F:zdistlens}.  The stellar mass subsamples
correspond roughly to luminosity samples with $r$-band luminosities of
$0.33$, $0.53$, $0.72$, $1.1$, $1.8$, $3.0$, and $4.7L_*$. 
The LRGs are red galaxies with typical luminosities of a few $L_*$,
and the maxBCG clusters are clusters selected from imaging data with
masses $\gtrsim 5\times 10^{13}h^{-1}M_{\odot}$. 

The average redshift
calibration biases
$\langle b_z\rangle$ (defined in Eq.~\ref{E:avgbz}) for the redshift
determination methods given 
in Fig.~\ref{F:biaszl} for these nine lens redshift distributions are
shown in Table~\ref{T:avgbz}.  As shown, for the stellar mass
subsamples, the bias gets more significant at higher stellar
mass because of the higher mean redshift.  The maxBCG sample gives
similar bias to sm7 because of the similar redshift range, and the LRG
sample gives the worst bias because it has the highest mean redshift.
The only method for which the trend is different is the template
photoz code, for which the trend of $b_z(z_l)$ changes sign with
redshift due to the different trends of photoz error with redshift.
\begin{table*}
\begin{center}
\caption{Average redshift bias $\langle b_z\rangle$ for nine lens
  redshift distributions described in the text.\label{T:avgbz}} 
\begin{tabular}{cc|c|c|c|c|c|c}
\hline\hline
    & $r<21$ & $r>21$ & LRG & template  & NN/CC2  & NN/D1
    & ZEBRA/SDSS \\
sm1$\!$ & $-0.033 \pm 0.008$ & $0.005 \pm 0.009$ & $0.004 \pm 0.003$ & $0.020 \pm 0.003$ & $-0.039 \pm 0.007$ & $-0.051 \pm 0.007$ & $-0.018 \pm 0.006$ \\
sm2$\!$ & $-0.043 \pm 0.009$ & $0.008 \pm 0.011$ & $0.005 \pm 0.004$ & $0.020 \pm 0.004$ & $-0.048 \pm 0.008$ & $-0.059 \pm 0.008$ & $-0.022 \pm 0.007$ \\
sm3$\!$ & $-0.057 \pm 0.011$ & $0.013 \pm 0.013$ & $0.006 \pm 0.005$ & $0.021 \pm 0.004$ & $-0.059 \pm 0.008$ & $-0.070 \pm 0.008$ & $-0.029 \pm 0.007$ \\
sm4$\!$ & $-0.077 \pm 0.012$ & $0.020 \pm 0.015$ & $0.007 \pm 0.006$ & $0.020 \pm 0.005$ & $-0.075 \pm 0.009$ & $-0.084 \pm 0.009$ & $-0.038 \pm 0.008$ \\
sm5$\!$ & $-0.104 \pm 0.014$ & $0.029 \pm 0.019$ & $0.010 \pm 0.008$ & $0.015 \pm 0.005$ & $-0.096 \pm 0.009$ & $-0.102 \pm 0.009$ & $-0.053 \pm 0.008$ \\
sm6$\!$ & $-0.136 \pm 0.016$ & $0.041 \pm 0.025$ & $0.014 \pm 0.011$ & $0.003 \pm 0.007$ & $-0.124 \pm 0.011$ & $-0.123 \pm 0.011$ & $-0.074 \pm 0.010$ \\
sm7$\!$ & $-0.169 \pm 0.018$ & $0.055 \pm 0.033$ & $0.022 \pm 0.016$ & $-0.014 \pm 0.009$ & $-0.155 \pm 0.015$ & $-0.146 \pm 0.015$ & $-0.099 \pm 0.012$ \\
LRG$\!$ & $-0.221 \pm 0.022$ & $0.069 \pm 0.045$ & $0.038 \pm 0.022$ & $-0.037 \pm 0.014$ & $-0.195 \pm 0.021$ & $-0.171 \pm 0.021$ & $-0.131 \pm 0.018$ \\
$\!\!$maxBCG$\!\!\!$ & $-0.171 \pm 0.018$ & $0.056 \pm 0.034$ & $0.023 \pm 0.016$ & $-0.015 \pm 0.009$ & $-0.158 \pm 0.015$ & $-0.147 \pm 0.015$ & $-0.101 \pm 0.013$ \\
\hline
\end{tabular}
\end{center}
\end{table*}


As shown, the NN/D1 photoz's give nominally worse calibration bias
than the NN/CC2 photoz's for
lower-redshift lens samples, and the reverse is true at higher
redshift.  This trend is consistent with the difference between the two
methods in Fig.~\ref{F:biaszl}.  We also performed the analysis with
the DR5 NN photoz's, and found the lensing calibration bias for these
lens redshift distributions to be similar to the NN/CC2 calibration
biases, well within the $1\sigma$ errors.  This result suggests that the
failure mode to $0.75<z_p<1$ in the DR5 version was not a
significant source of lensing calibration bias, and the overall positive
photoz bias (present in all NN photoz's tested in this paper) is the
main cause. 

Finally, we consider what happens if we correct for the mean
photoz bias when estimating $\Sigma_c$ for each source.  For the
template photoz's, this correction causes the mean calibration bias for sm7
to go from $-0.014$ to $-0.14$.  This result may be puzzling until we
consider the effects of photoz bias and scatter separately
(section~\ref{SS:biasscatter}).  We know that photoz scatter causes a
negative calibation bias, and a negative photoz error like this
method has causes a positive calibration bias.  When we did not
correct for the mean photoz bias, these two effects apparently
cancelled out.  This cancellation is a non-trivial result that depends
on our sample selection. With a different cut on apparent magnitude,
for example, it is not clear that the effects would balance  as
precisely.  Now that we have corrected for the effects of mean photoz 
bias, we are left with the suppression of the lensing
signal due to the photoz scatter.  For the NN/CC2 and NN/D1 photoz's,
the correction for the mean photoz bias decreases calibration bias
from $-0.16$ and $-0.15$ to
$-0.10$ and $-0.07$, respectively, for sm7 (since the positive photoz
bias and the scatter change the lensing calibration in the same
direction).  For ZEBRA/SDSS, the photoz bias was slightly 
negative, so correcting for it worsens the lensing calibration bias
as for the template photoz's, but only slightly: from $-0.099$ to
$-0.125$ for sm7.

From these results, we can conclude that once the effects of the mean
photoz bias are removed, the effects on the lensing calibration due to
scatter in the photoz's are the smallest for the SDSS NN/D1
photoz's, followed by  SDSS NN/CC2, ZEBRA/SDSS, and finally are the 
largest for the template photoz's. This trend is consistent with the
trends in Table~\ref{T:photozerror} for the photoz scatter. We
therefore have two possible procedures for handling calibration bias
in the lensing signal: (1) to correct for the mean photoz bias before
computing the lensing signal, and apply a correction to the
lensing signal afterwards to account for residual calibration bias due
to photoz scatter; or (2) to apply a correction to the lensing signal
due to the combined effects of photoz bias and scatter at once.  In
either case, we must depend on the fact that our calibration subsample
has the same sample properties as the full source catalog, so that
corrections derived using this subsample will apply to the full catalog.

\subsection{Implications for previous work}\label{SS:prevwork}

Here we determine the implications of Table~\ref{T:avgbz} for previous
work with this lensing source catalog. 

First, we consider the results for \cite{2006MNRAS.368..715M}, in
which we divided  
the sample into stellar mass and luminosity subsamples with the seven
redshift distributions sm1--sm7 shown in Fig.~\ref{F:zdistlens}.  For
that work, the signal presented was an average over the signal using
the $r<21$ and $r>21$ source sample with $1/\sigma^2$ weighting.  To
determine the average bias on this signal, we use our
bootstrap-resampled $b_z(z_l)$ and $w(z_l)$, averaging the bias as a
function of redshift for each resampling using the weights for these
two samples, then find the average over all the resampled datasets.
The average biases for sm1--sm7 are
shown in Table~\ref{T:prevwork}.
\begin{table}
\begin{center}
\caption{Average redshift bias $\langle b_z\rangle$ in previous works
  using this source catalog when combining source samples.\label{T:prevwork}} 
\begin{tabular}{cc}
\hline\hline
Lens sample   & $\langle b_z\rangle$ \\
sm1 & $-0.016\pm0.008$ \\ 
sm2 & $-0.020\pm0.009$ \\
sm3 & $-0.025\pm0.011$ \\
sm4 & $-0.032\pm0.013$ \\
sm5 & $-0.039\pm0.016$ \\
sm6 & $-0.045\pm0.020$ \\
sm7 & $-0.046\pm0.026$ \\
LRG & $+0.021\pm0.038$ \\
\hline
\end{tabular}
\end{center}
\end{table}

We also consider the spectroscopic LRG lens redshift
distribution, which was used for lensing in
\cite{2006MNRAS.372..758M} and \cite{2007JCAP...06...24M}.  In that case, we
detected a $\sim 15$\% suppression of the 
lensing signal for the $r<21$ source sample relative to the $r>21$ and
LRG source samples.  Table~\ref{T:avgbz} makes it clear that this
suppression was, in fact, real.  To account for this suppression, we
had multiplied the signal and its error by a factor of $1.18$.  This is equivalent
to multiplying $\tilde{\Sigma}_c$ by $1.18$ when computing both the weights
($\propto \tilde{\Sigma}_c^{-2}$) and the lensing signal.  We thus incorporate
this factor into the computation of the bias in Eq.~(\ref{E:defbz})
before taking the weighted average with the $r>21$ sample. 
The average bias once the correction factor is incorporated is shown
in Table~\ref{T:prevwork}.  Because of this suppression of the weight
in the $r<21$ sample due to the calibration factor, and because of its 
already low weight relative to $r>21$
for $z_l>0.22$ (see Fig.~\ref{F:biaszl}), the uncertainty on the
calibration bias is actually 
dominated by the larger $r>21$ sample uncertainty, which is why it is
larger than one might naively expect from combining the results in
Table~\ref{T:avgbz} for $r<21$ and $r>21$. It is clear that this
way of combining the signal for $r<21$ and $r>21$ is non-optimal from
the perspective of constraining calibration bias.

No results are shown for the maxBCG
lensing sample because none of the previous works using this source
catalog have used it.

It is clear from this table that there was statistically significant
redshift calibration bias in previous works using this source
catalog.  However,  the absolute value of the error is below the
statistical error on the lensing signal in those works, and is smaller
than the generous $8$\% ($1\sigma$) systematic 
error that was used for those science results.  We conclude that
there is no cause for concern in using results in our previous work
with this catalog without applying a correction.  

\subsection{Systematics: targeting and redshift failure}

In the previous sections, all quoted calibration errors were
statistical.  Here, we 
consider the size of systematic errors.

First, we include the DEEP2 redshift failures in
 the sample, once putting them all at $z=0$ and then all at $z=1.5$
 (with an LSS weight of $1$).
 We have already shown in  section~\ref{SS:matching} that the failures have a
 similar SDSS magnitude and colour distribution to the remainder of
 the sample.  This statement is
 also true in the DEEP2 $BRI$ photometry, placing these galaxies
 without spectroscopic redshifts in the $0<z<0.7$ colour locus (like
 those with successful redshift determination).
 Consequently, placing them all at
 $z=0$ and $z=1.5$ gives extremely conservative bounds on the
 systematic error due to these redshift failures.  
 Table~\ref{T:faildeep} shows the new $\langle b_z\rangle$ and the
 change in $\langle b_z\rangle$ compared to 
 table~\ref{T:avgbz} for all methods of source redshift determination,
 including the combined $r<21$ and $r>21$ method used in our previous
 work (Sec.~\ref{SS:prevwork}), for four lens redshift distributions: sm1, sm4, sm7,
 and LRG, which are at progressively higher redshifts.
\begin{table*}
\begin{center}
\caption{Change in redshift bias $\langle b_z\rangle$
 for all
 methods of source redshift determination (including combined methods
 for $r<21$ and $r>21$ as in 
 previous work, Sec.~\ref{SS:prevwork}) when putting all DEEP2
 failures at $z=0$ and $z=1.5$ as 
 shown.  The number given is the resulting redshift bias, and the number in
 parenthesis is the fractional change in the bias from Table~\ref{T:avgbz}
 relative to the 
 statistical error.\label{T:faildeep}} 
\begin{tabular}{cccccccc}
\hline\hline
 & $r<21$ & $r>21$ & LRG & template & NN/CC2 & ZEBRA/SDSS & Previous work \\
\hline
\multicolumn{8}{c}{Fail to $z=0$} \\
$\!$sm1$\!\!$ & $ -0.036 \,( -0.38)$ & $ -0.017 \,( -2.4)$ & $ -0.002 \,( -2.0)$ & $  0.008 \,( -4.0)$ & $ -0.062 \,( -1.9)$ & $ -0.029 \,( -1.8)$ & $ -0.028 \,( -1.5)$ \\
$\!$sm4$\!\!$ & $ -0.081 \,( -0.33)$ & $ -0.003 \,( -1.5)$ & $  0.002 \,( -0.8)$ & $  0.007 \,( -2.6)$ & $ -0.094 \,( -1.6)$ & $ -0.050 \,( -1.5)$ & $ -0.044 \,( -0.9)$ \\
$\!$sm7$\!\!$ & $ -0.173 \,( -0.22)$ & $  0.032 \,( -0.7)$ & $  0.017 \,( -0.3)$ & $ -0.027 \,( -1.4)$ & $ -0.166 \,( -1.0)$ & $ -0.111 \,( -1.0)$ & $ -0.060 \,( -0.5)$ \\
$\!$LRG$\!\!$ & $ -0.224 \,( -0.14)$ & $  0.045 \,( -0.5)$ & $  0.033 \,( -0.2)$ & $ -0.050 \,( -0.9)$ & $ -0.218 \,( -0.8)$ & $ -0.143 \,( -0.67)$ & $  0.005 \,( -0.4)$ \\
\hline
\multicolumn{8}{c}{Fail to $z=1.5$} \\
$\!$sm1$\!\!$ & $ -0.032 \,(  0.13)$ & $  0.009 \,(  0.4)$ & $  0.004 \,(  0.00)$ & $  0.022 \,(  0.7)$ & $ -0.046 \,(  0.43)$ & $ -0.015 \,(  0.50)$ & $ -0.013 \,(  0.38)$ \\
$\!$sm4$\!\!$ & $ -0.075 \,(  0.17)$ & $  0.027 \,(  0.5)$ & $  0.008 \,(  0.17)$ & $  0.024 \,(  0.8)$ & $ -0.075 \,(  0.56)$ & $ -0.034 \,(  0.50)$ & $ -0.026 \,(  0.46)$ \\
$\!$sm7$\!\!$ & $ -0.166 \,(  0.17)$ & $  0.074 \,(  0.6)$ & $  0.024 \,(  0.13)$ & $ -0.005 \,(  1.0)$ & $ -0.140 \,(  0.73)$ & $ -0.089 \,(  0.83)$ & $ -0.033 \,(  0.50)$ \\
$\!$LRG$\!\!$ & $ -0.217 \,(  0.18)$ & $  0.097 \,(  0.6)$ & $  0.042 \,(  0.18)$ & $ -0.025 \,(  0.9)$ & $ -0.186 \,(  0.71)$ & $ -0.118 \,(  0.72)$ & $  0.043 \,(  0.58)$ \\
\hline
\end{tabular}
\end{center}
\end{table*}

As shown in Table~\ref{T:faildeep}, these extreme 
assumptions  change our estimated calibration bias at the $<
3\sigma$ level, in most cases $<1\sigma$.  If we consider
that the real effect is likely many factors smaller than this (since
the failures roughly follow the magnitude and colour distribution of
the successes, and therefore likely the redshift distribution), this
systematic is far below our $1\sigma$ 
uncertainty on the calibration bias, from which we can conclude that
systematic effects due to the excluded DEEP2 redshift failures are negligible.

We next consider the effects of using the zCOSMOS photoz for 
 their redshift failures.  As shown in Fig.~\ref{F:fail}, the failures
 have similar colours and magnitudes as the successes, so we do not
 anticipate that 
 they will have a significantly different photoz error distribution
 from the successes shown at the bottom of that figure.  To 
 test the effect of using ZEBRA photoz's for this 8\% of the sample, we
 randomly replace the photoz's for the spectroscopic redshifts in 
 another 8\% of the sample that are redshift successes.  We then
 compare the resulting calibration 
 biases $\langle b_z\rangle$ to the original ones.  These results
 (shown in Table~\ref{T:failzcosmos}) indicate that for all methods of
 source redshift distribution 
determination and lens redshift distributions, the use of zCOSMOS
photoz's for the 8\% of the zCOSMOS sample that
lacks redshifts  changes the results well below the $1\sigma$
 statistical error. 
We conclude that  systematic error in our
results due to redshift failures in either survey are unimportant,
 with the caveat that if the redshift failures are a systematically
 different population than the successes, this test would
 not uncover any resulting systematic error (however, we have no
 evidence that this is the case).
\begin{table*}
\begin{center}
\caption{Change in redshift bias $\langle b_z\rangle$
 for all
 methods of source redshift determination when replacing 8\% of the
 redshifts for zCOSMOS 
 successes with their photoz's. The number given is the resulting
 redshift  bias, and the number in
 parenthesis is the fractional change in the bias from
 Table~\ref{T:avgbz} relative to the
 statistical error. \label{T:failzcosmos}} 
\begin{tabular}{cccccccc}
\hline\hline
 & $r<21$ & $r>21$ & LRG & template & NN/CC2 & ZEBRA/SDSS & Previous work \\
\hline
sm1 & $ -0.033 \,\,(  0.00)$ & $  0.005 \,\,(  0.00)$ & $  0.004 \,\,(  0.00)$ & $  0.019 \,\,( -0.33)$ & $ -0.049 \,\,(  0.00)$ & $ -0.018 \,\,(  0.00)$ & $ -0.016 \,\,(  0.00)$ \\
sm4 & $ -0.078 \,\,( -0.08)$ & $  0.019 \,\,( -0.07)$ & $  0.008 \,\,(  0.17)$ & $  0.020 \,\,(  0.00)$ & $ -0.080 \,\,(  0.00)$ & $ -0.039 \,\,( -0.13)$ & $ -0.032 \,\,(  0.00)$ \\
sm7 & $ -0.170 \,\,( -0.06)$ & $  0.055 \,\,(  0.00)$ & $  0.024 \,\,(  0.13)$ & $ -0.013 \,\,(  0.11)$ & $ -0.151 \,\,(  0.00)$ & $ -0.098 \,\,(  0.08)$ & $ -0.046 \,\,(  0.00)$ \\
LRG & $ -0.221 \,\,(  0.00)$ & $  0.070 \,\,(  0.02)$ & $  0.041 \,\,(  0.14)$ & $ -0.035 \,\,(  0.14)$ & $ -0.201 \,\,(  0.00)$ & $ -0.130 \,\,(  0.06)$ & $  0.022 \,\,(  0.03)$ \\
\hline
\end{tabular}
\end{center}
\end{table*} 

One final systematic is that in DEEP2 EGS, roughly 4\% of our
source catalog at bright magnitudes ($r<18.5$) was not targeted. We
must assess whether 
properly including these galaxies would significantly change the
results. However, the small photoz error for bright objects, and the
low mean redshift, makes this unlikely.  In the SDSS, only a
subset of these galaxies have spectroscopy, those with $r\lesssim
17.7$ (flux-limited) and fainter ones that are very red.  Since
including these SDSS spectroscopic redshifts will create a sample with
strange selection 
(lacking blue galaxies at $17.7\lesssim r < 18.5$), we instead take
the spectroscopic galaxies from zCOSMOS at $r<18.5$, choose a random
subset to 
account for the smaller size of the DEEP2 sample, and add the
resulting $42$ galaxies to the DEEP2 sample.  We then refit the
redshift histogram for DEEP2, getting new redshift distribution
parameters $z_*= 0.312\pm 0.048$, $\alpha = 2.14\pm 0.39$, and
$\langle z \rangle = 0.400\pm 0.025$.  We see that the change in mean
source redshift is well within the errors in 
Table~\ref{T:zdistparam}.    When computing the mean redshift bias
using this augmented sample, we find that the changes are even smaller
than those shown in Table~\ref{T:faildeep}. 
This is not surprising, because in that table we have taken redshift
failures and put them at very extreme redshifts, whereas here we have
added a comparable number of redshifts but with very good photoz's.

\subsection{Agreement between the two surveys}\label{SS:agreement}

As an additional  systematics test, we compare the results when
doing the full analysis separately for each survey.  In this case, we
use LSS weights derived using the redshift histograms for each survey
separately instead of using the combined histogram.  In 
Table~\ref{T:difftable}, we show the results for each survey
separately, with the bottom section showing the
statistical significance of the difference.
\begin{table*}
\begin{center}
\caption{Redshift bias $\langle b_z\rangle$ for each survey
 separately. The number given is the resulting 
 redshift  bias with statistical error.  The bottom section gives the
 statistical significance on the difference in units of standard
 deviations. \label{T:difftable}}  
\begin{tabular}{cccccccc}
\hline\hline
 & $r<21$ & $r>21$ & LRG & template & NN/CC2 & ZEBRA/SDSS & Previous work \\
\hline
\multicolumn{8}{c}{zCOSMOS} \\
sm1 & $-0.045 \pm 0.013$ & $-0.001 \pm 0.012$ & $-0.005 \pm 0.007$ & $0.020 \pm 0.005$ & $-0.051 \pm 0.009$ & $-0.021 \pm 0.009$ & $-0.026 \pm 0.011$ \\
sm4 & $-0.101 \pm 0.020$ & $0.005 \pm 0.024$ & $-0.009 \pm 0.014$ & $0.019 \pm 0.007$ & $-0.089 \pm 0.011$ & $-0.044 \pm 0.011$ & $-0.053 \pm 0.018$ \\
sm7 & $-0.222 \pm 0.029$ & $0.013 \pm 0.063$ & $-0.016 \pm 0.033$ & $-0.026 \pm 0.019$ & $-0.179 \pm 0.027$ & $-0.109 \pm 0.025$ & $-0.097 \pm 0.044$ \\
LRG & $-0.295 \pm 0.034$ & $0.011 \pm 0.090$ & $-0.012 \pm 0.045$ & $-0.059 \pm 0.030$ & $-0.242 \pm 0.040$ & $-0.146 \pm 0.038$ & $-0.048 \pm 0.069$ \\
\hline
\multicolumn{8}{c}{DEEP2 EGS} \\
sm1 & $-0.013 \pm 0.006$ & $0.007 \pm 0.016$ & $0.013 \pm 0.004$ & $0.018 \pm 0.004$ & $-0.048 \pm 0.010$ & $-0.012 \pm 0.007$ & $-0.003 \pm 0.010$ \\
sm4 & $-0.036 \pm 0.009$ & $0.028 \pm 0.026$ & $0.025 \pm 0.009$ & $0.022 \pm 0.006$ & $-0.070 \pm 0.012$ & $-0.031 \pm 0.009$ & $-0.003 \pm 0.018$ \\
sm7 & $-0.071 \pm 0.017$ & $0.091 \pm 0.053$ & $0.062 \pm 0.021$ & $0.003 \pm 0.013$ & $-0.112 \pm 0.022$ & $-0.085 \pm 0.018$ & $0.023 \pm 0.040$ \\
LRG & $-0.075 \pm 0.025$ & $0.122 \pm 0.072$ & $0.089 \pm 0.030$ & $-0.007 \pm 0.019$ & $-0.145 \pm 0.032$ & $-0.111 \pm 0.025$ & $0.113 \pm 0.060$ \\
\hline
\multicolumn{8}{c}{Statistical significance of difference (in units of
 $\sigma$)} \\
sm1 & $   2.23 $ & $   0.40 $ & $   2.23 $ & $   0.31 $ & $   0.22 $ & $   0.79 $ & $   1.55 $ \\
sm4 & $   2.96 $ & $   0.65 $ & $   2.04 $ & $   0.33 $ & $   1.17 $ & $   0.91 $ & $   1.96 $ \\
sm7 & $   4.49 $ & $   0.95 $ & $   1.99 $ & $   1.26 $ & $   1.92 $ & $   0.78 $ & $   2.02 $ \\
LRG & $   5.21 $ & $   0.96 $ & $   1.87 $ & $   1.46 $ & $   1.89 $ & $   0.77 $ & $   1.76 $ \\
\hline
\end{tabular}
\end{center}
\end{table*} 

The results in this table show apparently significant
discrepancies between the results with 
zCOSMOS and with DEEP2 separately.  The fact that the statistical
significance of the difference is $\lesssim 2\sigma$ for the last four
columns, which use the full catalog, but $>2\sigma$ for the first
column (which uses $r<21$ only) and $\lesssim 1\sigma$ for the second
column (which uses $r>21$ only) suggests that we should focus on
the $r<21$  sample to find the source of the discrepancy. 
 We must 
understand  this discrepancy in order to assess
whether our results are biased or our 
errorbars are significantly underestimated on the final, combined
analysis. 

\begin{figure*}
\begin{center}
\includegraphics[width=5.8in,angle=0]{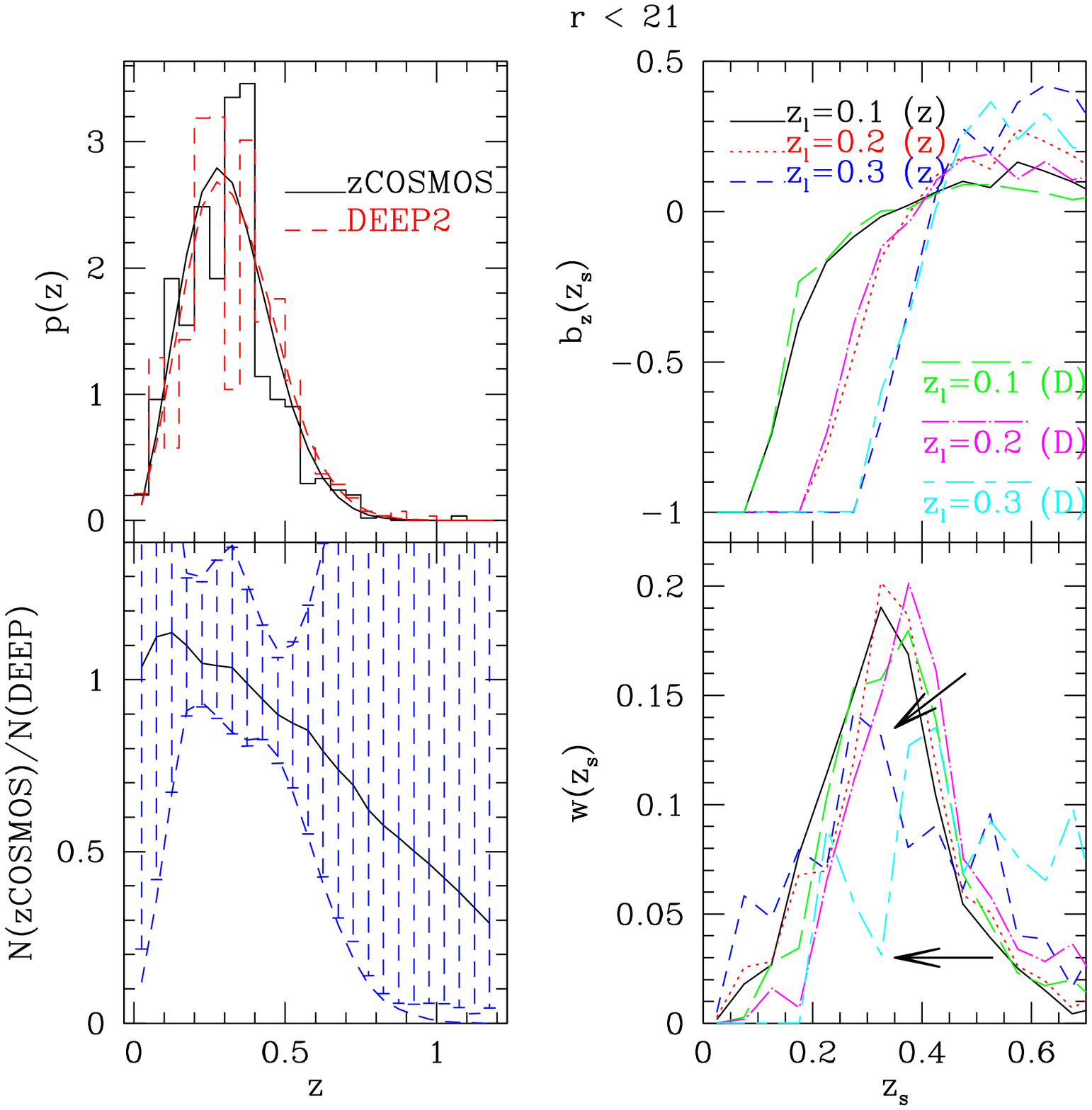}
\caption{\label{F:reconcile}Results for $r<21$ only for each survey
  separately, as described in more detail in the text of
  section~\ref{SS:agreement}.  We show the best-fit and observed
  redshift histograms (upper left); the ratio of the best-fit redshift
  distributions, with shaded 68\% CL region (lower left); and the
  lensing calibration bias $b_z$ (upper right) and the lensing weight as a function
  of source redshift (lower right) for three lens redshifts.}
\end{center}
\end{figure*}

In Fig.~\ref{F:reconcile} we show plots for $r<21$ that will shed
light on this discrepancy.  The upper left plot shows $p(z)$
for $r<21$ for both surveys.  As shown, the best-fit histograms are
very similar, but the LSS fluctuations are more pronounced than for the full
sample.  The lower left panel shows the ratio of the best-fit number predicted in
zCOSMOS to the number in DEEP2 (normalized to the same total
numbers of galaxies), with the 68\% confidence region shown with
dashed lines. This confidence region, including both Poisson and
sampling variance error, was determined as follows: for each survey,
$200$ bootstrap-resampled redshift histograms were created, and used
to fit for the $\rmd N/\rmd z$.  We then pair up the 200 best-fit
$N_i^{\rm (model)}$ from zCOSMOS and from DEEP2 EGS, and determine the
ratio of these values for each survey. The $200$ ratios are ranked,
and the middle 68\% are chosen to determine the 68\% confidence region. It is
reassuring that for all redshifts, this shaded 
region includes a ratio of $1$.  It is apparent that the scarcity of
redshifts at $z>0.6$ causes the errorbars on the ratio to become
extremely large (well off the limits of the plot).

The top right panel in Fig.~\ref{F:reconcile} shows $b_z(z_s)$ for several
lens redshifts.  As shown, these results are very
similar for the two surveys.  The bottom right plot shows the
fractional weight $w(z_s)$ for each 
lens redshift and 
survey.  In principle, the LSS weighting was designed to ensure that
these curves would not have structure due to LSS fluctuations in
number density as a function of redshift.  We can see (particularly for
$z_l=0.3$) that the curves for each survey are quite 
different and have significant LSS fluctuations, so we must understand
why this is the case.  We have 
ascertained that if we use $b_z(z_s)$ from DEEP2 with the weight
$w(z_s)$ from zCOSMOS, we recover the same $\langle b_z\rangle$ as
when we use $b_z(z_s)$ and $w(z_s)$ from zCOSMOS, implying that the
weight differences 
cause the discrepancy in $\langle b_z\rangle$.

To solve this problem, we consider only sources with $0.3\le
z_s<0.35$.  As shown with arrows, for $z_l=0.3$, the weight in this bin is a
factor of $\sim 
4$ higher in zCOSMOS as in DEEP2.  We have confirmed that this bin
alone is a significant reason why the average calibration bias is on
average more negative for zCOSMOS as for DEEP2.  There are $179$ and
$21$ galaxies at $r<21$ in this bin in zCOSMOS and DEEP2
respectively.   Using the LSS weights derived for each survey 
separately, we weight zCOSMOS and DEEP2 by factors of $0.8$ and
$2.25$, giving weighted numbers of galaxies of $143$ and $63$.  Thus,
the weighted ratio $N(\textrm{zCOSMOS})/N(\textrm{DEEP2})\sim 2.3$,
where the expected 
value is $1.85$ given the total number of galaxies in each survey.
This ratio of $2.3$ therefore represents a $23$\% enhancement of zCOSMOS
relative to DEEP2, due 
to the fact that the LSS weights were derived using all
galaxies in each survey, not just those at $r<21$ that we use here.
While we can therefore conclude that LSS 
weighting may need to be done as a function of apparent magnitude,
this $23$\% enhancement in source number does not account for a factor of $4$
enhancement in the weights.

\begin{figure}
\begin{center}
\includegraphics[width=3in,angle=0]{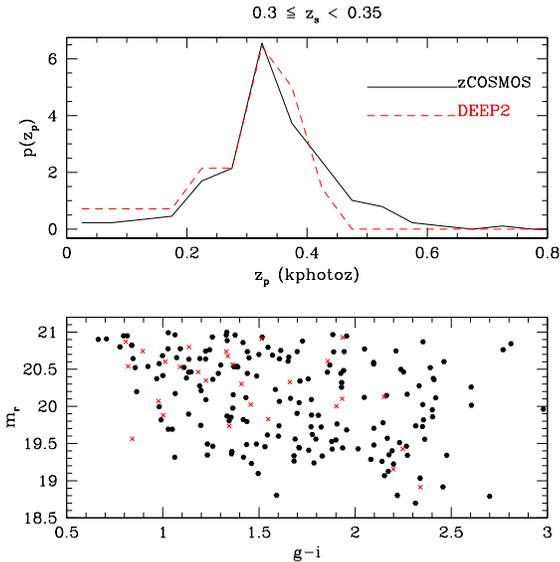}
\caption{\label{F:zpdetails}Photoz distribution (top) and
  colour-magnitude information for DEEP2 and zCOSMOS sources with
  $0.3\le z_s<0.35$ (kphotoz).  In the bottom panel, we show the $g-i$
  colour and $r$-band magnitude, where the red
  crosses are DEEP2 and the black hexagons are zCOSMOS.}
\end{center}
\end{figure}

Figure~\ref{F:zpdetails} shows the photoz distribution $p(z_p)$
for kphotoz for the $r<21$ sources in this narrow redshift slice
in each survey.  It is important to note that our past analyses have
required $z_p>z_l+0.1$.  The photoz distributions for the zCOSMOS and
DEEP2 galaxies in this redshift slice are quite different, with the
DEEP2 distribution being skewed to lower photoz, and the zCOSMOS one
to higher photoz.  Consequently, forty of the $179$ zCOSMOS galaxies
pass this photoz cut (23\%), as compared with two of the $29$ DEEP2 galaxies
(7\%).  In terms of raw numbers, this gives an additional factor of
$23/7\sim 3.2$ enhancement of the weight in zCOSMOS on top of the
previous factor of $1.2$.  Thus, the two factors together give
nearly the factor of four enhancement in weight that we noticed on
Fig.~\ref{F:reconcile} as 
the source of the discrepancy. 

Having accounted for the source of the problem, we must understand why
the photoz distributions look so different for the two surveys.  The
bottom panel of Fig.~\ref{F:zpdetails} gives colour-magnitude
information for these $r<21$, $0.3\le z_s<0.35$ galaxies in the two
surveys.  As shown, the 
DEEP2 galaxies are both fainter and bluer on average than those in
zCOSMOS at this redshift.  This is consistent with the fact that the
redshift histograms show a local underdensity in
DEEP2 and a significant overdensity in zCOSMOS at this redshift.  We have
found that for this photoz method, the  photoz's are biased low for {\it
  blue} galaxies, but not red galaxies.  Hence, the different photoz
distributions in the top panel of
Fig.~\ref{F:zpdetails} reflect the different mixes of spectral types
and different $S/N$ detections of the galaxies in the two surveys at
this source redshift,
rather than some more ominous effect such as differences in
photometric calibration across the SDSS survey area.

We have confirmed that similar effects are at play in other parts of
the source redshift distribution (e.g. $0.6\le z_s<0.65$) that show
significant differences in weight between the two surveys in
Fig.~\ref{F:reconcile}.  In short, the cause of the different redshift
biases in the two surveys is the interplay between large-scale structure and photoz
errors, where LSS emphasizes certain spectral types that have
different photoz error properties.  (Explicit demonstration of how
this effect can come about 
will be shown in Section~\ref{SS:fluxlim}, where we show photoz error
distributions for ZEBRA/SDSS as a function of colour and magnitude.)
Even in the absence of our 
$z_p>z_l+0.1$ cut, the mean estimated $\Sigma_c^{-1}$ would have been
much higher in zCOSMOS than in DEEP2, giving the same sign of the
discrepancy between the surveys as we have  now (except in that case,
both $b_z(z_s)$ and $w(z_s)$ would be different, not just $w(z_s)$).
This interplay between photoz's and LSS is a problem
when trying to estimate the bias due to redshift calibration with a
reasonably small subsample of redshifts ($\sim 1000$) on a small area
of the sky.  It is also
avoidable in principle, if we use our sample with spectroscopic
redshifts to derive photoz
error distributions as a function of colour and magnitude, which may be
used to obtain accurate $\rmd p/\rmd z$ for each object.

\begin{figure}
\begin{center}
\includegraphics[width=3in,angle=0]{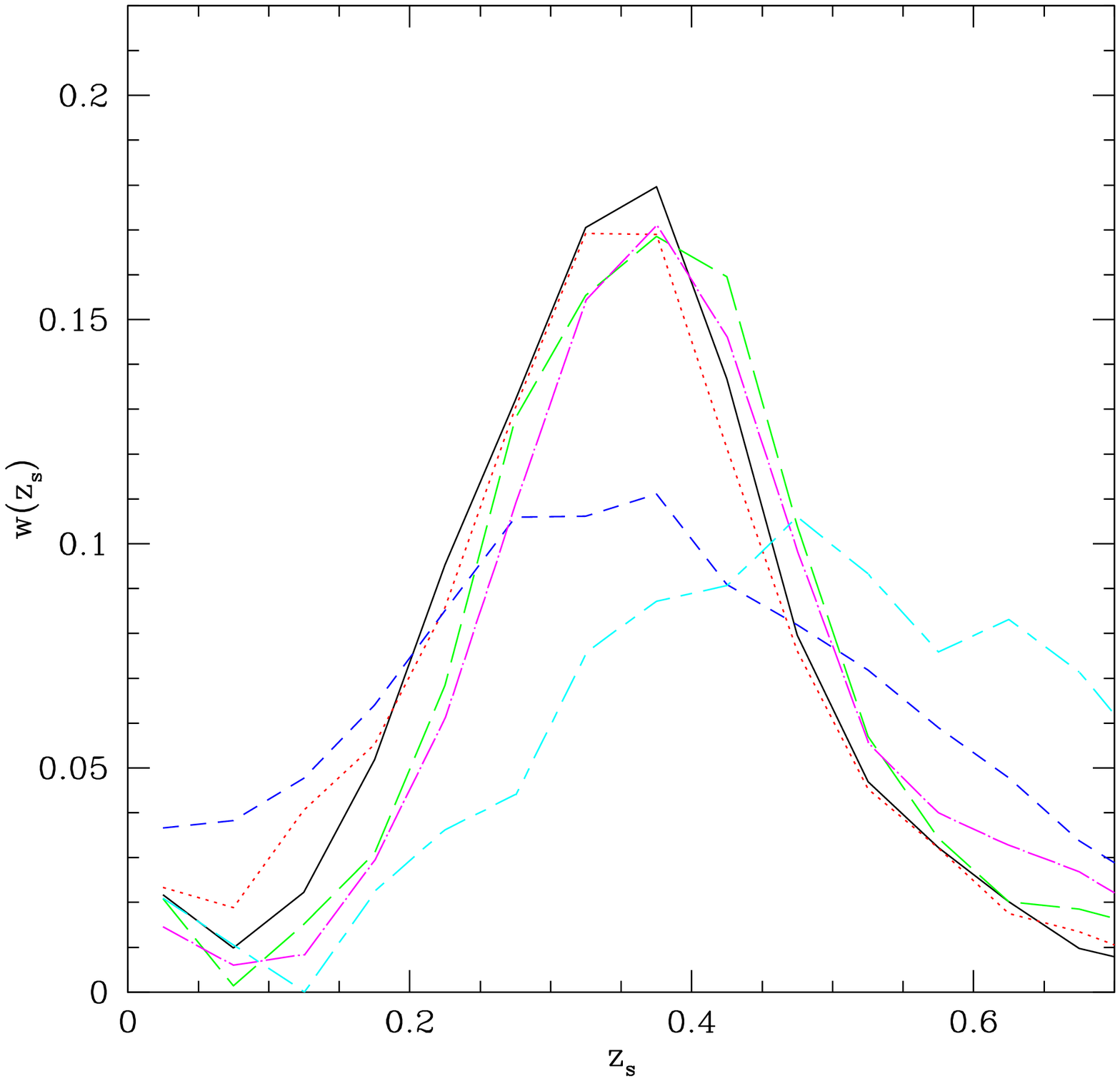}
\caption{\label{F:smoothwt}
Smoothed weight as a function of source redshift for several lens
redshifts in DEEP2 and zCOSMOS, to minimize the effects of LSS.  This
plot is a smoothed version of the lower right panel of
Fig.~\ref{F:reconcile}, with the same line types and colours as in
that plot.  
The smoothing algorithm is described in the text.}
\end{center}
\end{figure}

To confirm these findings, we have boxcar-smoothed the weights
$\tilde{w}_s(z_s)$ shown in Fig.~\ref{F:reconcile} with smoothing lengths of
$\Delta z_s=0.1$, $0.15$, and $0.2$ for $z_l=0.1$, $0.2$, and $0.3$
(larger smoothing lengths chosen for higher $z_l$ because the LSS
fluctuations in $w(z_s)$ are more 
significant there).  The resulting weight functions are
reasonably smooth, as shown in Fig.~\ref{F:smoothwt}, but include
some apparent mean offset in the redshift distributions for the two
surveys.  We find that the discrepancy between $\langle b_z\rangle$
for the two surveys is 5\%,
15\%, and 50\% smaller for $z_l=0.1$, $0.2$, and $0.3$ respectively
than when using the unsmoothed $w(z_s)$.
Most of the change arises from the DEEP2 mean calibration bias going to
lower (more negative) values, with the zCOSMOS mean calibration bias
changing only slightly.   
The apparent $5\sigma$ discrepancy in Table~\ref{T:difftable} for LRG
lenses is thus reduced due to this smoothing to a $2.5\sigma$
discrepancy, with the remaining discrepancy presumably due to the offset in the
weight histograms shown in Fig.~\ref{F:smoothwt}.  

We now ask if the LSS fluctuations are the cause of the $2\sigma$ discrepancy with the
other photoz methods.  As we will show later for ZEBRA/SDSS and have
confirmed for the template and neural net photoz algorithms (but
do not show here), it is a general tendency of these photoz algorithms to
underestimate the photoz's for blue galaxies, and slightly
overestimate them for red galaxies.  Consequently the same effect
occurs when the mixes of spectral types are different in the two
surveys, even when we are using another photoz algorithm, and this is
evident in $w(z_s)$ for each survey.  We therefore estimate using the
same method of boxcar smoothing the weight as a function of redshift
for each survey that
the $2\sigma$ discrepancies for these methods are really $1\sigma$.

We now address another unusual feature of the calibration
uncertainties in Table~\ref{T:difftable}: the uncertainties are
actually {\em smaller} for DEEP2 than for zCOSMOS (only slightly
larger than for the combined sample), despite the fact
that  sampling variance is $\sim 20$\% larger for DEEP2 EGS as for
zCOSMOS!  This result is also due  
to the LSS fluctuations in the weights for both surveys.  The DEEP2
mean calibration bias was, as we saw 
previously, significantly affected by this problem, and it is also 
responsible for making the errorbars artificially small (since 
our method of getting the errors does not allow $w(z_s)$ to vary as
much as it should in reality).  So, our worst-case $2.5\sigma$ and
$1\sigma$ calibration differences for LRG lenses (with kphotoz and
with the other photoz methods, respectively) is actually much 
less significant than these numbers suggest, and therefore not a
problem. 

We must ask whether this effect means that our mean results are biased
or our errorbars are too
optimistic when using the combined sample of galaxies for the two
surveys.  However, we are fortunate to be able to combine large samples at
completely 
different points on the sky.  The total (sample variance $+$ Poisson)
errors when using two uncorrelated fields with $N_1$ and $N_2$
galaxies are smaller than if we simply had a single field on the sky
with $N_1+N_2$ galaxies (which would be correlated with each other).

A comparison of
Fig.~\ref{F:reconcile} with Fig.~\ref{F:biaszs} can help us answer
this question.  In Fig.~\ref{F:reconcile}, it is clear that the weight
as a function of source redshift $w(z_s)$ for $z_l=0.2$ is not smooth
at all due to 
LSS-photoz error correlation in each survey.  The fluctuations
are at times $\sim 30$\% off from the value one might
expect if the curve is smooth.  However, in
Fig.~\ref{F:biaszs}, these curves for the combined sample are
significantly smoother, with 
fluctuations that are at most $20$\% for the LRG sources (the smallest
and most highly clustered sample) and even less for the other samples, $\sim
10$\%.  We thus conclude that the effect is reduced by a
factor of $\sim 3$, and is therefore negligible for the combined
sample. To verify this conclusion, we
have performed the same boxcar smoothing of the weight functions in
Fig.~\ref{F:biaszs} with
the same smoothing lengths as for the two survey subsamples, and found
that the resulting redshift calibration biases $\langle b_z\rangle$
for the combined sample changed by $< 0.5$\% for sm1--sm5, $<1$\% for sm6, sm7,
LRGs, and maxBCG lenses.  These changes are well within the $1\sigma$
errors on the calibration bias for these lens samples.

Finally, we  notice in the top panel of Fig.~\ref{F:zpdetails} that
our naive requirement that 
$z_p>z_l+0.1$ has required us to ignore a significant majority of the
galaxies in this redshift slice, all of which are actually lensed.
Since $z_l=0.3$ and the sources are all at true redshifts $z_s>0.3$,
we could conceivably use them all for lensing; using the subset at
$z_p>0.4$ eliminates a large fraction of these sources.  We return
to this point in sections~\ref{SS:puritycomp} and~\ref{SS:physassoc}.

\subsection{Size of errorbars on calibration bias}\label{SS:errorsize}

While we have previously asserted (section~\ref{SS:lss}) that
correlations between the bins in the redshift histograms should be
negligible, we now present tests of this assertion, which (if
violated) could cause the errorbars to be underestimated.  One reason
why they might be violated is the existence of a supercluster that happens to
lie partially within two histogram bins instead of entirely within one.  While
such a large LSS fluctuation is unlikely in an area of such small
comoving volume, we nonetheless present tests of this possibility.

As an example of a candidate supercluster, we find a large overdensity with
$0.34<z<0.38$ in zCOSMOS.    By plotting the detailed redshift distribution
in this region, we see that there are, in fact, $\sim 3$
large overdensities with line of sight separations of $\sim 80h^{-1}$Mpc
between them. Clusters that are separated by such a large separation are 
unlikely to be correlated: the correlation function for dark matter at 
this separation is 
$10^{-3}$, so the clusters would need to have bias of 
$\sim 30$ to have the correlation probability to become appreciable relative to 
a random distribution. There should be fewer 
than one cluster with such a high bias in an observable universe. While
magnification bias may increase the probability by a factor of a few
\citep{2007PhRvD..76j3502H}, it  
does so by invoking the cross-correlation between mass and galaxies,  so 
one loses one power of the bias, which therefore cannot bring the
correlations to a level comparable to unity.  These galaxy bias and
magnification bias effects are difficult to simulate realistically, so
we cannot turn to simulations to solve this problem. 
 
To test the effects
on the errorbars of the best-fit redshift distribution and on the
final calibration bias, we redo the analysis using bins of size
$\Delta z=0.1$, which will then include these structures all in one
bin.  We find that for zCOSMOS, this procedure increases the errors on the
final results by $30$\%, whereas the size of the errors for DEEP2 and
the combined sample (DEEP2 $+$ zCOSMOS) are essentially unaffected.

As an additional test, we shift the original histogram bins by $-0.02$ in redshift, so
that all three structures fall into the bin from $0.33\le z<0.38$.
We find that while the best-fit redshift histogram is unaffected, the
errors on it are significantly increased (by nearly a factor of $2$ in the
bins near this LSS fluctuation, and a smaller factor further away from
it).  To understand why it has such a large effect, we consider that
it adds an additional number of galaxies $\Delta N$ to the histogram in
that one bin.  The penalty on the fit $\Delta^2$
(Eq.~\ref{E:defdelta}) is therefore 
$(\Delta N)^2$.  When we consider splitting the fluctuation equally into
two bins (as we had effectively been doing before), the excess number
of galaxies in each bin is $0.5\Delta N$, leading to a $\Delta^2$
penalty of $2(0.5\Delta N)^2=0.5(\Delta N)^2$, half as much as if the entire
overdensity is in one bin.  The effect when fitting to the shifted
histogram using both surveys
together is nearly the same as when fitting zCOSMOS alone, whereas the
errors for DEEP2 
alone are unaffected (because our contrived bin-shifting did not
correlate with any LSS fluctuations in DEEP2).  

Given that these structures are likely to be uncorrelated, our 
bin-shifting that treated them as correlated leads to over-estimated
errors. On the other hand, 
our default binning puts one of them into one histogram bin,
and left the other two together; we may therefore suppose that our
errors for zCOSMOS and the combined sample are, in fact, slightly
over-estimated (since we effectively treated two of the structures as
correlated).   It is clear that the limited number of independent patches 
makes the error estimate from the bootstrap noisy, and while
our final results may be treated as having conservative errorbars, we cannot 
exclude the possibility that they may be  a factor of two larger.
However, this finding that the zCOSMOS errorbars may be overestimated may also
explain the fact that in the previous section, we found the
calibration of the lensing signal in DEEP2 to be constrained more
tightly than in zCOSMOS despite the fact that DEEP2 is smaller.

Finally, we note that bootstrapping $M$ data points $\gg M$ times will
in general lead to statistical uncertainty in the determined errors at the
$1/\sqrt{M}$ level.  For the case where we bootstrap a redshift histogram with
$24$ bins to get the best-fit redshift distribution, and use those
results to get errors on the lensing signal calibration uncertainty,
the errors are therefore reliable at the $\sim 20$\% level.  This
uncertainty is due to noise, rather than violation of the bootstrap
assumptions as in the rest of this section.

\subsection{Purity and completeness}\label{SS:puritycomp}

Here we address questions of purity and completeness of the source
sample for each  photoz method.  We define purity as
the fraction of the total estimated lensing weight that is attributed
to sources with spectroscopic redshift above the lens redshift (i.e.,
that are truly lensed).  Low
purity would be associated with a strong negative calibration bias.
Completeness can be defined by constructing the analogues of the
lensing weights in Eq.~(\ref{E:wj}), but using the true $\Sigma_c$ rather
than the estimated one.  We then define a ``true'' $w_j$ for each
object, and find the fraction of the total summed ``true'' weights
that is actually used by lensed sources defined using any given photoz
method.  Low completeness can occur because photoz's are scattered
low, so that we assume they are below the lens redshift.

These two issues, purity and completeness, are two of the three factors that
determine the statistical error on the lensing signal $\Delta\Sigma$
for a given photoz method as compared with the statistical error in
the optimal case where all lens and source redshifts are known.  The final
factor is how much a photoz method causes the weighting scheme to
deviate from optimal weighting.  We would like to estimate the total
increase in the error on the lensing signal due to all three factors
combined.

To do so, we consider the lensing signal estimator in the optimal
case where all lenses and sources are known.  In that case, we have a
shear $\gamma$, a critical surface density $\Sigma_c$, and weights
$w=1/(\Sigma_c\sigma_{\gamma})^2$.  (These weights are analogous to those
defined in Eq.~\ref{E:wj}, where $\sigma_{\gamma}$ comes from 
shape noise and measurement error added in quadrature.)  In this ideal case,
the lensing signal is
\begin{equation}
\Delta\Sigma = \frac{\sum w (\Sigma_c \gamma)}{\sum w}
\end{equation}
and its variance is 
\begin{equation}\label{E:idealvar}
\mbox{Ideal var}(\Delta\Sigma) = \frac{\sum w^2 \Sigma_c^2
  \sigma_{\gamma}^2}{(\sum w)^2} = \frac{\sum w}{(\sum w)^2} = \frac{1}{\sum w}.
\end{equation}

In reality, we have an estimated critical surface density
$\tilde{\Sigma}_c$, an estimated weight
$\tilde{w}=1/(\tilde{\Sigma}_c\sigma_{\gamma})^2$, and a calibration
bias defined via Eq.~(\ref{E:defbz}).  We can relate it to the true
lensing signal
\begin{equation}
\Delta\Sigma = \frac{\sum \tilde{w}(\tilde{\Sigma}_c\gamma)}{(1+b_z)\sum
  \tilde{w}},
\end{equation}
so its variance is
\begin{equation}\label{E:realvar}
\mbox{Real var}(\Delta\Sigma) = \frac{\sum \tilde{w}^2
  \tilde{\Sigma}_c^2 \sigma_{\gamma}^2}{(1+b_z)^2 (\sum \tilde{w})^2} =
  \frac{1}{(1+b_z)^2 (\sum \tilde{w})}.
\end{equation}
We then rearrange the definition of $b_z$ as follows:
\begin{equation}
1+b_z = \frac{\sum \tilde{w}(\tilde{\Sigma}_c\Sigma_c^{-1})}{\sum
  \tilde{w}} = \frac{\sum \sqrt{\tilde{w}w}}{\sum \tilde{w}}.
\end{equation}
Inserting this form for $1+b_z$ into equation~\ref{E:realvar}, we find
that
\begin{equation}\label{E:realvarsimple}
\mbox{Real var}(\Delta\Sigma) = \frac{\sum\tilde{w}}{(\sum\sqrt{\tilde{w}w})^2}
\end{equation}

Comparing equations~\ref{E:idealvar} and~\ref{E:realvarsimple}, we
find that
\begin{equation}\label{E:varratio}
\frac{\mbox{Ideal var}(\Delta\Sigma)}{\mbox{Real var}(\Delta\Sigma)} =
\frac{(\sum\sqrt{\tilde{w}w})^2}{(\sum w)(\sum\tilde{w})}.
\end{equation}
This ratio has the form of a correlation coefficient between
the square roots of the real and ideal weights for each lens-source
pair, and therefore is 
constrained to lie between 0 and 1 (not between -1 and 1 as for
correlation coefficients in general, since the
weights are strictly $\ge 0$).  It is only equal to one in the case
where the estimated weight $\tilde{w}$ is strictly proportional to the
ideal weight $w$.  This is as it should be: the
measured (``real'') variance of the lensing signal using a given
photoz method is always greater than or equal to the ideal variance.
This expression encodes all three possible ways the real measurement
can be degraded relative to the ideal one: via loss of lensed sources,
inclusion of sources that are not lensed, and non-optimal weighting.
This statistic is therefore another lensing-optimized metric than can
be used to classify photoz algorithms for g-g lensing purposes.

\begin{figure*}
\begin{center}
\includegraphics[width=5.8in,angle=0]{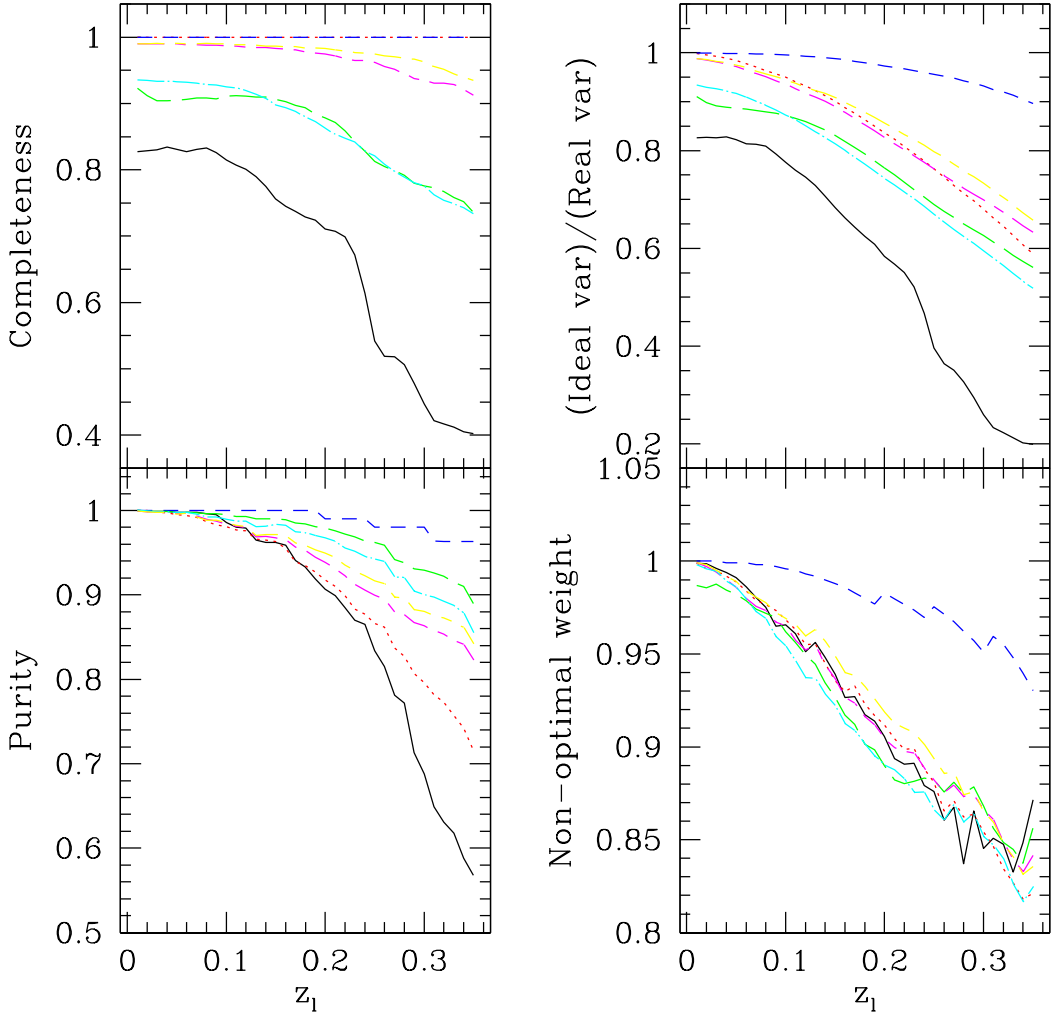}
\caption{\label{F:wtfrac}
Left: Completeness (top) and purity (bottom) as defined in the text
as a function of lens redshift.  Top right: The resulting ratio of
ideal to real variance for each method of source redshift
determination.  Bottom right: the derived change in variance due to
the non-optimal weighting.  Redshift determination methods are as
follows: solid black $=$ kphotoz ($r<21$), dotted red $=$ $r>21$ redshift
distribution, dashed blue $=$ high-redshift LRGs, long-dashed green $=$
template photoz's, long-short-dashed magenta $=$ NN/CC2 photoz's,
long-short-dashed yellow $=$ NN/D1 photoz's, and
dot-dashed cyan $=$ ZEBRA/SDSS.}
\end{center}
\end{figure*}

Fig.~\ref{F:wtfrac} shows the purities (bottom left), completenesses
(top left), the 
variance ratio (top right), and the implied change in variance due to
non-optimal weighting (bottom right) as a function
of lens redshift for each method.  
We first consider the completeness as a function of lens redshift in
the top left panel of Fig.~\ref{F:wtfrac}.  The results for kphotoz verify our
previous findings that the combination of a broad photoz error
distribution with our requirement that $z_p>z_l+0.1$ causes us to lose
a significant fraction of the available lensing weight.  The results
for the LRG source sample verify our previous assertions that the
photoz's for these sources are able to correctly put them all at high
redshift, so that we do not lose essentially any of them.  The
template photoz completeness is $\sim 80$\% on average, which is not
surprising given the significant failure mode to $z_p=0$ that causes
us to lose some sources.  The neural net photoz's (CC2 and D1) give the highest
completeness of all the photoz methods considered here (except the
highly specialized LRG source sample), in part due to the positive
mean photoz error.

In the lower left panel of Fig.~\ref{F:wtfrac}, we see the purity as a
function of lens redshift.   The swiftly declining purity above
$z_l=0.2$ for kphotoz is the main cause of the large negative
calibration bias for this method for higher redshift lens samples, and
is a result of large photoz error coupled with a lower mean redshift
for $r<21$ than the full samples used for the other photoz methods.  The LRG
source sample purity is 
uniformly high, dropping from $1$ at $z_l=0$ to a minimum of $0.96$ at
$z_l=0.35$.  This result attests to the efficiency of the colour cuts in
selecting only high-redshift sources, and the small size of the photoz
error distribution.  Of the other photoz methods, the template photoz
has the highest purity; the tendency towards a positive photoz error
seen previously for the NN and ZEBRA/SDSS photoz's cause a decline in
purity with redshift (though it is also the cause of their relatively high
completeness) just as it causes a negative calibration bias in the
lensing signal. 

The upper right panel of Fig.~\ref{F:wtfrac} shows the variance in the
ideal case relative to the true variance that results from using a
given photoz method.  For kphotoz, this number drops as low as $0.2$
for $z_l>0.3$, implying that the errors are a factor of
$\sqrt{1/0.2}\sim 2.2$ larger when using this photoz method than in the ideal
case.  ZEBRA/SDSS and the template photoz's give similar results for this
parameter, from $0.85$ at $z_l=0$ to $0.5$ at $z_l=0.35$, implying
errors ranging from $1.1$ to $1.4$ times the ideal.  The NN photoz's
give slightly better results than that, as does using a redshift
distribution for $r>21$ galaxies.  The high-redshift LRGs naturally
give nearly identical errors in reality than in the ideal case,
because the sources are at redshifts significantly higher than the
lenses, so any photoz errors cannot cause a significant deviation from
optimal weighting.  

Finally, the lower right panel shows the estimated change in variance
due to non-optimal weighting, obtained by taking the variance ratio
and dividing out the effects of impurity and incompleteness.  The
results suggest that for all source samples except the 
high-redshift LRGs, the non-optimal weighting is non-optimal has a
similar effect on the errors independent of photoz method, increasing
them by $\sim 7$\% at worst for this range of lens redshifts.


\subsection{Using $p(z)$ distributions}\label{SS:posterior}

Here we consider the possibility of using a full redshift probability
distribution, $p(z)$, for each
object, with two different sources of this distribution.  The first is the
posterior $p(z)$ from
the ZEBRA/SDSS method.  For
this method, $p(z)$ is 
determined by marginalizing over templates $T$ using the joint
redshift-template prior $P(z,T)$ and the likelihood $L(z,T)$ from the
fit $\chi^2$:
\begin{equation}
p(z) \propto \sum_T L(z,T) P(z,T)
\end{equation}

The second is a $p(z)$ distribution determined using some of the
machinery described in \cite{2007arXiv0708.0030O} but independently of
the photoz determination in that paper.  The photoz-independent
estimate of $p(z)$ (Cunha et al. 2007, in prep.) is calculated as
follows:  the training set comprised of 639~915 spectroscopic
objects from a variety of surveys is reweighted using the procedures
in \cite{2007arXiv0708.0030O} and Lima et al. (2007), in prep. to match
the joint, 5-dimensional probability distribution of the source
catalog for which we would like to obtain photoz's.  The five
parameters used to create this distribution are $u-g$, $g-r$, $r-i$,
$i-z$ colours and the $r$-band apparent magnitude.  The redshift
distribution of the weighted training set provides an estimate of the
true underlying distribution of the photometric sample. The estimate
of $p(z)$ for each galaxy in the photometric sample is given by the
weighted $z_{\rm spec}$ distribution of the 100 nearest training set
  neighbors in colour/magnitude space (the same 4-colours and
  $r$ band-magnitude mentioned above). Finally, to reduce the effects
  of Poisson noise, large-scale structure, and magnitude errors in the
  training sample, we adopt a "moving window" smoothing technique.  
We calculate $p(z)$ in 140 bins in the redshift range $0<z<2$ 
with a constant bin width of $0.067$. The $p(z)$ derived in this way
will be referred to as the NN $p(z)$, where NN in this context refers
to ``nearest neighbor'' rather than ``neural net.''

\begin{figure}
\begin{center}
\includegraphics[width=3in,angle=0]{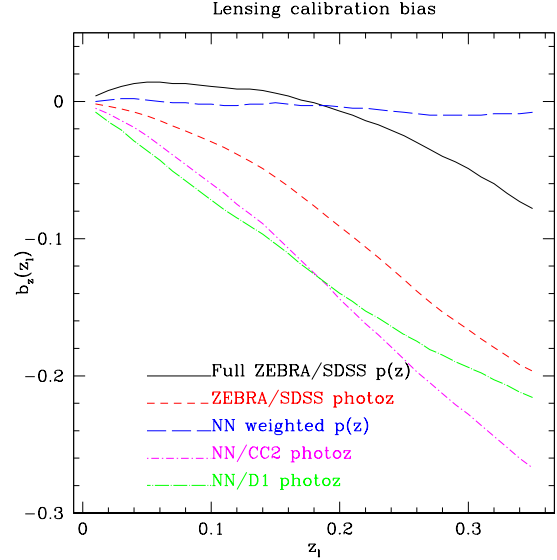}
\caption{\label{F:posterior}
Lensing calibration bias $b_z(z_l)$ using photoz's alone versus using
the full $p(z)$ to compute 
$\Sigma_c$ as described in the text.}
\end{center}
\end{figure}

In this section, we recompute $b_z(z_l)$ and $\langle
b_z\rangle$ for various lens redshift distributions, but instead of
using the photoz $z_p$ to get $\tilde{\Sigma}_c(z_l, z_s=z_p)$, we
integrate over the full $p(z)$ (normalized to integrate to
unity):
\begin{equation}
\tilde{\Sigma}_c^{-1}(z_l | p(z)) = \int_0^{\infty} p(z)
\Sigma_c^{-1}(z_l,z)\,\rmd z.
\end{equation}
We then compare the results using the two estimates of $p(z)$ to the results
using the photoz alone.  Figure~\ref{F:posterior} shows the
calibration bias $b_z$ as a function of $z_l$ using the 
photozs directly (as in Fig.~\ref{F:biaszl}) and the full estimates of
$p(z)$.   In Table~\ref{T:posterior}, we show the calibration bias 
averaged over various lens redshift distributions (as in
Table~\ref{T:avgbz}) using the full $p(z)$.   As
shown in both the figure and the table, most of the calibration bias
is eliminated when using the full $p(z)$ from either method.
\begin{table}
\begin{center}
\caption{Average calibration bias $\langle b_z\rangle$ for several
  lens redshift distributions using the full posterior $p(z)$ to get
  $\Sigma_c$.  The errors are approximately the same on the two
  columns.\label{T:posterior}}   
\begin{tabular}{ccc}
\hline\hline
Lenses & ZEBRA/SDSS $p(z)$ & NN $p(z)$ \\
\hline
sm1 & $0.013\pm 0.006$ & $-0.001$ \\
sm2 & $0.012\pm 0.007$ & $-0.001$ \\
sm3 & $0.011\pm 0.007$ & $-0.002$ \\
sm4 &  $0.009\pm 0.008$ & $-0.002$ \\
sm5 &  $0.005\pm 0.008$ & $-0.002$ \\
sm6 & $-0.002\pm 0.010$ & $-0.003$ \\
sm7 & $-0.013\pm 0.014$ & $-0.005$ \\
LRG & $-0.032\pm 0.018$ & $-0.007$ \\
maxBCG  & $-0.014\pm 0.013$ & $-0.006$ \\
\end{tabular}
\end{center}
\end{table}

The fact that the bias is nearly eliminated by using the full
posterior $p(z)$ is not a trivial result; when integrating over a
$p(z)$, there are many effects that will change the $\Sigma_c$
estimation in opposing directions.  We have determined that the reason
the negative calibration bias was nearly eliminated is the change in
$\tilde{\Sigma}_c$ for sources
with photoz near the lens redshift but slightly above it.  When using
the photoz alone, $\Sigma_c$ was on average underestimated due to the
way it varies with source redshift near the lens.  Integrating over
the full $p(z)$  raises it to a more reasonable value, which
both increases the signal calibration and lowers the weight given to
these sources.  

\begin{figure}
\begin{center}
\includegraphics[width=3in,angle=0]{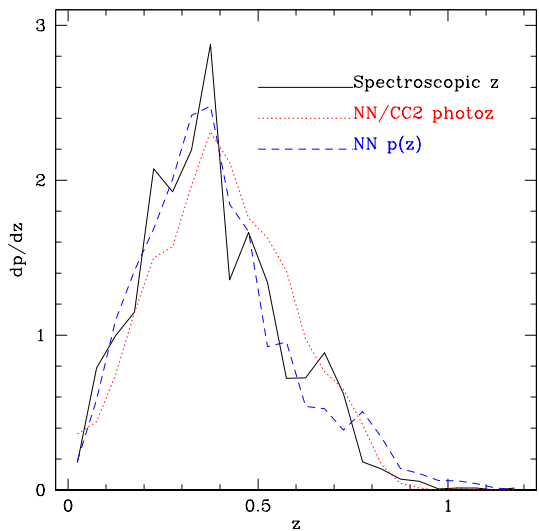}
\caption{\label{F:pzhist}
Redshift distribution $\rmd p/\rmd z$ for the full calibration sample
using spectroscopic redshifts, NN/CC2 photoz's, and the NN $p(z)$ for
each object.}
\end{center}
\end{figure}

To understand this result in more detail, we consider
Fig.~\ref{F:pzhist}, which shows the full spectroscopic sample
redshift distributions from spectroscopy, from the NN/CC2 photoz, and
from the summation of the $p(z)$ for each object.  As shown, the use
of $p(z)$ gives a mean redshift that is quite close to the mean
redshift of the full sample, unlike for the photoz's which gives a
higher mean redshift.  There is a slight suggestion that the $p(z)$ for
objects at $z\sim 0.6$ is getting spread to higher redshift, but these
objects are such a small fraction of the sample and the critical
surface density is not varying strongly with source redshift at these
high redshifts, so this
effect is not very important for lensing calibration with $z_l\lesssim
0.35$.  It is this correction to the mean redshift, in combination
with an inclusion of a realistic estimate of the scatter for each
object when estimating $\Sigma_c$, that eliminates the non-negligible
calibration bias when using NN/CC2 photoz's alone. 

\subsection{Avoiding physically-associated pairs}\label{SS:physassoc}

One benefit of using photoz's instead of a source redshift
distribution is that it is possible to eliminate some fraction of
the ``source'' galaxies that are physically associated with the lenses.  This
is important because of intrinsic alignments which can suppress the
lensing signal \citep{2006ApJ...644L..25A,2006MNRAS.372..758M}.  

In the absence of detailed calibration of the photoz error
distribution, we can simply 
require $z_s>z_l+\epsilon$ for some $\epsilon$, with the best chance
of success if the photoz method does not have a mean positive bias 
 $\langle z_p-z\rangle > 0$ for all redshifts for
which there are lenses.  Our current method  (kphotoz), the neural
network photoz's, and the ZEBRA/SDSS photoz's clearly fail this
criterion.  Of the methods under consideration here, only the SDSS
template photoz's are optimal for avoiding the inclusion of
physically-associated sources with this simple scheme.  This is due to
their negative photoz bias, which may be a liability in some other
applications and which may cause us to exclude so many sources that
the statistical error on the signal is strongly degraded.   

In the context of our previous work, the  plots in
  section~\ref{SS:agreement} make it quite apparent that our naive
  $z_s>z_l+0.1$ cut, while 
  the best we could do with only $162$ spectroscopic redshifts with
  which to determine the photoz error distribution, was causing us to
  eliminate a significant fraction of true, lensed sources from the
  analysis, without even fulfilling our purpose of excluding nearly all
  the physically associated sources.  

However, the existence of this analysis will help us fix this problem
for the future.  With detailed understanding of the photoz error distribution
from several thousand sources, we can simply construct a redshift 
distribution (see section~\ref{SS:fluxlim}) as a function of photoz,
source colour and 
magnitude.  This distribution will tell us $p(z | z_p, r,
\textrm{colour})$.  We can then choose to only use sources with  
\begin{equation}\label{E:pthres}
 \int_{z_l}^{\infty} p( z|z_p, r,
\textrm{colour})\,\rmd z > p_{thres}
\end{equation}
for some threshold probability $p_{thres}$.  The choice of $p_{thres}$
will depend on the situation: it should be large
for lens samples such as LRGs and clusters in which intrinsic
alignments of satellite ellipticities have been detected
\citep{2006ApJ...644L..25A,2006MNRAS.372..758M,2007ApJ...662L..71F},
and at small 
transverse separations ($\lesssim 200$kpc) where the effect is similar
to or larger than the statistical error.  In other scenarios,
such as at larger transverse separations, we
may find that we can afford a lower $p_{thres}$, even zero 
(because we are only using it to remove physically-associated sources
to avoid intrinsic 
alignment contamination, not those with zero shear).  A simpler
alternative to this procedure for ZEBRA/SDSS and other similar methods
that return a full posterior $p(z)$ is to perform the integral in
Eq.~\ref{E:pthres} using that $p(z)$, provided that it is found to
accurately describe the redshift distribution for galaxies of a given
magnitude and colour.

Note that once we have applied such a cut on the source sample, the
true redshift distribution of those sources is changed, so we must
re-estimate the lensing calibration bias, and if we had chosen to
deconvolve the photoz error distribution for more accurate estimation
of the critical surface density, we would have to redo this
procedure.  This is one major reason we have chosen to estimate the
calibration bias using photoz's directly.

Fig.~\ref{F:zperr} suggests that the optimal methods for the purpose of
excluding physically-associated sources with this more sophisticated
method are the NN and  ZEBRA/SDSS methods, because
of the lack of failure modes that will complicate this procedure
(i.e., because their error distributions are more 
compact, and therefore easier to sample fully using a spectroscopic
sample of limited size, and because the $p(z)$ will not be multimodal
as for the other methods).  This statement applies to samples of
galaxies reasonably similar to those presented here, but would need to
be re-evaluated for samples that are much deeper, bluer, and/or at
significantly higher redshift. 

\subsection{Without lensing selection}\label{SS:fluxlim}

Here we show some results for a full flux-limited sample of redshifts
from zCOSMOS and DEEP2.  The difference between these and the previous
results is that here, we do not imposed the lensing selection cuts.
Instead, we have simply required that there be a match in the SDSS
reductions (rerun 137) within 1'' of the spectrum from zCOSMOS or
DEEP2. 

\begin{figure}
\begin{center}
\includegraphics[width=3.0in,angle=0]{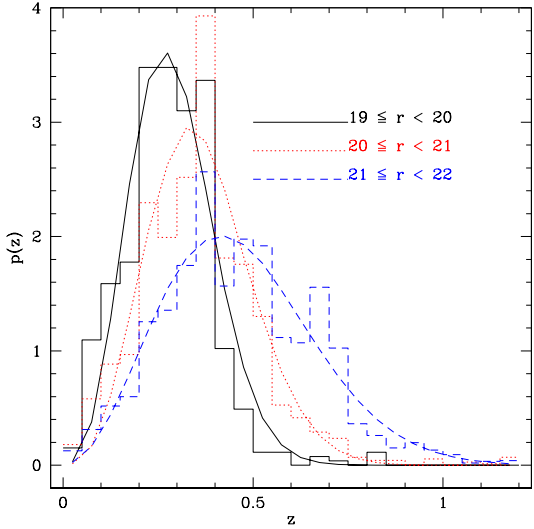}
\caption{\label{F:fluxdndz}Redshift distributions for all photometric
  galaxies without lensing selection.}
\end{center}
\end{figure}

For this test, we use $3415$ photometric galaxies from SDSS with
$r<22$ that have spectra from zCOSMOS (or zCOSMOS photoz's for the 8\%
with redshifts with reliability $<99\%$), and
$1761$ from DEEP2.  Figure~\ref{F:fluxdndz} shows the redshift
histograms $p(z)$ in magnitude bins one magnitude wide,  with
best-fit redshift distributions using 
the functional form in equation~\ref{E:zdist}.  The best-fit
parameters are tabulated in Table~\ref{T:fluxdndz}.  For these
results, we have again included the DEEP2 selection probabilities;
however the selection is so flat for the magnitude range shown here
that the effect on the final results is negligible.
\begin{table}
\begin{center}
\caption{Parameters of fits to redshift distribution from
  Eq.~(\ref{E:zdist}) for all photometric galaxies.\label{T:fluxdndz}} 
\begin{tabular}{ccccc}
\hline\hline
Sample & $N_{\rm gal}$ & $z_*$ & $\alpha$ & $\langle z \rangle$ \\
\hline
$\!\!19\le r<20\!\!\!$ & $529$ & $\!\!0.157\pm0.021\!\!$ & $4.04\pm 1.03$ & $\!\!0.290\pm0.015\!\!\!$ \\
$\!\!20\le r<21\!\!\!$ & $1446$ & $\!\!0.196\pm0.031\!\!$ & $4.15\pm 1.20$ & $\!\!0.363\pm0.013\!\!\!$ \\
$\!\!21\le r<22\!\!\!$ & $2996$ & $\!\!0.290\pm0.022\!\!$ & $3.08\pm 0.33$ & $\!\!0.467\pm0.017\!\!\!$ \\
\hline
\end{tabular}
\end{center}
\end{table}

\begin{figure}
\begin{center}
\includegraphics[width=3.0in,angle=0]{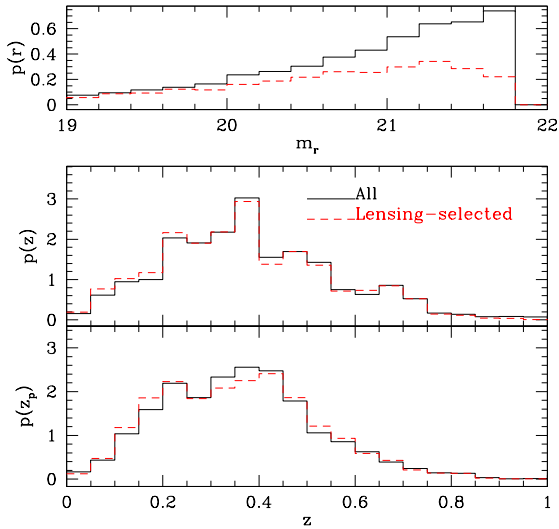}
\caption{\label{F:lensingsel}Magnitude (top), redshift (middle), and
  ZEBRA/SDSS photoz
  (bottom) histograms for the full flux-limited sample and for the lensing
  sources.  For the magnitude histogram, we have normalized both to
  the same number of galaxies so the fraction that pass our cuts as a
  function of magnitude will be apparent.  For the redshift and photoz
  histograms, the histograms for both the full and the
  lensing-selected sample are normalized to integrate to unity.}
\end{center}
\end{figure}

We also use these results to test the effects of lensing
selection.  As an example, we use the ZEBRA/SDSS
photoz's for this comparison.  Figure~\ref{F:lensingsel} shows the
effects of lensing selection on apparent magnitude, redshift, and
photoz histograms.  Here we require $r<21.8$ rather than $r<22$ in
order to compare more readily against our source catalog; this cut
reduces the number of matches in the flux-limited sample by 13\%.
The magnitude distribution in the flux-limited sample does not rise as
sharply as expected at the very faint end because of difficulties with
star/galaxy separation in SDSS.  A previous comparison with HST data
\citep{2001ASPC..238..269L} 
found that the default SDSS star/galaxy separation tends to err
on the side of putting more galaxies as stars rather than vice versa,
causing the galaxy counts to flatten for $r\gtrsim 21.5$ in a way that
depends on the seeing (more flattening in worse seeing).  

As shown, the lensing selection rate is a strong function of $r$-band
magnitude, ranging from nearly one around $r\sim 19$ to $\sim 0.3$
around the flux limit of $21.8$.  Nonetheless, the redshift
distribution 
is nearly the same for the full and the lensing-selected sample.  This
non-trivial result requires some explanation, since we have
already established (a) in 
the top panel of Fig.~\ref{F:lensingsel} that the flux-limited sample
is fainter on average than the lensing-selected sample, and (b) in
Fig.~\ref{F:fluxdndz}, that fainter samples are on average at higher
redshift.  A reconcilation of these facts would require that at a
given apparent magnitude, the lensing-selected sample is at higher
redshift than the flux-limited sample.

To explain this result, we consider two early-type galaxies at the same apparent
magnitude but different redshifts $z_1$ and $z_2>z_1$, in the limit
that the differences in 
their redshifts is small enough that the $k$-correction connecting the
bandpasses at the two redshifts is 
negligible.  In that case, the more distant galaxy is more luminous by
a factor of $[D_L(z_2)/D_L(z_1)]^2$ (where $D_L$ here is the
luminosity distance).  For early type galaxies, the physical size of
the galaxy is related to luminosity via $R \propto L^{1.4}$ (e.g.,
\citealt{2007AJ....133.1741B}), so the 
more distant galaxy is intrinsically larger than the more nearby one by a factor of
$[D_L(z_2)/D_L(z_1)]^{2.8}$.  The angular size of the more distant
galaxy relative to the more nearby one is smaller by a factor of
$D_A(z_1)/D_A(z_2)$ ($D_A$ is the 
angular diameter distance).  We therefore conclude that before
convolution with the PSF, the factor due to the intrinsic luminosity
and size difference wins out over the factor due to the decreased angular size,
so the more distant galaxy is actually larger.  This argument suggests
that if one of the galaxies will be eliminated due to our apparent size cut, it is the
one at {\em lower} redshift.  This counter-intuitive argument (which may
explain our finding above, that the lensing-selected redshift
distribution is the same as the flux-limited one despite being
brighter on average) is not
nearly the full story, because (a) in many situations, the
$k$-corrections or luminosity evolution will change the outcome of this
result, and (b) not all galaxies are early types following this
scaling relation between luminosity and size, but it appears to be a
strong enough effect that it balances out the difference in mean depth
between the samples.  One must also consider the effects of the
luminosity function, which means that the galaxies at the same
magnitude but higher redshift will be fewer in number, so while they
are less likely to be eliminated by an apparent size cut, they will
also be rarer to begin with.

As a test of this unexpected finding, we fit redshift distributions to
the lensing-selected galaxies as a function of apparent magnitude, and
compared to the mean redshifts in Table~\ref{T:fluxdndz}.  For
flux-limited samples, when using $19\le r<20$, $20\le r<21$, and
$21\le r<22$, we find mean redshifts of $0.290\pm0.015$,
$0.363\pm0.013$, and $0.467\pm0.017$.  For the lensing-selected
samples with the same cuts on apparent magnitude, we find mean
redshifts of $0.287\pm 0.015$ (well within $1\sigma$ of the
flux-limited sample), $ 0.372\pm 0.015$ ($0.5\sigma$ higher than the
flux-limited sample), and $0.484\pm 0.015$ ($0.7\sigma$ higher than
the flux-limited sample).  The results for the faintest sample are
most remarkable, because the flux-limited sample used for the fits is
cut at $r=22$, whereas the lensing-selected sample is cut at $r=21.8$,
so its mean magnitude is $0.2$ magnitudes brighter yet it is at
slightly higher redshift.  The effect is fortuituously of just the
right size that, despite the full lensing-selected sample being
brighter, the redshift distribution is nearly the same as for the
flux-limited sample.

Next, we present photoz error distributions as a function of colour and
magnitude for the full and the lensing-selected sample.  We split the
sample by colour because of the fact that photoz's are easier to
compute for red galaxies than for blue ones due to their clearer
colour-redshift relation.  Our colour separator is redshift-dependent
and purely empirical
based on the sample properties, $g-i = 0.7+2.67z$. The slope was
chosen to roughly trace the observed colour of the red ridge, with 40\%
of the galaxies classified as red.  Within each colour,
we then split into roughly equal numbers of galaxies based on
magnitude, so the magnitude bins are different for each colour.  While we tabulate
the mean photoz bias, $\langle z_p-z\rangle$ in analogy to earlier in
this paper, the plots show $p(z-z_p)$ since that can be used in
combination with $p(z_p)$ to reconstruct $p(z|r, g-i)$.  

\begin{figure}
\begin{center}
\includegraphics[width=3.0in,angle=0]{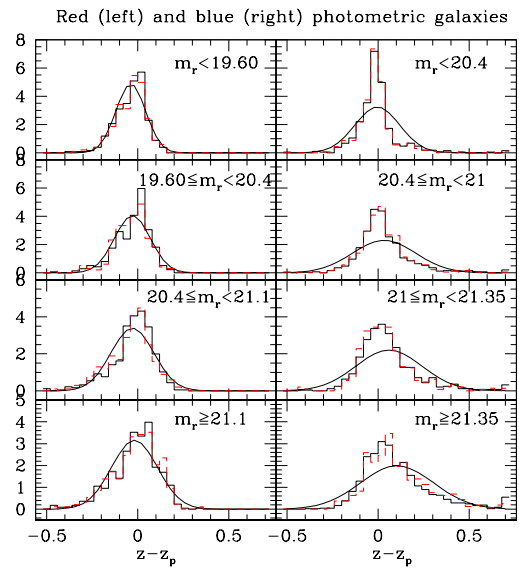}
\caption{\label{F:allzperr}Photoz error distributions for the
  SDSS/ZEBRA method as a function of
  colour and magnitude for the full flux-limited sample (black, solid) and for the lensing
  sources (red, dashed). We have also shown the Gaussians with the
  mean and scatter from Table~\ref{T:allzperr}.}
\end{center}
\end{figure}

Because it
would take a significant amount of space to present the distributions as a
function of photoz, we average them over all values of photoz.
Table~\ref{T:allzperr} shows the mean bias and scatter as a function
of colour and magnitude.  Figure~\ref{F:allzperr} shows the error
distributions as a function of colour and magnitude, and a 
Gaussian with the sample mean bias and scatter, to make any non-Gaussianity
apparent.
\begin{table}
\begin{center}
\caption{Mean photoz bias and scatter for the ZEBRA/SDSS algorithm as a function of
  colour and magnitude for all photometric and lensing-selected galaxies.\label{T:allzperr}} 
\begin{tabular}{cccccc}
\hline\hline
& & \multicolumn{2}{c}{Flux-limited} & \multicolumn{2}{c}{Lensing-selected} \\
$\!\!\!$Colour$\!\!\!$ & Magnitude & bias & $\!\!$scatter$\!\!$ & bias & $\!\!$scatter$\!\!$ \\
\hline
Red & $r<19.6$ & $0.038$ & $0.082$ & $0.039$ & $0.085$ \\
Red & $\!\!19.6\le r<20.4\!\!$  & $0.029$ & $0.098$ & $0.035$ & $0.101$ \\
Red & $\!\!20.4\le r<21.1\!\!$ & $0.029$ & $0.118$ & $0.039$ & $0.119$ \\
Red & $r\ge 21.1$ & $0.017$ & $0.126$ & $0.013$ & $0.126$ \\
Blue & $r<20.4$ & $0.004$ & $0.123$ & $0.008$ & $0.110$ \\
Blue & $20.4\le r<21.0$  & $-0.034$ & $0.173$ & $-0.025$ & $0.143$ \\
Blue & $\!\!21.0\le r<21.35\!\!$ & $-0.060$ & $0.181$ & $-0.043$ & $0.154$ \\
Blue & $r\ge 21.35$ & $-0.104$ & $0.201$ & $-0.114$ & $0.187$ \\
\hline
\end{tabular}
\end{center}
\end{table}

As shown in the table~\ref{T:allzperr}, the imposition of lensing selection seems to slightly
decrease the scatter for blue galaxies, but has little effect for red
galaxies.  Figure~\ref{F:allzperr} shows that for red galaxies, the
photoz error distributions are slightly non-Gaussian, whereas for blue
galaxies they are significantly non-Gaussian.  We also see the same
pattern as for kphotoz, a positive photoz bias for red galaxies and
negative for blue ones, and different sizes for the scatter.  These
trends will emphasize the correlation we have previously noted between
LSS and photoz error.  We have not attempted 
any more complex functional modeling, e.g. double Gaussians, but
future work will use the true distributions (smoothed) rather than the
Gaussians.

\subsection{Star/galaxy separation results}

We also matched our source catalog against a catalog of objects from
COSMOS with stellarity information.  Their space-based photometry
allows a more reliable 
star/galaxy classification than in SDSS.  Here we use their stellarity
information that is determined using both the Sextractor CLASS\_STAR
parameter and visual inspection, as follows:
\begin{itemize}
\item Those with CLASS\_STAR $\ge 0.8$ are automatically counted as
  stars, without visual inspection.
\item Those with CLASS\_STAR $<0.8$ are visually inspected,
  with the decision about star/galaxy classification made based on the
  inspection.  
\end{itemize}

Of the
7028 matches between the COSMOS catalog and our source catalog, 67 are
identified in COSMOS as 
stars, or $0.95$\%.  This number is constrained to be within $[0.74,
1.21]$\% at the 95\% CL assuming Poisson errors.  To check whether
this number is typical compared to the rest of the survey, we compute
the mean $r$-band seeing in the COSMOS area compared to the entire
SDSS survey area, and find that the mean seeing in the area that
overlaps with COSMOS is 1.20'' (PSF FWHM), compared to 1.18'' in the
rest of the survey.  We therefore conclude that this number is
fairly typical and may be applied as a correction to the entire source
catalog, provided that the stellar contamination fraction is not an
extremely strong function of the PSF FWHM.  

To test for this
possibility, we have used three SDSS runs that overlap the COSMOS
region and have $r$-band PSF FWHM ranging from $0.9$ to $1.4$, a range
that includes $\sim 85$\% 
of the source sample across the SDSS survey area.  We
then determined the stellar contamination fraction in bins of PSF FWHM
after application of all lensing selection criteria.  For the four
bins with median PSF FWHM of $1.02$, $1.14$, $1.21$, and $1.3$'', the
stellar contamination fractions are $1.04$\%, $0.92$\%, $0.79$\%, and
$0.56$\%.  The trend of decreasing stellar contamination in poorer
seeing is not well understood; however, the mean source number density
also decreases in poor seeing, so it seems that our cuts may be overly
conservative in regions of poor seeing.  This trend, when including
Poisson errorbars, is not quite significant at the $2\sigma$ level.
However, it is apparent that the stellar contamination fraction does
not shoot up rapidly in any part of this range of PSF FWHM including
nearly all the source sample, so we conclude that our value of
$0.95$\% should apply to the rest of the source catalog.

To properly apply this number to the rest of the source sample, we
must take into account that the number 
density of stars depends on galactic latitude in some complex way.
The average $\langle 1/\sin{b}\rangle$ for the whole source catalog is
1.40, and for the COSMOS region is 1.43, so we conclude that no
correction for the variation of stellar density with galactic latitude
is necessary.  While this calculation would not work if we included
regions where $\sin{b}\sim 0$ due to the strong increase in stellar
number density there, our requirement that $r$-band extinction be less
than $0.2$ magnitudes effectively eliminates these regions from the
source catalog.

\begin{figure}
\begin{center}
\includegraphics[width=3.0in,angle=0]{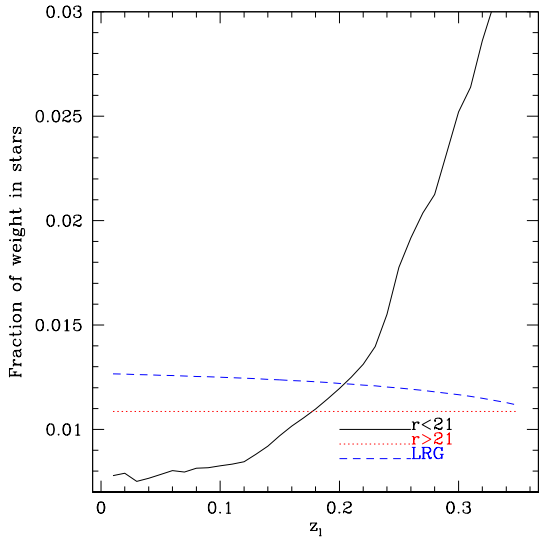}
\caption{\label{F:starweight}Fraction of the weight for our three
  source samples that is attributed to stellar contamination as a
  function of lens redshift.}
\end{center}
\end{figure}

However, we cannot conclude that  the fractional contamination in the lensing signal is
$-0.0095$, because it depends on the weight given to these
sources. The total fraction of the weight attributed to the 
stellar contamination as a function of lens redshift is shown in
Fig.~\ref{F:starweight} for the three source redshift determination
methods 
used in our current catalog.  As shown, the fraction of the weight
attributed to stars is in general larger than the actual stellar
contamination fraction.  This fraction rises significantly with
redshift for the $r<21$ sample because the stellar contamination tends
to be given relatively high photoz.  This is because the stellar
contamination is predominantly M stars that masquerade as
red galaxies at the high end of the redshift range for this sample.
However, as shown in 
Fig.~\ref{F:biaszl}, the $r>21$ sample has four times as much weight at
these lens redshifts, so the contamination to the signal is not strongly
affected by this increase in the contamination fraction for the $r<21$
sample.

\section{Discussion}\label{S:discussion}

In this paper, we have proposed a method for precision calibration of
the source redshift
distribution for g-g lensing with lens spectroscopy using
representative subsamples of the source catalog with
spectroscopy.  The key components of this method are an estimator for
the g-g lensing calibration bias (Eq.~\ref{E:defbz}) and for the
degradation of the statistical error due to non-optimal weighting
(Eq.~\ref{E:varratio}).  This method includes techniques for handling complications
such as large-scale structure in the spectroscopic redshift sample,
and redshift failure.   We then demonstrated its implementation by matching an 
SDSS lensing catalog used for many previous science works against a
sample of spectroscopic redshifts from DEEP2 and zCOSMOS.  We have
also used this method to assess the utility of three more recent photoz
algorithms that have been proposed for use with SDSS data.  In
Appendix~\ref{S:extension} we discuss the extension of these
techniques to g-g lensing with lens photoz's; with redshift
distributions for the lenses; and to cosmic shear. 

Our results in section~\ref{SS:prevwork} show that the galaxy-galaxy
lensing calibration bias  can be as high as 20--30\% for some of the
photoz methods, especially for higher lens redshifts.  
This is despite the fact that for all of the photoz methods,
the average redshift bias is well below the scatter. The 
reason for this finding is the nonlinear dependence of the critical 
surface density on the source redshift, which amplifies the photoz
errors in a highly asymmetric way: while an underestimate of photoz
to a value below the lens redshift leads to a rejection of the source
galaxy and does not produce lensing bias, an overestimate leads to 
an enhancement of lensing weight and can produce a significant bias. 
One of the main lessons of present work is that 
lensing applications require a dedicated photoz calibration, which can
give very different results from the general photoz calibration tests.  

Our analysis demonstrates that the calibration bias in the lensing
signal due to redshift distribution uncertainty in previous 
works using the SDSS source catalog used for several previous science
projects was well within the quoted systematic
error of $8$\%.  Future lensing work using this source catalog will
use the results in this paper to obtain a highly accurate lensing
calibration with a smaller uncertainty than in our previous work.
The decreased systematic error budget due to redshift calibration
uncertainty, which is now known to $\sim 2$\% due to this work, is a
timely improvement to SDSS g-g lensing measurements: results coming
out in the next year will have total statistical error of $\sim 5$\%,
so the reduction in the systematic error is necessary to ensure that
it does not exceed the statistical error.

For the three new photoz methods tested here, we have measured the
lensing calibration bias using a statistic $b_z$ (Eq.~\ref{E:defbz})
which is optimized 
for characterization of photoz's for galaxy-galaxy lensing purposes.
Another statistic, in Eq.~\ref{E:varratio}, can be used to determine
how much a photoz method causes a deviation from optimal weighting,
affecting the statistical error of the measurement.  We have also
carefully identified important aspects of the photoz 
error distribution.  We found that for our source sample, 
using the SDSS template photoz's (without any corrections for mean
photoz bias) led to the smallest lensing 
calibration bias. This result is due to a fortuitous cancellation of
lensing calibration biases due to photoz bias and scatter, and would
not necessarily happen with a sample with different selection
criteria.  While for some
applications, the presence of a failure mode that sends sources to
zero redshift would be quite problematic, it does not cause any bias
for lensing (though as we have already shown, it leads to increased
statistical error on the lensing signal).   The SDSS neural net 
photoz's and the ZEBRA/SDSS photoz's both cause significant lensing
calibration bias, despite having a 
reasonable scatter, because of a significant positive photoz bias for
$0<z<0.4$.  This calibration bias can be corrected for after
computation of the lensing signal using a calibration factor, since
our spectroscopic sample has 
the same selection as the full catalog.  If the mean photoz bias is
corrected for before computing the lensing signal, the SDSS neural net
photoz's lead to smaller lensing calibration bias than the other two
new methods, implying that the effects of photoz scatter are smaller
for this method.  On some level, once a reliable calibration of the
photoz's for lensing is known for a given source sample, the fact that
a photoz method causes calibration bias is unimportant: the
deterioration of the statistical error due to the non-optimal
weighting, and the inability to properly remove physically-associated
sources, are both more important.  In that sense, the negative photoz
bias of the template photoz code, which is the cause of its low
lensing calibration bias, may in fact be a liability for its practical
use. 


We have isolated ways that sampling variance can
complicate the estimation of redshift calibration bias  using a
small subsample of galaxies.  Because LSS tends to change the fractions
of blue and red galaxies, which generally have different photoz error
distributions, it can bias the estimated lensing calibration bias
$\langle b_z\rangle$, and can also artificially reduce the error.  We
have verified that our use of two 
degree-scale uncorrelated redshift samples drastically reduces this
effect, making it negligible for our analysis.

We have also assessed the level of stellar contamination in our source
catalog using COSMOS data, and have placed stringent limits on the
systematic error due to this contamination.

We have tested the use of a full
$p(z)$ for estimation of the critical surface density, and
find that it tends to give superior results to the use of the photoz alone,
with calibration biases consistent with zero for all lens redshift
distributions considered in this paper.  Because of this success, we
advocate further work 
exploring the use of a full $p(z)$ for lensing rather than a single
photoz for each object.  

We have learned that  the details
of the photoz bias and scatter as a function of redshift are
important. For example, the mean bias for sources with redshift within
$\Delta z\sim 0.2$ of the lenses is more important than the overall
mean photoz bias.  In the extension of this formalism to higher
redshift, it is important to consider that both the size of the photoz
error and the derivative $\rmd \Sigma_c/\rmd z_s$ determine the
redshift calibration bias, so deeper surveys that can ensure a larger
separation between the lenses and sources may find smaller redshift
calibration bias even with comparable or larger photoz errors than for
the methods demonstrated here.  However, these deeper surveys may have
a larger systematic uncertainty due to spectroscopic redshift failure:
our high redshift success rate meant that we were not very sensitive
to this problem, but that high success rate was also a product of the
relatively bright magnitude of the source sample.  

For deeper surveys with a
higher redshift failure rate, one can imagine two possible scenarios.
The first is that the higher failure rate is due to the lower $S/N$ of
the spectra.  In that case, the failure rate as a function of apparent
magnitude and colour can be quantified, and included as a weight in the
lensing calibration bias calculation.  We would assume that for a given
magnitude and colour the redshift distribution is properly being
sampled despite redshift failure, so we upweight those in regions of
parameter space where failure is more likely.  The second case is more
pernicious: if there is a region of colour and magnitude space for
which essentially all the redshifts are failures, then no amount of
reweighting will be able to account for this.  Consequently, for
proper redshift calibration, one would need to either remove those
sources entirely due to the impossibility of calibration, or get
external information from some other spectrograph that is capable of
obtaining redshifts for that region of colour space.

In summary, the results in this work resoundingly verify our claim
that the 
spectroscopic sample used to assess photoz error for lensing purposes
must have the same selection as the source catalog, or selection close
enough that it can be made comparable by a reweighting scheme (see
Section~\ref{SS:failures}). The photoz 
error is a strong function of galaxy type and apparent magnitude, and
the lensing calibration is a very sensitive to details of the photoz
error distribution.  We have also shown 
that at least two independent degree-scale patches of the sky must be surveyed in
order  
to suppress the sampling variance effects on photoz calibration (this
choice would have to be re-evaluated for deeper surveys, as would our
choice of redshift histogram bins $\Delta z=0.05$). 
Having two independent spectroscopic surveys, DEEP2 and zCOSMOS, 
with nearly 3000 galaxies in total, allowed us to provide photoz
calibration of the galaxy-galaxy lensing signal at a percent level,
depending on the lens sample.  
As more spectroscopic redshift surveys become available, it will
become easier for weak lensing measurements to be carried out with
tight constraints on the redshift calibration bias using this method.
This is one more important step on the way towards galaxy-galaxy
lensing becoming a high-precision tool for addressing questions of
astrophysical and cosmological importance. 
Similar calibration methods must be developed and applied also to other 
weak lensing applications, most notably 
 galaxy-galaxy lensing in the case where lens redshifts
are not known, and shear-shear autocorrelations; we discuss the steps
that would be needed for such a process in Appendix~\ref{S:extension}.

\section*{Acknowledgments}
\label{sec:acknowledgments}

R.M. is supported by NASA 
through Hubble Fellowship grant \#HST-HF-01199.02-A awarded by the
Space Telescope Science Institute, which is operated by the
Association of Universities for Research in Astronomy, Inc., for NASA, 
under contract NAS 5-26555.  U.S. is supported by the
Packard Foundation 
and NSF CAREER-0132953. We thank Josh Frieman, Marcos
Lima, Huan Lin, Hiro Oyaizu, Nikhil Padmanabhan, and Erin Sheldon for
useful discussion 
about a variety of topics addressed in this paper.

Funding for the SDSS and SDSS-II has been provided by the Alfred
P. Sloan Foundation, the Participating Institutions, the National
Science Foundation, the U.S. Department of Energy, the National
Aeronautics and Space Administration, the Japanese Monbukagakusho, the
Max Planck Society, and the Higher Education Funding Council for
England. The SDSS Web Site is http://www.sdss.org/. 

The SDSS is managed by the Astrophysical Research Consortium for the
Participating Institutions. The Participating Institutions are the
American Museum of Natural History, Astrophysical Institute Potsdam,
University of Basel, University of Cambridge, Case Western Reserve
University, University of Chicago, Drexel University, Fermilab, the
Institute for Advanced Study, the Japan Participation Group, Johns
Hopkins University, the Joint Institute for Nuclear Astrophysics, the
Kavli Institute for Particle Astrophysics and Cosmology, the Korean
Scientist Group, the Chinese Academy of Sciences (LAMOST), Los Alamos
National Laboratory, the Max-Planck-Institute for Astronomy (MPIA),
the Max-Planck-Institute for Astrophysics (MPA), New Mexico State
University, Ohio State University, University of Pittsburgh,
University of Portsmouth, Princeton University, the United States
Naval Observatory, and the University of Washington. 

Funding for the DEEP2 survey has been provided by NSF grants
AST-0071048, AST-0071198, AST-0507428, and AST-0507483.  

Some of the data presented herein were obtained at the W. M. Keck
Observatory, which is operated as a scientific partnership among the
California Institute of Technology, the University of California and
the National Aeronautics and Space Administration. The Observatory was
made possible by the generous financial support of the W. M. Keck
Foundation. The DEEP2 team and Keck Observatory acknowledge the very
significant cultural role and reverence that the summit of Mauna Kea
has always had within the indigenous Hawaiian community and appreciate
the opportunity to conduct observations from this mountain. 

\bibliography{apjmnemonic,cosmo,cosmo_preprints}

\begin{thebibliography}{}

\bibitem[\protect\citeauthoryear{{Abazajian} et~al.}{2003}]{2003AJ....126.2081A}
{Abazajian} K. et~al. 2003, \aj, 126, 2081

\bibitem[\protect\citeauthoryear{{Abazajian} et~al.}{2004}]{2004AJ....128..502A}
{Abazajian} K. et~al. 2004, \aj, 128, 502

\bibitem[\protect\citeauthoryear{{Abazajian} et~al.}{2005}]{2005AJ....129.1755A}
{Abazajian} K. et~al. 2005, \aj, 129, 1755

\bibitem[\protect\citeauthoryear{{Abdalla} et~al.}{2007}]{2007arXiv0705.1437A}
{Abdalla} F.~B.,  {Amara} A.,  {Capak} P.,  {Cypriano} E.~S.,  {Lahav} O.,
  {Rhodes} J.,  2007, preprint (arXiv:0705.1437)

\bibitem[\protect\citeauthoryear{{Adelman-McCarthy} et~al.}{2006}]{2006ApJS..162...38A}
{Adelman-McCarthy} J.~K. et~al. 2006, \apjs, 162, 38

\bibitem[\protect\citeauthoryear{{Adelman-McCarthy} et~al.}{2007a}]{2007ApJS..172..634A}
{Adelman-McCarthy} J.~K. et~al.  2007a, \apjs, 172, 634

\bibitem[\protect\citeauthoryear{{Adelman-McCarthy} et~al.}{2007b}]{2007arXiv0707.3413A}
{Adelman-McCarthy} J.~K. et~al. 2007b, preprint (arXiv:0707.3413)

\bibitem[\protect\citeauthoryear{{Agustsson} \& {Brainerd}}{{Agustsson} \&
  {Brainerd}}{2006}]{2006ApJ...644L..25A}
{Agustsson} I.,  {Brainerd} T.~G.,  2006, \apjl, 644, L25

\bibitem[\protect\citeauthoryear{{Altay}, {Colberg} \& {Croft}}{{Altay}
  et~al.}{2006}]{2006MNRAS.370.1422A}
{Altay} G.,  {Colberg} J.~M.,    {Croft} R.~A.~C.,  2006, \mnras, 370, 1422

\bibitem[\protect\citeauthoryear{{Bartelmann} \& {Schneider}}{{Bartelmann} \&
  {Schneider}}{2001}]{2001PhR...340..291B}
{Bartelmann} M.,  {Schneider} P.,  2001, \physrep, 340, 291

\bibitem[\protect\citeauthoryear{{Bernardi} et~al.}{2007}]{2007AJ....133.1741B}
{Bernardi} M.,  {Hyde} J.~B.,  {Sheth} R.~K.,  {Miller} C.~J.,    {Nichol}
  R.~C.,  2007, \aj, 133, 1741

\bibitem[\protect\citeauthoryear{{Bernstein}}{{Bernstein}}{2006}]{2006ApJ...63%
7..598B}
{Bernstein} G.,  2006, \apj, 637, 598

\bibitem[\protect\citeauthoryear{{Bernstein} \& {Jain}}{{Bernstein} \&
  {Jain}}{2004}]{2004ApJ...600...17B}
{Bernstein} G.,  {Jain} B.,  2004, \apj, 600, 17

\bibitem[\protect\citeauthoryear{{Bernstein} \& {Ma}}{{Bernstein} \&
  {Ma}}{2007}]{2007arXiv0712.1562B}
{Bernstein} G.,  {Ma} Z.,  2007, preprint (arXiv:0712.1562)

\bibitem[\protect\citeauthoryear{{Blanton}
  et~al.}{2003a}]{2003AJ....125.2348B}
{Blanton} M.~R. et~al.  2003a, \aj,  125, 2348

\bibitem[\protect\citeauthoryear{{Blanton} et~al.}{2003b}]{2003AJ....125.2276B}
{Blanton} M.~R.,  {Lin} H.,  {Lupton} R.~H.,  {Maley} F.~M.,  {Young} N.,
  {Zehavi} I.,    {Loveday} J.,  2003b, \aj, 125, 2276

\bibitem[\protect\citeauthoryear{{Brainerd}, {Blandford} \& {Smail}}{{Brainerd}
  et~al.}{1996}]{1996ApJ...466..623B}
{Brainerd} T.~G.,  {Blandford} R.~D.,    {Smail} I.,  1996, \apj, 466, 623

\bibitem[\protect\citeauthoryear{{Brodwin}
    et~al.}{2006}]{2006ApJS..162...20B}
{Brodwin} M., {Lilly} S.~J., {Porciani} C., {McCracken} H.~J., 
        {Le F{\`e}vre} O., {Foucaud} S., {Crampton} D., 
        {Mellier} Y., 2006, \apjs, 162, 20

\bibitem[\protect\citeauthoryear{{Budav{\' a}ri} et~al.}{2000}]{2000AJ....120.1588B}
{Budav{\' a}ri} T.,  {Szalay} A.~S.,  {Connolly} A.~J.,  {Csabai} I.,
  {Dickinson} M.,  2000, \aj, 120, 1588

\bibitem[\protect\citeauthoryear{{Capak} et~al.}{2007}]{2007ApJS..172...99C}
{Capak} P. et~al. 2007, \apjs, 172, 99

\bibitem[\protect\citeauthoryear{{Coil}
  et~al.}{2004}]{2004ApJ...609..525C}
{Coil} A.~L. et~al. 2004, \apj, 609,
  525

\bibitem[\protect\citeauthoryear{{Coleman}, {Wu} \& {Weedman}}{{Coleman}
  et~al.}{1980}]{1980ApJS...43..393C}
{Coleman} G.~D.,  {Wu} C.-C.,    {Weedman} D.~W.,  1980, \apjs, 43, 393

\bibitem[\protect\citeauthoryear{{Collister} \& {Lahav}}{{Collister} \&
  {Lahav}}{2004}]{2004PASP..116..345C}
{Collister} A.~A.,  {Lahav} O.,  2004, \pasp, 116, 345

\bibitem[\protect\citeauthoryear{{Csabai}
  et~al.}{2003}]{2003AJ....125..580C}
{Csabai} I. et~al. 2003, \aj, 125, 580

\bibitem[\protect\citeauthoryear{{Davis} et~al.}{2003}]{2003SPIE.4834..161D}
{Davis} M. et~al.  2003, in Discoveries and Research Prospects
  from 6- to 10-Meter-Class Telescopes II. Edited by Guhathakurta, Puragra.
  Proceedings of the SPIE, Volume 4834, pp. 161-172 (2003).

\bibitem[\protect\citeauthoryear{{Davis} et~al.}{2005}]{2005ASPC..339..128D}
{Davis} M.,  {Gerke} B.~F.,  {Newman} J.~A.,    {the Deep2 Team}, 2005,
  {Constraining Dark Energy with the DEEP2 Redshift Survey}

\bibitem[\protect\citeauthoryear{{Davis} et~al.}{2007}]{2007ApJ...660L...1D}
{Davis} M. et~al., \apjl, 660, L1

\bibitem[\protect\citeauthoryear{{Eisenstein} et~al.}{2001}]{2001AJ....122.2267E}
{Eisenstein} D.~J. et~al.  2001, \aj, 122, 2267

\bibitem[\protect\citeauthoryear{{Faber} et~al.}{2003}]{2003SPIE.4841.1657F}
{Faber} S.~M. et~al.  2003, in {Iye} M.,
  {Moorwood} A.~F.~M.,  eds, Instrument Design and Performance for
  Optical/Infrared Ground-based Telescopes,  Proceedings of the SPIE,
  4841, 1657-1669

\bibitem[\protect\citeauthoryear{{Faltenbacher}
  et~al.}{2007}]{2007ApJ...662L..71F}
{Faltenbacher} A.,  {Li} C.,  {Mao} S.,  {van den Bosch} F.~C.,  {Yang} X.,
  {Jing} Y.~P.,  {Pasquali} A.,    {Mo} H.~J.,  2007, \apjl, 662, L71

\bibitem[\protect\citeauthoryear{{Feldmann} et~al.}{2006}]{2006MNRAS.372..565F}
{Feldmann} R. et~al. 
  2006, \mnras, 372, 565

\bibitem[\protect\citeauthoryear{{Finkbeiner} et~al.}{2004}]{2004AJ....128.2577F}
{Finkbeiner} D.~P. et~al.  2004, \aj, 128, 2577

\bibitem[\protect\citeauthoryear{{Fischer} et~al.}{2000}]{2000AJ....120.1198F}
{Fischer} P. et~al. 2000, \aj, 120,
  1198

\bibitem[\protect\citeauthoryear{{Fukugita} et~al.}{1996}]{1996AJ....111.1748F}
{Fukugita} M.,  {Ichikawa} T.,  {Gunn} J.~E.,  {Doi} M.,  {Shimasaku} K.,
  {Schneider} D.~P.,  1996, \aj, 111, 1748

\bibitem[\protect\citeauthoryear{{Gunn} et~al.}{1998}]{1998AJ....116.3040G}
{Gunn} J.~E. et~al. 1998, \aj,
  116, 3040

\bibitem[\protect\citeauthoryear{{Gunn} et~al.}{2006}]{2006AJ....131.2332G}
{Gunn} J.~E. et~al.  2006, \aj, 131, 2332

\bibitem[\protect\citeauthoryear{{Heymans} et~al.}{2006a}]{2006MNRAS.371L..60H}
{Heymans} C. et~al.  2006a, \mnras, 371, L60

\bibitem[\protect\citeauthoryear{{Heymans} et~al.}{2006b}]{2006MNRAS.368.1323H}
{Heymans} C. et~al.  2006b,
  \mnras, 368, 1323

\bibitem[\protect\citeauthoryear{{Heymans} et~al.}{2006c}]{2006MNRAS.371..750H}
{Heymans} C.,  {White} M.,  {Heavens} A.,  {Vale} C.,    {van Waerbeke} L.,
  2006c, \mnras, 371, 750

\bibitem[\protect\citeauthoryear{{Hirata} \& {Seljak}}{{Hirata} \&
  {Seljak}}{2003}]{2003MNRAS.343..459H}
{Hirata} C.,  {Seljak} U.,  2003, \mnras, 343, 459

\bibitem[\protect\citeauthoryear{{Hirata} et~al.}{2004}]{2004MNRAS.353..529H}
{Hirata} C.~M. et~al.  2004, \mnras, 353, 529

\bibitem[\protect\citeauthoryear{{Hoekstra} et~al.}{2005}]{2005ApJ...635...73H}
{Hoekstra} H.,  {Hsieh} B.~C.,  {Yee} H.~K.~C.,  {Lin} H.,    {Gladders} M.~D.,
   2005, \apj, 635, 73

\bibitem[\protect\citeauthoryear{{Hoekstra}, {Yee} \& {Gladders}}{{Hoekstra}
  et~al.}{2004}]{2004ApJ...606...67H}
{Hoekstra} H.,  {Yee} H.~K.~C.,    {Gladders} M.~D.,  2004, \apj, 606, 67

\bibitem[\protect\citeauthoryear{{Hogg} et~al.}{2001}]{2001AJ....122.2129H}
{Hogg} D.~W.,  {Finkbeiner} D.~P.,  {Schlegel} D.~J.,    {Gunn} J.~E.,  2001,
  \aj, 122, 2129

\bibitem[\protect\citeauthoryear{{Hudson} et~al.}{1998}]{1998ApJ...503..531H}
{Hudson} M.~J.,  {Gwyn} S. D.~J.,  {Dahle} H.,    {Kaiser} N.,  1998, \apj,
  503, 531+

\bibitem[\protect\citeauthoryear{{Hui}, {Gazta{\~n}aga} \& {Loverde}}{{Hui}
  et~al.}{2007}]{2007PhRvD..76j3502H}
{Hui} L.,  {Gazta{\~n}aga} E.,    {Loverde} M.,  2007, \prd, 76, 103502

\bibitem[\protect\citeauthoryear{{Huterer} et~al.}{2006}]{2006MNRAS.366..101H}
{Huterer} D.,  {Takada} M.,  {Bernstein} G.,    {Jain} B.,  2006, \mnras, 366,
  101

\bibitem[\protect\citeauthoryear{{Ilbert} et~al.}{2006}]{2006A&A...457..841I}
{Ilbert} O. et~al.  2006, \aap, 457, 841

\bibitem[\protect\citeauthoryear{{Ivezi{\' c}} et~al.}{2004}]{2004AN....325..583I}
{Ivezi{\' c}} {\v Z}. et~al. 2004, Astronomische Nachrichten, 325, 583

\bibitem[\protect\citeauthoryear{{Jain} \& {Taylor}}{{Jain} \&
  {Taylor}}{2003}]{2003PhRvL..91n1302J}
{Jain} B.,  {Taylor} A.,  2003, PRL, 91, 141302

\bibitem[\protect\citeauthoryear{{Koester} et~al.}{2007a}]{2007ApJ...660..239K}
{Koester} B.~P. et~al. 2007a, \apj, 660, 239

\bibitem[\protect\citeauthoryear{{Koester}
  et~al.}{2007b}]{2007ApJ...660..221K}
{Koester} B.~P. et~al.  2007b, \apj, 660, 221

\bibitem[\protect\citeauthoryear{{LeFevre} et~al.}{2003}]{2003SPIE.4841.1670L}
{LeFevre} O. et~al.  2003, in {Iye} M.,  {Moorwood} A.~F.~M.,  eds,
  Instrument Design and Performance for Optical/Infrared Ground-based
  Telescopes. Proceedings of the
  SPIE, Volume 4841, pp. 1670-1681 (2003).

\bibitem[\protect\citeauthoryear{{Lilly} et~al.}{2007}]{2007ApJS..172...70L}
{Lilly} S.~J. et~al. 2007, \apjs, 172, 70

\bibitem[\protect\citeauthoryear{{Limousin}
  et~al.}{2007}]{2007A&A...461..881L}
{Limousin} M.,  {Kneib} J.~P.,  {Bardeau} S.,  {Natarajan} P.,  {Czoske} O.,
  {Smail} I.,  {Ebeling} H.,    {Smith} G.~P.,  2007, \aap, 461, 881

\bibitem[\protect\citeauthoryear{{Lupton} et~al.}{2001}]{2001ASPC..238..269L}
{Lupton} R.~H.,  {Gunn} J.~E.,  {Ivezi{\' c}} Z.,  {Knapp} G.~R.,  {Kent} S.,
   {Yasuda} N.,  2001, in ASP Conf. Ser. 238: Astronomical Data Analysis
  Software and Systems X {The SDSS Imaging Pipelines}.
pp 269--+

\bibitem[\protect\citeauthoryear{{Ma}, {Hu} \& {Huterer}}{{Ma}
  et~al.}{2006}]{2006ApJ...636...21M}
{Ma} Z.,  {Hu} W.,    {Huterer} D.,  2006, \apj, 636, 21

\bibitem[\protect\citeauthoryear{{Madgwick} et~al.}{2003}]{2003ApJ...599..997M}
{Madgwick} D.~S. et~al. 
  2003, \apj, 599, 997

\bibitem[\protect\citeauthoryear{{Mandelbaum}
  et~al.}{2005}]{2005MNRAS.361.1287M}
{Mandelbaum} R. et~al.  2005,
  \mnras, 361, 1287

\bibitem[\protect\citeauthoryear{{Mandelbaum} et~al.}{2006a}]{2006MNRAS.370.1008M}
{Mandelbaum} R.,  {Hirata} C.~M.,  {Broderick} T.,  {Seljak} U.,    {Brinkmann}
  J.,  2006a, \mnras, 370, 1008

\bibitem[\protect\citeauthoryear{{Mandelbaum} et~al.}{2006b}]{2006MNRAS.372..758M}
{Mandelbaum} R.,  {Seljak} U.,  {Cool} R.~J.,  {Blanton} M.,  {Hirata} C.~M.,
   {Brinkmann} J.,  2006b, \mnras, 372, 758

\bibitem[\protect\citeauthoryear{{Mandelbaum} et~al.}{2006c}]{2006MNRAS.368..715M}
{Mandelbaum} R.,  {Seljak} U.,  {Kauffmann} G.,  {Hirata} C.~M.,    {Brinkmann}
  J.,  2006c, \mnras, 368, 715

\bibitem[\protect\citeauthoryear{{Mandelbaum} \& {Seljak}}{{Mandelbaum} \&
  {Seljak}}{2007}]{2007JCAP...06...24M}
{Mandelbaum} R.,  {Seljak} U.,  2007, JCAP, 6, 24

\bibitem[\protect\citeauthoryear{{Massey} et~al.}{2007}]{2007MNRAS.376...13M}
{Massey} R. et~al.  2007, \mnras, 376, 13

\bibitem[\protect\citeauthoryear{{McKay} et~al.}{2001}]{2001astro.ph..8013M}
{McKay} T.~A. et~al.  2001, astro-ph/0108013

\bibitem[\protect\citeauthoryear{{Mobasher} et~al.}{2007}]{2007ApJS..172..117M}
{Mobasher} B. et~al. 
   2007, \apjs, 172, 117

\bibitem[\protect\citeauthoryear{{Natarajan}, {Kneib} \& {Smail}}{{Natarajan}
  et~al.}{2002}]{2002ApJ...580L..11N}
{Natarajan} P.,  {Kneib} J.-P.,    {Smail} I.,  2002, \apjl, 580, L11

\bibitem[\protect\citeauthoryear{{Oyaizu} et~al.}{2007}]{2007arXiv0708.0030O}
{Oyaizu} H.,  {Lima} M.,  {Cunha} C.~E.,  {Lin} H.,  {Frieman} J.,    {Sheldon}
  E.~S.,  2007, arXiv:0708.00300

\bibitem[\protect\citeauthoryear{{Padmanabhan} et~al.}{2005}]{2005MNRAS.359..237P}
{Padmanabhan} N. et~al. 2005, \mnras, 359, 237

\bibitem[\protect\citeauthoryear{{Pier} et~al.}{2003}]{2003AJ....125.1559P}
{Pier} J.~R.,  {Munn} J.~A.,  {Hindsley} R.~B.,  {Hennessy} G.~S.,  {Kent}
  S.~M.,  {Lupton} R.~H.,    {Ivezi{\' c}} {\v Z}.,  2003, \aj, 125, 1559

\bibitem[\protect\citeauthoryear{{Richards} et~al.}{2002}]{2002AJ....123.2945R}
{Richards} G.~T. et~al.  2002, \aj, 123,
  2945

\bibitem[\protect\citeauthoryear{{Scoville} et~al.}{2007a}]{2007ApJS..172...38S}
{Scoville} N. et~al.  2007a, \apjs, 172, 38

\bibitem[\protect\citeauthoryear{{Scoville} et~al.}{2007b}]{2007ApJS..172....1S}
{Scoville} N. et~al. 2007b, \apjs, 172, 1

\bibitem[\protect\citeauthoryear{{Seljak} et~al.}{2005}]{2005PhRvD..71d3511S}
{Seljak} U. et~al.  2005, \prd, 71, 043511

\bibitem[\protect\citeauthoryear{{Sheldon} et~al.}{2004}]{2004AJ....127.2544S}
{Sheldon} E.~S. et~al. 2004, \aj, 127, 2544

\bibitem[\protect\citeauthoryear{{Smith} et~al.}{2001}]{2001ApJ...551..643S}
{Smith} D.~R.,  {Bernstein} G.~M.,  {Fischer} P.,    {Jarvis} M.,  2001, \apj,
  551, 643

\bibitem[\protect\citeauthoryear{{Smith} et~al.}{2002}]{2002AJ....123.2121S}
{Smith} J.~A. et~al.  2002, \aj, 123, 2121

\bibitem[\protect\citeauthoryear{{Stoughton} et~al.}{2002}]{2002AJ....123..485S}
{Stoughton} C. et~al. 2002, \aj, 123, 485

\bibitem[\protect\citeauthoryear{{Strauss} et~al.}{2002}]{2002AJ....124.1810S}
{Strauss} M.~A. et~al. 
  2002, \aj, 124, 1810

\bibitem[\protect\citeauthoryear{{Taniguchi} et~al.}{2007}]{2007ApJS..172....9T}
{Taniguchi} Y. et~al. 2007,
  \apjs, 172, 9

\bibitem[\protect\citeauthoryear{{Tucker} et~al.}{2006}]{2006AN....327..821T}
{Tucker} D.~L. et~al.  2006, Astronomische Nachrichten, 327, 821

\bibitem[\protect\citeauthoryear{{Tyson} et~al.}{1984}]{1984ApJ...281L..59T}
{Tyson} J.~A.,  {Valdes} F.,  {Jarvis} J.~F.,    {Mills} A.~P.,  1984, \apjl,
  281, L59

\bibitem[\protect\citeauthoryear{{York} et~al.}{2000}]{2000AJ....120.1579Y}
{York} D.~G. et~al.  2000, \aj,
  120, 1579

\bibitem[\protect\citeauthoryear{{Zehavi}
    et~al.}{2002}]{2002ApJ...571..172Z} {Zehavi} I. et~al. 2002, \apj,
    571, 172

\bibitem[\protect\citeauthoryear{{Zehavi}
    et~al.}{2005}]{2005ApJ...630....1Z} {Zehavi} I. et~al. 2005, \apj,
    630, 1

\end{thebibliography}
\bibliographystyle{mn2e}

\appendix

\section{Extension to other lensing measurements}\label{S:extension}

In this paper we have demonstrated the lensing calibration using SDSS g-g
lensing data 
with lens redshifts.  Here we discuss the extension of this
analysis to other lensing scenarios, particularly
\begin{enumerate}
\item Galaxy-galaxy lensing with lens photoz's instead of
  spectroscopic redshifts; 
\item Galaxy-galaxy lensing with redshift distributions for both
  lenses and sources; and 
\item Cosmic shear (shear-shear autocorrelations) with photoz's or
  redshift distributions for the source sample.
\end{enumerate}

We discuss the first case on its own, and the second and third
together.

\subsection{G-g lensing with lens photoz's}

The first case, g-g lensing with photoz's for the lenses, involves the
same lensing formalism as for g-g lensing with spectroscopic
redshifts.  We simply require an additional spectroscopic calibration sample for
the lenses to trace their photoz error distribution.  However, in
addition to the multiplicative calibration 
bias $b_z$ (Eqs.~\ref{E:defbz} and~\ref{E:avgbz}) which will now
include contributions from the lens photoz error distribution, the
increased variance due to non-optimal weighting
(Eq.~\ref{E:varratio}), and the systematic calibration uncertainty to
the sampling variance in the calibration sample, there is one
additional effect to consider.

The conversion to transverse separation $R$, used to bin the stacked sources
for comparison against theoretical predictions, depends on the lens
redshift.  In our formalism, which uses comoving coordinates,
$R=\theta_{ls}D_A(z_l)(1+z_l)$, where $\theta_{ls}$ is the angular
separation between the lens and source in radians.  When using
photoz's for lenses, we can define an estimated separation $\tilde{R}$
determined using the lens photoz.  Consequently, the
measured lensing signal $\widetilde{\Delta\Sigma}(\tilde{R})$ can be
expressed as an integral over the photoz error distribution:
\begin{equation}
\widetilde{\Delta\Sigma}(\tilde{R}) = \int_{0}^{\infty} \widetilde{\Delta\Sigma}(R)
p_L(\tilde{R}|R) \,dR 
\end{equation}
where $p_L(\tilde{R}|R)$ represents the probability, given the lens
photoz error distribution, that a source at separation $R$ will be put
at estimated separation $\tilde{R}$.  This probability can be obtained
trivially from the lens photoz error distribution expressed as $p_L(z_p|z)$
using the transformation from redshift to transverse separation and
the derivative $\rmd R/\rmd z$.  Even for relatively simple models for
$\Delta\Sigma$ and $p_L(z_p|z)$ (e.g., power-law and Gaussian,
respectively) this integral does not reduce to a simple analytic
expression. 

Note that this effect is more 
pernicious in some ways than a pure calibration error, since the
effect depends on the scale-dependence of the true lensing signal
$\Delta\Sigma$.  This error must be treated differently than a
pure calibration error: rather than changing the computation of the
signal by incorporating a calibration factor, this error must be
incorporated at the interpretation step of the analysis, when some
model is used to predict $\Delta\Sigma$.  At that stage, the
additional step of numerically convolving the prediction with
$p_L(\tilde{R}|R)$  
can be included before comparing against the data.  The convolution
will change the prediction, and also induce some theoretical
uncertainty depending on the statistical $+$ sampling variance
uncertainty on $p_L(\tilde{R}|R)$.  That theoretical uncertainty in
the model prediction can be determined by using $p_L(\tilde{R}|R)$
from many realizations of the data to get
$\widetilde{\Delta\Sigma}(\tilde{R})$ and fit for the model parameters
on each realization.

\subsection{Redshift distributions for g-g lensing and cosmic shear}

The case of galaxy-galaxy lensing with a redshift distribution used
for both lenses and sources, and the case of cosmic shear, are similar
in several important aspects.  In both cases, the observed signal is
typically expressed as a function of shears as a function of angular
separation (angle $\theta$ or multipole $\ell$).   Most work either
does not incorporate redshift information, or uses tomographic cosmic
shear in which the photoz's are used to separate the source sample
into several bins, with shear-shear autocorrelation functions measured
in each bin (and cross-correlation functions measured between
bins). The full redshift information ($\rmd N/\rmd z$, or $\rmd N/\rmd
z$ for each
bin) is then incorporated at the interpretation stage of the analysis,
when a model for the 
signal (i.e., $\Delta\Sigma(R)$ in case 2 or the convergence power
spectrum in case 3) is transformed to the form of the observable to
fit for the model parameters.  In general, errors in the redshift
distributions can lead to nontrivial changes in this prediction -- not pure
calibration bias, but some change with scale dependence.  The choice
of the wrong redshift distribution therefore leads to the selection of
the wrong model parameters because the theoretical predictions have
been computed in the wrong way.  Here we assume that a spectroscopic
training sample is being used to obtain the proper source redshift
distribution in the mean, but we would like to determine the uncertainty in the
model parameters due to Poisson $+$ sampling variance uncertainty in
the source redshift distribution.

In practice, this uncertainty can be trivially included in the
analysis using
modifications of the procedures described for galaxy-galaxy lensing
with lens redshifts.  
For example, for g-g lensing without lens or source redshift, one can
use spectroscopic training samples with the same selection as the
lens and source samples to create redshift histograms and fit
them to some functional form for many bootstrap resamplings of the
redshift histogram pairs ($z_i,N_i$). One can
then generate the theoretical prediction for each of the many
realizations of the 
best-fit redshift histogram, and fit for the model parameters
on each one to see how much they vary due to the changes in the
redshift histogram from realization to realization.  For cosmic shear,
this procedure can be adopted using a single spectroscopic calibration
sample that is comparable to the source sample.  The Poisson and
LSS uncertainty in the redshift histograms will therefore be
propagated to uncertainties on the model parameters. 

\end{document}